\documentclass[preprint]{imsart}

\RequirePackage[OT1]{fontenc}
\RequirePackage{amsthm,amsmath}
\RequirePackage[numbers]{natbib}
\RequirePackage[colorlinks,citecolor=blue,urlcolor=blue]{hyperref}

\startlocaldefs
\usepackage{amsfonts,amsmath,amssymb,amsthm}
\usepackage{mathrsfs}
\usepackage[active]{srcltx}
\usepackage{enumitem} 
\usepackage{aliascnt,bbm}
\usepackage[textwidth=3cm,textsize=footnotesize]{todonotes}
\usepackage{xargs}
\usepackage[Symbolsmallscale]{upgreek}
\usepackage{array,multirow,makecell}
\setcellgapes{4pt}
\makegapedcells
\newcolumntype{R}[1]{>{\raggedleft\arraybackslash }b{#1}}
\newcolumntype{L}[1]{>{\raggedright\arraybackslash }b{#1}}
\newcolumntype{C}[1]{>{\centering\arraybackslash }b{#1}}

\usepackage[]{algorithm2e}

\usepackage{cleveref}
\makeatletter
\newtheorem{theorem}{Theorem}
\crefname{theorem}{theorem}{Theorems}
\Crefname{Theorem}{Theorem}{Theorems}

\newaliascnt{lemma}{theorem}
\newtheorem{lemma}[lemma]{Lemma}
\aliascntresetthe{lemma}
\crefname{lemma}{lemma}{lemmas}
\Crefname{Lemma}{Lemma}{Lemmas}

\newaliascnt{corollary}{theorem}
\newtheorem{corollary}[corollary]{Corollary}
\aliascntresetthe{corollary}
\crefname{corollary}{corollary}{corollaries}
\Crefname{Corollary}{Corollary}{Corollaries}

\newaliascnt{proposition}{theorem}
\newtheorem{proposition}[proposition]{Proposition}
\aliascntresetthe{proposition}
\crefname{proposition}{proposition}{propositions}
\Crefname{Proposition}{Proposition}{Propositions}

\newaliascnt{definition}{theorem}

\aliascntresetthe{definition}
\crefname{definition}{definition}{definitions}
\Crefname{Definition}{Definition}{Definitions}

\newaliascnt{definition-proposition}{theorem}

\aliascntresetthe{definition-proposition}
\crefname{definition-proposition}{definition-proposition}{definitions-propositions}
\Crefname{Definition-Proposition}{Definition-Proposition}{Definitions-Propositions}

\newaliascnt{remark}{theorem}

\aliascntresetthe{remark}
\crefname{remark}{remark}{remarks}
\Crefname{Remark}{Remark}{Remarks}

\crefname{example}{example}{examples}
\Crefname{Example}{Example}{Examples}

\crefname{figure}{figure}{figures}
\Crefname{Figure}{Figure}{Figures}

\newtheorem{assumption}{\textbf{H}\hspace{-3pt}}
\Crefname{assumption}{\textbf{H}\hspace{-3pt}}{\textbf{H}\hspace{-3pt}}
\crefname{assumption}{\textbf{H}}{\textbf{H}}

\Crefname{condition}{\textbf{C}\hspace{-3pt}}{\textbf{C}\hspace{-3pt}}
\crefname{condition}{\textbf{C}}{\textbf{C}}

\Crefname{probleme}{\textbf{Problem}\hspace{-3pt}}{\textbf{Problem}\hspace{-3pt}}
\crefname{probleme}{\textbf{Problem}}{\textbf{Problem}}

\Crefname{assumptionG}{\textbf{G}\hspace{-3pt}}{\textbf{G}\hspace{-3pt}}
\crefname{assumptionG}{\textbf{G}}{\textbf{G}}

\Crefname{assumptionV}{\textbf{V}\hspace{-3pt}}{\textbf{G}\hspace{-3pt}}
\crefname{assumptionV}{\textbf{V}}{\textbf{V}}

\newtheorem{assumptionL}{\textbf{L}\hspace{-3pt}}
\Crefname{assumptionL}{\textbf{L}\hspace{-3pt}}{\textbf{L}\hspace{-3pt}}
\crefname{assumptionL}{\textbf{L}}{\textbf{L}}

\makeatother

\def\convSet{\mathsf{K}}

\newcommand{\ball}[2]{\operatorname{B}(#1,#2)}
\def\vol{\operatorname{Vol}}

\def\complementaire{\operatorname{c}}
\def\borelSet{\mathcal{B}}
\def\Ccal{\mathcal{C}}

\def\e{\mathrm{e}}

\def\Z{Z}
\def\Zhat{\hat{\Z}}
\newcommand{\suite}[3]{\{#1_{i}\}_{i=#2}^{#3}}
\newcommand{\suiteD}[3]{\{#1\}_{#2}^{#3}}
\newcommand{\m}{m}
\newcommand{\Llip}{L}

\def\mi{\m_i}
\def\Ui{\U[i]}
\def\Zi{\Z_i}
\def\pii{\pi_i}
\def\Li{\Llip_i}
\def\gi{g_i}
\def\ai{a_i}
\def\gami{\gamma_i}
\newcommand{\XEi}[1]{X_{i,#1}}
\def\sig2{\sigma^2}
\def\Ri{R_i}
\def\pihati{\hat{\pi}_{i}}
\def\ni{n_i}
\def\Ni{N_i}
\def\MSEi{\MSE_i}
\def\kappai{\kappa_i}
\def\sigDi{\sigma^2_i}
\def\alphai{\alpha_i}
\def\betai{\beta_i}
\def\gamibar{\bar{\gamma}_i}
\def\nibar{\bar{n}_i}
\def\Nibar{\bar{N}_i}

\def\simu{\mathcal{S}}
\def\Asimu{\boreleanA_{\simu,\epsilon}}

\def\gM{g_{M-1}}

\def\mMs{\m_{\Ms-1}}
\def\nMs{n_{\Ms-1}}

\def\NMs{N_{\Ms-1}}
\def\kappaMs{\kappa_{\Ms-1}}
\def\gamMs{\gamma_{\Ms-1}}
\def\LMs{L_{\Ms-1}}
\def\alphaMs{\alpha_{\Ms-1}}
\def\betaMs{\beta_{\Ms-1}}

\def\gamMsbar{\bar{\gamma}_{\Ms-1}}
\def\nMsbar{\bar{n}_{\Ms-1}}
\def\NMsbar{\bar{N}_{\Ms-1}}

\def\mMc{\m_{\Mc-1}}
\def\nMc{n_{\Mc-1}}

\def\NMc{N_{\Mc-1}}
\def\kappaMc{\kappa_{\Mc-1}}
\def\gamMc{\gamma_{\Mc-1}}
\def\LMc{L_{\Mc-1}}

\def\gamMcbar{\bar{\gamma}_{\Mc-1}}
\def\nMcbar{\bar{n}_{\Mc-1}}
\def\NMcbar{\bar{N}_{\Mc-1}}

\newcommand{\LLip}[1]{L_{#1}}
\newcommand{\pihat}{\hat{\pi}_{n}^{N}}

\def\tildL{\tilde{L}}

\newcommand{\U}[1][]{U_{#1}}

\newcommand{\fracUn}[1]{(#1)^{-1}}
\newcommand{\g}[1][]{g_{#1}}

\def\Fc{F}

\newcommand{\Aci}[1]{A_{i,#1}}
\newcommand{\Bci}[1]{B_{i,#1}}
\newcommand{\Cci}[1]{C_{i,#1}}

\newcommand{\AcMc}[1]{A_{\Mc-1,#1}}
\newcommand{\BcMc}[1]{B_{\Mc-1,#1}}

\newcommand{\AcMs}[1]{A_{\Ms-1,#1}}

\newcommand{\CcMs}[1]{C_{\Ms-1,#1}}

\newcommand{\Ac}[1]{A_{#1}}
\newcommand{\Bc}[1]{B_{#1}}

\def\piV{\pi}

\def\rkerv{R_{\gamma}}

\newcommand{\rhoUn}{\rho_1}
\newcommand{\rhoDeux}{\rho_2}
\def\J{J}
\def\rayon{D}
\def\gbarM{\bar{g}_{M-1}}
\def\gbarMc{\bar{g}_{\Mc-1}}
\def\pihatM{\hat{\pi}_{M-1}}

\def\piM{\pi_{M-1}}
\newcommand{\Lip}[1]{\norm{#1}_{\operatorname{Lip}}}

\def\cost{\operatorname{cost}}
\def\inte{\operatorname{int}}

\newcommand{\nbDeux}[1][]{K_{#1}}
\newcommand{\morc}[1][]{\mathcal{I}_{#1}}
\newcommand{\cmorc}[1][]{\absolute{\mathcal{I}_{#1}}}

\newcommand{\ray}{\mathsf{R}}
\def\Ms{M}
\def\Mc{M}

\newcommand{\intercept}{\operatorname{intercept}}
\newcommand{\mo}{\mathcal{M}}
\newcommand{\BF}{BF}
\newcommand{\diag}{\operatorname{diag}}
\newcommand{\NP}{\operatorname{NP}}
\newcommand{\PGC}{\operatorname{PGC}}
\newcommand{\BP}{\operatorname{BP}}
\newcommand{\TST}{\operatorname{TST}}
\newcommand{\BMI}{\operatorname{BMI}}
\newcommand{\DP}{\operatorname{DP}}
\newcommand{\AGE}{\operatorname{AGE}}
\newcommand{\XUn}{X^{(1)}}
\newcommand{\XDeux}{X^{(2)}}
\newcommand{\Xk}{X^{(k)}}
\newcommand{\transpose}{\operatorname{T}}

\def\gaStep{\gamma}
\def\GaStep{\Gamma}
\def\RKer{R}
\def\QKer{Q}

\def\Tr{\operatorname{T}}
\def\trace{\operatorname{Tr}}
\newcommandx{\functionspace}[2][1=+]{\mathbb{F}_{#1}(#2)}

\newcommand{\Var}[1]{\operatorname{Var}\left[#1 \right]}
\newcommand{\estimateur}[1]{\hat{\pi}_n^N(#1)}

\newcommandx{\VarDeux}[3][3=]{\operatorname{Var}^{#3}_{#1}\left[#2 \right]}
\newcommand{\VarDeuxLigne}[2]{\operatorname{Var}_{#1}[#2 ]}

\newcommand{\1}{\mathbbm{1}}

\newcommand{\LeftEqNo}{\let\veqno\@@leqno}

\newcommand{\ceil}[1]{\left\lceil #1 \right\rceil}

\newcommand{\N}{\ensuremath{\mathbb{N}}}

\newcommand{\PE}{\mathbb{E}}
\newcommand{\PP}{\mathbb{P}}

\newcommand{\absolute}[1]{\left\vert #1 \right\vert}
\newcommand{\abs}[1]{\left\vert #1 \right\vert}

\newcommandx{\Vnorm}[2][1=V]{\| #2 \|_{#1}}
\newcommandx{\VnormEq}[2][1=V]{\left\| #2 \right\|_{#1}}
\newcommandx{\norm}[2][1=]{\ifthenelse{\equal{#1}{}}{\left\Vert #2 \right\Vert}{\left\Vert #2 \right\Vert^{#1}}}

\newcommandx{\normLigne}[2][1=]{\ifthenelse{\equal{#1}{}}{\Vert #2 \Vert}{\Vert #2\Vert^{#1}}}

\newcommand{\parenthese}[1]{\left(#1 \right)}

\newcommand{\defEns}[1]{\left\lbrace #1 \right\rbrace }

\newcommand{\ps}[2]{\left\langle#1,#2 \right\rangle}
\newcommand{\eqdef}{\overset{\text{\tiny def}} =}

\newcommand{\proba}[1]{\mathbb{P}\left( #1 \right)}

\newcommandx\probaMarkovTilde[2][2=]
{\ifthenelse{\equal{#2}{}}{{\widetilde{\mathbb{P}}_{#1}}}{\widetilde{\mathbb{P}}_{#1}\left[ #2\right]}}

\newcommand{\expe}[1]{\PE \left[ #1 \right]}

\newcommand{\expeMarkov}[2]{\PE_{#1} \left[ #2 \right]}

\newcommand{\bigO}{\ensuremath{\mathcal O}}
\newcommand{\softO}{\Tilde{\ensuremath{\mathcal O}}}

\newcommand{\couplage}[2]{\Pi(#1,#2)}

\newcommand{\plusinfty}{+\infty}

\newcounter{hypoconbis}
\newcounter{saveconbis}
\newcommand\debutH{\begin{list}
{\textbf{H\arabic{hypoconbis}}}{\usecounter{hypoconbis}}\setcounter{hypoconbis}{\value{saveconbis}}}
\newcommand\finH{\end{list}\setcounter{saveconbis}{\value{hypoconbis}}}

\def\ie{\textit{i.e.}}
\def\eqsp{\;}
\newcommand{\coint}[1]{\left[#1\right)}
\newcommand{\ocint}[1]{\left(#1\right]}
\newcommand{\ooint}[1]{\left(#1\right)}
\newcommand{\ccint}[1]{\left[#1\right]}

\newcommandx{\weight}[2][2=n]{\omega_{#1,#2}^N}

\def\rmd{\mathrm{d}}
\newcommandx\sequence[3][2=,3=]
{\ifthenelse{\equal{#3}{}}{\ensuremath{\{ #1_{#2}\}}}{\ensuremath{\{ #1_{#2}, \eqsp #2 \in #3 \}}}}
\newcommandx{\sequencen}[2][2=n\in\N]{\ensuremath{\{ #1, \eqsp #2 \}}}
\newcommandx\sequenceDouble[4][3=,4=]
{\ifthenelse{\equal{#3}{}}{\ensuremath{\{ (#1_{#3},#2_{#3}) \}}}{\ensuremath{\{  (#1_{#3},#2_{#3}), \eqsp #3 \in #4 \}}}}
\newcommandx{\sequencenDouble}[3][3=n\in\N]{\ensuremath{\{ (#1_{n},#2_{n}), \eqsp #3 \}}}
\newcommand{\wrt}{w.r.t.}

\def\iid{i.i.d.}
\def\rme{\mathrm{e}}

\def\eg{e.g.}

\def\rset{\mathbb{R}}

\def\nset{\mathbb{N}}

\def\MSE{\operatorname{MSE}}

\newcommandx{\CPE}[3][1=]{{\mathbb E}^{#3}_{#1}\left[#2 \right]} 
\newcommandx{\CPVar}[3][1=]{\mathrm{Var}^{#3}_{#1}\left[ #2 \right]}
\newcommand{\CPP}[3][]
{\ifthenelse{\equal{#1}{}}{{\mathbb P}\left(\left. #2 \, \right| #3 \right)}{{\mathbb P}_{#1}\left(\left. #2 \, \right | #3 \right)}}

\def\generator{\mathscr{A}}

\newcommandx{\osc}[2][1=]{\mathrm{osc}_{#1}(#2)}

\def\mcg{\mathcal{G}}

\newcommand{\chunk}[4][]%
{\ifthenelse{\equal{#1}{}}{\ensuremath{{#2}_{#3:#4}}}{\ensuremath{#2^#1}_{#3:#4}}
}

\def\Id{\operatorname{Id}}

\def\boreleanA{\mathrm{A}}

\def\martInc{\Phi}
\def\martIncF{\Psi}

\endlocaldefs

\begin{document}

\begin{frontmatter}

\title{Normalizing constants of log-concave densities}
\runtitle{Normalizing constants of log-concave densities}


\author{\fnms{Nicolas} \snm{Brosse}\ead[label=e1]{nicolas.brosse@polytechnique.edu}}
\address{Centre de Math\'ematiques Appliqu\'ees, UMR 7641, Ecole Polytechnique, France. \\ \printead{e1}}
\author{\fnms{Alain} \snm{Durmus}\ead[label=e2]{alain.durmus@cmla.ens-cachan.fr}}
\address{Ecole Normale Sup\'erieure CMLA, 61 Av. du Pr\'esident Wilson 94235 Cachan Cedex, France. \\ \printead{e2}}

\and
\author{\fnms{\'Eric} \snm{Moulines}\ead[label=e3]{eric.moulines@polytechnique.edu}}
\address{Centre de Math\'ematiques Appliqu\'ees, UMR 7641, Ecole Polytechnique, France. \\ \printead{e3}}

\runauthor{N.Brosse, A.Durmus, \'E.Moulines}

\begin{abstract}
  We derive explicit bounds for the computation of normalizing
  constants $\Z$ for log-concave densities $\pi =
  \e^{-\U}/\Z$ \wrt~the Lebesgue measure on $\rset^d$. Our approach
  relies on a Gaussian annealing combined with recent and precise
  bounds on the Unadjusted Langevin Algorithm
  \cite{durmusSampling}. Polynomial bounds in the dimension $d$ are
  obtained with an exponent that depends on the assumptions made on
  $\U$.
  The algorithm also provides a theoretically grounded choice of the
  annealing sequence of variances.
  A numerical
  experiment supports our findings.  Results of independent interest
  on the mean squared error of the empirical average of locally Lipschitz functions are
  established.
\end{abstract}

\begin{keyword}[class=MSC]
\kwd[Primary ]{65C05}
\kwd{60F25}
\kwd{62L10}
\kwd[; secondary ]{65C40}
\kwd{60J05}
\kwd{74G10}
\kwd{74G15}
\end{keyword}

\begin{keyword}
\kwd{Normalizing constants}
\kwd{Bayes Factor}
\kwd{Annealed Importance Sampling}
\kwd{Unadjusted Langevin Algorithm}
\end{keyword}



\end{frontmatter}

\section{Introduction}
\label{sec:introduction}

Let $\U :\rset^d \to \rset$ be a continuously differentiable convex function such that $\Z = \int_{\rset^d} \e^{-\U (x)} \rmd x < \plusinfty$.
$\Z$ is the normalizing constant of the probability density $\pi$ associated with the potential $\U$, defined for $x\in\rset^d$ by $\pi(x)=\Z^{-1} \rme^{-U(x)}$.
We discuss in this paper a method to estimate $\Z$ with polynomial complexity in the dimension $d$.

Computing the normalizing constant is a challenge which has applications in Bayesian inference and statistical physics in particular. In statistical physics, $\Z$ is better known under the name of partition function or free energy \cite{balian2007microphysics}, \cite{lelievre2010free}. Free energy differences allow to quantify the relative likelihood of different states (microscopic configurations) and are linked to thermodynamic work and heat exchanges. In Bayesian inference, the models can be compared by the computation of the Bayes factor which is the ratio of two normalizing constants (see \eg~\cite[chapter 7]{robert2007bayesian}). This problem has consequently attracted a wealth of contribution; see for example \cite[chapter 5]{chenmonte}, \cite{marin2009importance}, \cite{friel2012estimating}, \cite{ardia:2012}, \cite{dutta2013bayes}, \cite{Knuth2015}, \cite{zhou2015towards} and, for a more specific molecular simulations flavor, \cite{lelievre2010free}.
Compared to the large number of proposed methods to estimate $\Z$, few theoretical guarantees have been obtained on the ouput of these algorithms; see below for further references and comments.
Our algorithm relies on a sequence of Gaussian densities with increasing variances, combined with the precise bounds of \cite{durmusSampling}.

The paper is organized as follows. The outline of the algorithm is first described, followed by the assumptions made on $\U$. Our main results are then stated and compared to previous works on the subject. The theoretical analysis of the algorithm is detailed in \Cref{sec:analysis-algorithm}.
In \Cref{sec:numerical-experiments}, a numerical experiment is provided to support our theoretical claims.
Finally, the proofs are gathered in \Cref{sec:proofs}.
In \Cref{sec:MSE-loc-lipschitz}, a result of independent interest on the mean squared error of the empirical average of locally Lipschitz functions is established.


%
%

\subsection*{Notations and conventions}

Denote by $\mathcal{B}(\rset^d)$ the Borel $\sigma$-field of $\rset^d$.
For $\mu$ a probability measure on $(\rset^d, \mathcal{B}(\rset^d))$ and $f$ a $\mu$-integrable function, denote by $\mu(f)$ the integral of $f$ \wrt~$\mu$.
We say that $\zeta$ is a transference plan of $\mu$ and $\nu$ if it is a probability measure on $(\rset^d \times \rset^d, \mathcal{B}(\rset^d \times \rset^d) )$  such that for all measurable sets $\boreleanA$ of $\rset^d$, $\zeta(\boreleanA \times \rset^d) = \mu(\boreleanA)$ and $\zeta(\rset^d \times \boreleanA) = \nu(\boreleanA)$. We denote by $\couplage{\mu}{\nu}$ the set of transference plans of $\mu$ and $\nu$.
Furthermore, we say that a couple of $\rset^d$-random variables $(X,Y)$ is a coupling of $\mu$ and $\nu$ if there exists $\zeta \in \couplage{\mu}{\nu}$ such that $(X,Y)$ are distributed according to $\zeta$.
For two probability measures $\mu$ and $\nu$, we define the Wasserstein distance of order $p \geq 1$ as
\begin{equation}
\label{eq:definition_wasserstein}
W_p(\mu,\nu) \eqdef \left( \inf_{\zeta \in \couplage{\mu}{\nu}} \int_{\rset \times \rset} \norm[p]{x-y}\rmd \zeta (x,y)\right)^{1/p} \eqsp.
\end{equation}
By \cite[Theorem 4.1]{villani2008optimal}, for all $\mu,\nu$ probability measure on $\rset^d$, there exists a transference plan $\zeta^\star \in \couplage{\mu}{\nu}$ such that the infimum in \eqref{eq:definition_wasserstein} is reached in $\zeta^\star$.
$\zeta^\star$ is called an optimal transference plan associated with $W_p$.

$f : \rset^d \to \rset$ is a Lipschitz function if there exists $C \geq 0$ such that for all $x,y \in \rset^d$, $ \abs{f(x) - f(y)} \leq C \norm{x-y}$. Then we denote
\[ \Lip{f} = \sup \{  \abs{f(x) - f(y)} \norm[-1]{x-y} \ | \ x,y \in \rset^d , x \not = y \} \eqsp. \]
For $k\in\nset$, $\Ccal^k(\rset^d)$ denotes the set of $k$-continuously differentiable functions $\rset^d \to \rset$, with the convention that $\Ccal^0(\rset^d)$ is the set of continuous functions. Let $\nset^* = \nset \setminus \{0\}$, $n,m \in \nset^*$ and $F : \rset^n \to \rset^m$ be a twice continuously differentiable function. Denote by $\nabla F$ and $\nabla^2 F$ the Jacobian and the Hessian of $F$ respectively.
For $m=1$, the Laplacian is defined by $\Delta F = \trace \nabla^2 F$ where $\trace$ is the trace operator.
In the sequel, we take the convention that for $n,p \in \nset$, $n <p$ then $\sum_{p}^n =0$ and $\prod_p ^n = 1$.
By convention, $\inf \defEns{\emptyset} = \plusinfty$, $\sup \defEns{\emptyset} = -\infty$ and for $j>i$ in $\mathbb{Z}$, $\defEns{j,\ldots,i}=\emptyset$.
For a finite set $E$, $\absolute{E}$ denotes the cardinality of $E$.
For $a,b\in\rset$, $a\wedge b = \min(a,b)$ and $a\vee b = \max(a,b)$.
Let $\psi, \phi:\rset_+^\star \to \rset_+^\star$. We write $\psi = \softO(\phi)$ if there exists $t_0>0$, $C,c>0$ such that $\psi(t) \leq C \phi(t) \abs{\log t}^c$ for all $t\in\ocint{0,t_0}$.
Denote by $\ball{x}{r}=\defEns{y\in\rset^d : \norm{y-x} \leq r}$.


\subsection*{Presentation of the algorithm}

Since $\Z<\plusinfty$ and $\U$ is convex, by \cite[Lemma 2.2.1]{brazitikos2014geometry}, there exist constants $\rhoUn>0$ and $\rhoDeux\in\rset$ such that $U(x) \geq \rhoUn \norm{x} -\rhoDeux $. Therefore, by continuity, $\U$ has a minimum $x^\star$. Without loss of generality, it is assumed in the sequel that $x^\star=0$ and $\U(x^\star)=0$.

Let $M \in \nset^{\star}$, $\suite{\sig2}{0}{M}$ be a positive increasing sequence of real numbers and set $\sig2_M = \plusinfty$.
Consider the sequence of functions $\suite{U}{0}{M}$ defined for all $i \in \{0,\ldots, M\}$ and $x \in \rset^d$ by
\begin{equation}\label{eq:defUi}
\U[i](x) = \frac{\norm{x}^2}{2\sigma_i^2}+\U(x) \eqsp,
\end{equation}
with the convention $1/\infty = 0$.
We define a sequence of probability densities $\suite{\pi}{0}{M}$ for $i \in\{0,  \ldots, M\}$ and $x \in \rset^d$ by
\begin{equation}\label{eq:defpii}
\pii(x) = \Zi^{-1}\e^{-\Ui(x)} \eqsp, \qquad \Zi = \int_{\rset^d} \e^{-\Ui(y)} \rmd y \eqsp.
\end{equation}
The dependence of $\Zi$ in $\sigDi$ is implicit.
By definition, note that $\U[M] = \U$, $\Z_M=\Z$ and $\pi_M = \pi$.
As in the multistage sampling method \cite[Section 3.3]{gelman1998simulating}, we use the following decomposition
\begin{equation}\label{eq:bridgeSampling}
\frac{\Z}{\Z_0} = \prod_{i=0}^{M-1}\frac{\Z_{i+1}}{\Z_i}  \eqsp .
\end{equation}
$\Z_0$ is estimated by choosing $\sigma_0^2$ small enough so that $\pi_0$ is sufficiently close to a Gaussian distribution of mean $0$ and covariance $\sig2_0 \Id$.
For $i \in\{0,  \ldots, M-1\}$, the ratio $\Z_{i+1}/\Z_i$ may be expressed as
\begin{equation}\label{eq:oneRatio}
\frac{\Z_{i+1}}{\Z_i} = \int_{\rset^d} g_i(x) \pii(x) \rmd x = \pii(\gi) \eqsp ,
\end{equation}
where $g_i:\rset^d \to \rset_{+}$ is defined for all $x \in \rset^d$ by
\begin{equation}\label{eq:defgi}
g_i(x) = \exp \left( a_i \norm{x}^2 \right) \eqsp ,
\qquad a_i = \frac{1}{2}\left(\frac{1}{\sigma_i^2}-\frac{1}{\sigma_{i+1}^2}\right) \eqsp .
\end{equation}
The quantity  $\pii(\gi)$
is estimated by the Unadjusted Langevin Algorithm (ULA) targeting $\pii$.
Introduced in \cite{ermak:1975} and \cite{parisi:1981} (see also \cite{roberts:tweedie:1996}),
the ULA algorithm can be described as follows. For $i \in\{0,  \ldots, M-1\}$, the (overdamped) Langevin stochastic differential equation (SDE) is given by
\begin{equation}
\label{eq:langevin}
\rmd Y_{i,t} = -\nabla \Ui (Y_{i,t}) \rmd t + \sqrt{2} \rmd B_{i,t} \eqsp, \quad
Y_{i,0}=0 \eqsp ,
\end{equation}
where $\{(B_{i,t})_{t\geq0}\}_{i=0}^{M-1}$ are independent $d$-dimensional Brownian motions.
The sampling method is based on the Euler discretization of the Langevin diffusion, which defines a discrete-time Markov chain, for $i\in\{0,\ldots,M-1\}$ and $k \in \nset$
\begin{equation}
\label{eq:euler-proposal}
\XEi{k+1}= \XEi{k} - \gami \nabla \Ui(\XEi{k}) + \sqrt{2 \gami} W_{i, k+1} \eqsp , \quad
\XEi{0}=0 \eqsp ,
\end{equation}
where 
$\{(W_{i, k})_{k\in \nset^{\star}}\}_{i=0}^{M-1}$ are independent \iid~sequences of standard Gaussian random variables and $\gami>0$ is the stepsize.
For $i \in \{0,\ldots, M-1\}$, consider the following estimator of $\Z_{i+1}/\Zi$,
\begin{equation}\label{eq:defEstimatorZiPlusUnZi}
\pihati(\gi) = \frac{1}{\ni}\sum_{k=\Ni+1}^{\Ni+\ni} \gi(\XEi{k}) \eqsp ,
\end{equation}
where $\ni\geq 1$ is the sample size and $\Ni \geq 0$ the burn-in period.
We introduce the following assumptions on $\U$.



\begin{assumption}\label{assumption:C1GradientLipschitz}
$\U :\rset^d \to \rset$ is continuously differentiable and $L$-gradient Lipschitz, \ie~there exists $L \geq 0$ such that for all $x,y \in \rset^d$,
\begin{equation}\label{eq:LGradientLipschitz}
\norm{\nabla \U (x) - \nabla \U (y)} \leq L \norm{x-y} \eqsp.
\end{equation}
\end{assumption}

\begin{assumption}[$m$]
\label{assumption:stronglyConvex}
$\U :\rset^d \to \rset$ is continuously differentiable and satisfies for all $x,y \in \rset^d$,
\begin{equation}\label{eq:mStronglyConvex}
 \U (y) \geq \U (x)+  \ps{ \nabla \U (x)}{y-x} +  (m/2) \norm[2]{x-y} \eqsp.
\end{equation}
\end{assumption}

\begin{assumption}\label{assumption:hC3}
The function $\U$ is three times continuously differentiable and there exists $\tildL \geq 0$ such that for all $x, y \in \rset^d$
\begin{equation}\label{eq:hC3tildeL}
\norm{\nabla^2 \U(x) - \nabla^2 \U(y)} \leq \tildL \norm{x-y} \eqsp.
\end{equation}
\end{assumption}


The strongly convex case (\Cref{assumption:stronglyConvex}($\m$) with $\m > 0$) is considered in \Cref{sec:normalizingConstantStronglyConvex} and the convex case (\Cref{assumption:stronglyConvex}($\m$) with $\m = 0$) is dealt with in \Cref{sec:NormalizingconstantConvexGradientLipU}.
Assuming \Cref{assumption:C1GradientLipschitz} and \Cref{assumption:stronglyConvex}($m$) for $m\geq 0$, for $i\in\defEns{0,\ldots,M}$, $\Ui$ defined in \eqref{eq:defUi} is $\Li$-gradient Lipschitz and $\mi$-strongly convex if $\mi>0$ (and convex if $\mi=0$) where
\begin{equation}\label{eq:def-Li-mi}
\Li = L + \frac{1}{\sig2_i} \eqsp , \quad \mi = m + \frac{1}{\sig2_i} \eqsp .
\end{equation}
Define also the following useful quantities,
\begin{equation}\label{eq:defkappa}
\kappa = \frac{2mL}{m+L} \eqsp , \quad \kappai = \frac{2\mi\Li}{\mi+\Li} \eqsp .
\end{equation}
\Cref{assumption:hC3} enables to have tighter bounds on the mean squared error of $\pihati(\gi)$ defined in \eqref{eq:defEstimatorZiPlusUnZi}. Under \Cref{assumption:hC3}, for all $i\in\defEns{0,\ldots,M}$, $\Ui$ satisfies \eqref{eq:hC3tildeL} with $\tildL$.
Finally, since $\Z<\plusinfty$ and by \cite[Lemma 2.2.1]{brazitikos2014geometry}, there exist $\rhoUn>0$ and $\rhoDeux\in\rset$ such that for all $x\in\rset^d$,
\begin{equation}\label{eq:linear-growth-U}
\U(x) \geq \rhoUn \norm{x} - \rhoDeux \eqsp.
\end{equation}



Denote by $\simu$ the set of simulation parameters,
\begin{equation}\label{eq:sim-parameters}
\simu = \defEns{M, \suite{\sig2}{0}{M-1}, \suite{\gamma}{0}{M-1}, \suite{n}{0}{M-1}, \suite{N}{0}{M-1}} \eqsp ,
\end{equation}
and by $\Zhat$ the following estimator of $\Z$,
\begin{equation}\label{eq:def-Zhat}
\Zhat = (2\pi\sigma_0^2)^{d/2}(1+\sigma_0^2 \m)^{-d/2} \left\{ \prod_{i=0}^{M-1} \pihati(g_i) \right\} \eqsp ,
\end{equation}
where $\pihati(\gi)$ is defined in \eqref{eq:defEstimatorZiPlusUnZi}.
The dependence of $\Zhat$ in $\simu$ is implicit.
Note that $\Zhat$ is a biased estimator of $\Z$ because $Z_0$ is approximated by $(2\pi\sigma_0^2)^{d/2}(1+\sigma_0^2 \m)^{-d/2}$.
We define the $\cost$ of the algorithm by the total number of iterations performed by the $M$ Markov chains $(\XEi{n})_{n \geq 0}$ for $i\in\defEns{0,\ldots,M-1}$, \ie
\begin{equation}\label{eq:def-cost}
\cost = \sum_{i=0}^{M-1} \{\Ni + \ni\} \eqsp .
\end{equation}
Observe that each step of the Markov chain takes time linear in $d$.
We state below a simplified version of our results; explicit bounds are given in \Cref{thm:costAlgorithmStronglyConvex,thm:costAlgorithmStronglyConvex-UC3,thm:costAlgorithmConvex,thm:costAlgorithmConvex-UC3}. 

\begin{theorem}
\label{thm:main-results-simplified}
Assume \Cref{assumption:C1GradientLipschitz}, \Cref{assumption:stronglyConvex}($m$) for $m\geq 0$. Let $\mu,\epsilon\in\ooint{0,1}$.
There exists an explicit choice of the simulation parameters $\simu$ such that the estimator $\Zhat$ defined in \eqref{eq:def-Zhat} satisfies 
\begin{equation}\label{eq:Zhat-Z-epsilon}
\PP\parenthese{\absolute{\Zhat/\Z -1}>\epsilon} \leq \mu \eqsp.
\end{equation}
Moreover, the $\cost$ of the algorithm \eqref{eq:def-cost} is upper-bounded by,

\begin{center}
\begin{tabular}{|c|c|}
\hline
• & \Cref{assumption:C1GradientLipschitz},\Cref{assumption:stronglyConvex}($m$) for $m>0$  \\
\hline
$\cost$ & $\frac{L^3}{\mu^2 m^3} \log(d) d^3 \times \softO (\epsilon^{-4})$ \\
\hline
 & \Cref{assumption:C1GradientLipschitz},\Cref{assumption:stronglyConvex}($m$) for $m>0$,\Cref{assumption:hC3} \\
\hline
$\cost$ & $\parenthese{\frac{\tildL}{\mu^{3/2} m^{3/2}}+\frac{L^2}{\mu^{3/2} m^2}} \log(d) d^{5/2} \times \softO(\epsilon^{-3})$ \\
\hline
•  & \Cref{assumption:C1GradientLipschitz},\Cref{assumption:stronglyConvex}($m$) for $m\geq 0$ \\
\hline
$\cost$ & $\frac{ L^2}{\mu^2 \rhoUn^4}(d+\rhoDeux)^4 \log(d) d^3 \times\softO(\epsilon^{-4}) $ \\
\hline
 & \Cref{assumption:C1GradientLipschitz},\Cref{assumption:stronglyConvex}($m$) for $m\geq 0$,\Cref{assumption:hC3} \\
 \hline
 $\cost$ & $\parenthese{\frac{L^2}{\mu^{3/2}\rhoUn^4}+\frac{\tildL}{\mu^{3/2}(d+\rhoDeux) \rhoUn^3}}(d+\rhoDeux)^4 \log(d) d^{5/2} \times \softO(\epsilon^{-3}) $ \\
 \hline
\end{tabular}
\end{center}
\end{theorem}

By the median trick (see \eg~\cite[Lemma 6.1]{JERRUM1986169} or \cite{niemiro2009}), the dependence in $\mu$ of the $\cost$ can be reduced to a logarithmic factor, see \Cref{corollary:thms-median-trick-strongly,corollary:thms-median-trick-convex}.

It is interesting to compare these complexity bounds with previously reported results.
In \cite{moral:2006:SMC} and \cite{beskos:cste:2014} (see also \cite{delmoral:2004:book}), the authors propose to use sequential Monte Carlo (SMC) samplers to estimate the normalizing constant $\Z$
of a probability distribution $\pi$.
In \cite{beskos:cste:2014}, $\pi$ is supported on a compact set $\convSet$ included in $\rset^d$ and
satisfies for $x=(x_1,\ldots,x_d)\in\convSet$, $\pi(x) = \Z^{-1} \prod_{i=1}^d \exp(g(x_i))$.
\cite[Theorem 3.2]{beskos:cste:2014} states that there exists an estimator $\Zhat$ of $\Z$ such that $\lim_{d\to\plusinfty} \PE[|\Zhat/\Z-1|^2] = C/N$ where $N$ is the number of particles and $C$ depends on $g$ and on the parameters of the SMC (choice of the Markov kernel and of the annealing schedule).
With our definition \eqref{eq:def-cost}, the computational cost of the SMC algorithm is $\bigO(N d)$ (there are $d$ phases and $N$ particles for each phase).
To obtain an estimator $\Zhat$ satisfying \eqref{eq:Zhat-Z-epsilon} implies a cost of $d\mu^{-1} \bigO(\epsilon^{-2})$.
However, the product form of the density $\pi$ is restrictive,
the result is only asymptotic in $d$ and the state space is assumed to be compact.
\cite{Moral:Jasra:2016} combines SMC with a multilevel approach and
\cite{Jasra2016} establishes results on a multilevel particle filter.

%

\cite{huber2015} deals with the case where $\pi(x) = \exp(-\beta H(x))/\Z(\beta)$ where $x\in\Omega$, a finite state space, $\beta \geq 0$ and $H(x)\in\defEns{0,\ldots,n}$. These distributions known as Gibbs distributions include in particular the Ising model.
To compute $Z(\beta)$, \cite{huber2015} relies on an annealing process on the parameter $\beta$, starting from $\Z(0)$. Let $q = \log(Z(0))/\log(Z(\beta))$. \cite[Theorem 1.1]{huber2015} states that there exists an estimator $\Zhat(\beta)$ of $Z(\beta)$ such that \eqref{eq:Zhat-Z-epsilon} is satisfied with $\mu=1/4$ and $q\log(n)\softO(\epsilon^{-2})$ draws from the Gibbs distribution.

Our complexity results can also be related to the computation of the volume of a convex body $\convSet$ (compact convex set with non-empty interior) on $\rset^d$. This problem has attracted a lot of attention in the field of computer science, starting with the breakthrough of \cite{dyer1991computing} until the most recent results of \cite{cousins2015bypassing}.
Define for $x\in\rset^d$, $\pi(x)=1_{\convSet}(x) / \vol(\convSet)$. Under the assumptions $\ball{0}{1} \subset \convSet$ and $\int_{\rset^d} \norm{x}^2 \pi(x) \rmd x = \bigO(d)$, \cite[Theorem 1.1]{cousins2015bypassing} states that there exists an estimator $\Zhat$ of $\Z = \vol(\convSet)$ such that \eqref{eq:Zhat-Z-epsilon} is satisfied with $\mu=1/4$ and a cost of $\log(d) d^3\softO(\epsilon^{-2})$.

Nonequilibrium methods have been recently developed and
studied in order to compute free energy differences or Bayes factors,
see \cite{jarzynski1997equilibrium} and \cite[Chapter
4]{lelievre2010free}. They are based on an inhomogeneous diffusion
evolving (for example) from $t=0$ to $t=1$ such that $\pi_0$ and
$\pi_1$ are the stationary distributions respectively for $t=0$ and
$t=1$.  Recently, \cite{christophe:2016} provided an asymptotic and
non-asymptotic analysis of the bias and variance for estimators
associated with this methodology.
The main aim of this paper is to obtain
polynomial complexity and inspection of their results suggests a cost
of order $d^{15}$ at most to compute an estimator $\Zhat$ satisfying \eqref{eq:Zhat-Z-epsilon}. However, this cost may be due to
the strategy of proofs.


Multistage sampling type algorithms
are widely used and known under different names:
multistage sampling \cite{Valleau:1972}, (extended) bridge sampling \cite{gelman1998simulating}, annealed importance sampling (AIS) \cite{Neal2001}, thermodynamic integration \cite{girolami:2016}, power posterior \cite{Friel:2012}.
For the stability and accuracy of the method, the choice of the parameters (in our case $\suite{\sig2}{0}{M-1}$)
is crucial and is known to be difficult. Indeed, the issue has been pointed out in several articles under the names of tuning tempered transitions \cite{Friel:2012}, temperature placement \cite{Friel2014}, annealing sequence \cite[Sections 3.2.1, 4.1]{beskos:cste:2014}, temperature ladder \cite[Section 3.3.2]{girolami:2016}, effects of grid size \cite{dutta2013bayes}, cooling schedule \cite{cousins2015bypassing}. In \Cref{sec:normalizingConstantStronglyConvex,sec:NormalizingconstantConvexGradientLipU}, we explicitly define the sequence $\suite{\sig2}{0}{M-1}$.



\section{Theoretical analysis of the algorithm}
\label{sec:analysis-algorithm}

In this Section, we analyse the algorithm outlined in \Cref{sec:introduction}.
The strongly convex and convex cases are considered in \Cref{sec:normalizingConstantStronglyConvex,sec:NormalizingconstantConvexGradientLipU}, respectively. The choice of the simulation parameters $\simu$ explicitly depends on the (strong) convexity of $\U$.
Throughout this Section, we assume that $\Llip > \m$; note that if $\Llip = \m$, $\pi$ is a Gaussian density and $\Z$ is known.
For $M\in\nset^\star$ and $i\in\defEns{0,\ldots,M-1}$, we first provide an upper bound on the mean squared error $\MSEi$ of $\pihati(\gi)$ defined by
\begin{equation}\label{eq:def-MSE}
\MSEi = \expe{\defEns{\pihati(\gi) - \pii(\gi)}^2} \eqsp ,
\end{equation}
where $\pii(\gi)$ and $\pihati(\gi)$ are given by \eqref{eq:oneRatio} and
\eqref{eq:defEstimatorZiPlusUnZi}  respectively.
The $\MSEi$ can be decomposed as a sum of the squared bias and variance,
\begin{equation}\label{eq:decompo-MSE}
\MSEi 
= \defEns{\PE[ \pihati(\gi) ] - \pii(\gi)}^2 + \Var{ \pihati(\gi) } \eqsp.
\end{equation}
\Cref{prop:bias,prop:bias2} give upper bounds on the squared bias and \Cref{prop:variance} on the variance.
The results are based on the non-asymptotic bounds of the Wasserstein distance for a strongly convex potential obtained in \cite{durmusSampling} (see also \cite{dalalyan2016theoretical}, \cite{durmusNonAsymp}).
We introduce the following conditions on the stepsize $\gami$ used in the Euler discretization and the variance $\sig2_{i+1}$
\begin{equation}\label{eq:conditions-variances}
\gami \in \ocint{0,\frac{1}{m+L+2/\sigDi}} \eqsp, \quad
\sig2_{i+1} \leq 2(d+4) \parenthese{\frac{2d+7}{\sigDi}-m}_{+}^{-1} \eqsp,
\end{equation}
where by convention $1/0=\plusinfty$. Note that the condition on $\sig2_{i+1}$ is equivalent to $\ai \in \ccint{0,\mi/\{4(d+4)\}\wedge (2\sig2_i)^{-1} }$ where $\ai$ is defined in \eqref{eq:defgi} and $\mi$ in \eqref{eq:def-Li-mi}.
Assuming that $\gami$ and $\sig2_{i+1}$ satisfy \eqref{eq:conditions-variances}, we define the positive quantities
\begin{equation}\label{eq:defC}
\Cci{0} = \exp \left(\frac{4\ai(d+2)}{\kappai-8\ai} \right) \eqsp , \quad
\Cci{1} = 2d\frac{1-8\ai\gami}{\kappai-8\ai} \eqsp , \quad
\Cci{2} = 4\frac{d}{\mi} \eqsp,
\end{equation}
where $\mi$, $\Li$ and $\kappai$ are defined in \eqref{eq:def-Li-mi} and \eqref{eq:defkappa}, respectively.
Denote by,
\begin{align}
\label{eq:defA0}
\Aci{0} &=  2\Li^2 \kappai^{-1} d \eqsp , \\
\label{eq:defA1}
\Aci{1} &=  2d \Li^2 + d\Li^4(\kappai^{-1} + (\mi+\Li)^{-1})(\mi^{-1}+6^{-1}(\mi+\Li)^{-1})  \:.
\end{align}

\begin{proposition}\label{prop:bias}
Assume \Cref{assumption:C1GradientLipschitz} and \Cref{assumption:stronglyConvex}($\m$) for some $m \geq 0$. For $\Ni \in \nset$, $\ni\in \nset^{*}$ and $\gami, \sig2_{i+1}$ satisfying \eqref{eq:conditions-variances}, we have
\begin{multline*}
\defEns{\PE[ \pihati(\gi) ] - \pii(\gi)}^2
\leq 4\ai^2 (\Cci{2}+\Cci{0} \Cci{1}) \\
\times \defEns{\frac{4d}{\ni\mi\kappai\gami}\exp\left( -\Ni \frac{\kappai\gami}{2}\right) + 2\kappai^{-1}(\Aci{0} \gami + \Aci{1} \gami^2) } \eqsp .
\end{multline*}
\end{proposition}
\begin{proof}
  The proof is postponed to \Cref{sec:proofs-sec-OneRatio}.
\end{proof}

The squared bias can thus be controlled by adjusting the parameters $\gami,\ni$ and $\Ni$.
If $\U$ satisfies \Cref{assumption:hC3}, the bound on the squared bias can be improved. Define,
\begin{align}
\label{eq:defB0}
\Bci{0} &= d \left( 2\Li^2 + \kappai^{-1}\{(d\tildL^2)/3 + 4\Li^4/(3\mi)\}\right)  \eqsp , \\
\label{eq:defB1}
\Bci{1} &= d \Li^4 \parenthese{ \kappai^{-1} + \{6(\mi+\Li)\}^{-1} + \mi^{-1}} \eqsp.
\end{align}

\begin{proposition}\label{prop:bias2}
Assume \Cref{assumption:C1GradientLipschitz}, \Cref{assumption:stronglyConvex}($\m$) for some $m \geq 0$, and \Cref{assumption:hC3}. For $\Ni \in \nset$, $\ni\in \nset^{*}$ and $\gami, \sig2_{i+1}$ satisfying \eqref{eq:conditions-variances}, we have
\begin{multline*}
\defEns{\PE[ \pihati(\gi) ] - \pii(\gi)}^2
\leq 4\ai^2 (\Cci{2}+\Cci{0} \Cci{1}) \\
\times \defEns{ \frac{4d}{\ni\mi\kappai\gami}\exp\left( -\Ni \frac{\kappai\gami}{2}\right) + 2\kappai^{-1}(\Bci{0} \gami^2 + \Bci{1} \gami^3) } \eqsp .
\end{multline*}
\end{proposition}

\begin{proof}
  The proof is postponed to \Cref{sec:proofs-sec-OneRatio}.
\end{proof}

Note that the leading term is of order $\gami^2$ instead of $\gami$.
We consider now the variance term in \eqref{eq:decompo-MSE}.

\begin{proposition}\label{prop:variance}
Assume \Cref{assumption:C1GradientLipschitz} and \Cref{assumption:stronglyConvex}($\m$) for some $m \geq 0$. For $\Ni \in \nset$, $\ni\in \nset^{*}$ and $\gami, \sig2_{i+1}$ satisfying \eqref{eq:conditions-variances}, we have
\begin{equation*}
\Var{\pihati(\gi)} \leq \frac{32 \ai^2 \Cci{0} \Cci{1}}{\kappai^2 \ni\gami}\left( 1 + \frac{2}{\kappai \ni \gami} \right) \eqsp .
\end{equation*}
\end{proposition}
\begin{proof}
  The proof is postponed to \Cref{sec:proofs-sec-OneRatio}.
\end{proof}

\subsection{Strongly convex potential $\U$}
\label{sec:normalizingConstantStronglyConvex}


\begin{theorem}\label{thm:costAlgorithmStronglyConvex}
 Assume \Cref{assumption:C1GradientLipschitz} and \Cref{assumption:stronglyConvex}($m$) for $m>0$ and let $\mu,\epsilon \in \ooint{0,1}$.
There exists an explicit choice of the simulation parameters $\simu$ \eqref{eq:sim-parameters} such that the estimator $\Zhat$ defined in \eqref{eq:def-Zhat} satisfies with probability at least $1-\mu$
\[ (1-\epsilon) \Z \leq \Zhat \leq (1+\epsilon)\Z \eqsp,
\]
and the $\cost$ \eqref{eq:def-cost} of the algorithm is upper-bounded by
\begin{equation}\label{eq:thm-stronglyConvex}
\cost \leq \parenthese{\frac{6272C}{\epsilon^2\mu} + \log(5C d^2)} \frac{(1088C)^2 d^2(d+4)}{\epsilon^2\mu} \parenthese{\frac{\m+\Llip}{2\m}}^3 (C + 3) \eqsp ,
\end{equation}
with
\begin{equation}\label{eq:thm-nbDeux-upper-bound}
C = \left\lceil \frac{1}{\log(2)}\log\parenthese{ d\parenthese{d+\frac{7}{2}}\parenthese{\frac{L}{m}-1}\frac{1}{\log(1+\epsilon/3)}}\right\rceil \eqsp.
\end{equation}
\end{theorem}

\begin{proof}
The proof is postponed to \Cref{sec:proofCostAlgorithmStronglyConvex}.
\end{proof}

\begin{theorem}\label{thm:costAlgorithmStronglyConvex-UC3}
 Assume \Cref{assumption:C1GradientLipschitz}, \Cref{assumption:stronglyConvex}($m$) for $m>0$, \Cref{assumption:hC3} and let $\mu,\epsilon \in \ooint{0,1}$.
 There exists an explicit choice of the simulation parameters $\simu$ \eqref{eq:sim-parameters} such that the estimator $\Zhat$ defined in \eqref{eq:def-Zhat} satisfies with probability at least $1-\mu$
\[ (1-\epsilon) \Z \leq \Zhat \leq (1+\epsilon)\Z \eqsp,
\]
and the $\cost$ \eqref{eq:def-cost} of the algorithm is upper-bounded by
\begin{multline}\label{eq:thm-stronglyConvex-UC3}
\cost \leq \parenthese{\frac{6272C}{\epsilon^2\mu} + \log(5C d^2)} \sqrt{\frac{7}{3}} \frac{512C d^{3/2}}{\epsilon\sqrt{\mu}} (d+4)(C+3) \\
\times \defEns{\tildL \frac{2^{3/2}}{m^{3/2}} + \sqrt{10}\parenthese{\frac{m+L}{2m}}^2} \eqsp,
\end{multline}
with $C$ defined in \eqref{eq:thm-nbDeux-upper-bound}.
\end{theorem}

\begin{proof}
The proof is postponed to \Cref{sec:proofCostAlgorithmStronglyConvex}.
\end{proof}
Using the median trick (see \eg~\cite[Lemma 6.1]{JERRUM1986169} or \cite{niemiro2009}), we have the following corollary,

\begin{corollary}
\label{corollary:thms-median-trick-strongly}
Let $\epsilon, \tilde{\mu}\in\ooint{0,1}$. Repeat $2\ceil{4\log(\tilde{\mu}^{-1})}+1$ times the algorithm of \Cref{thm:costAlgorithmStronglyConvex,thm:costAlgorithmStronglyConvex-UC3} with $\mu=1/4$ and denote by $\Zhat$ the median of the output values. We have with probability at least $1-\tilde{\mu}$,
\[ (1-\epsilon) \Z \leq \Zhat \leq (1+\epsilon)\Z \eqsp.
\]
\end{corollary}

\begin{proof}
The proof is postponed to \Cref{sec:proofCostAlgorithmStronglyConvex}.
\end{proof}

The proof of \Cref{thm:costAlgorithmStronglyConvex,thm:costAlgorithmStronglyConvex-UC3,corollary:thms-median-trick-strongly} relies on several lemmas which are stated below.
These lemmas explain how the simulation parameters $\simu$ must be chosen. The details of the proofs are gathered in \Cref{sec:proofsStronglyConvex}.
Set
\begin{equation}
\label{eq:def-sig20}
\sig2_0=\{2\log(1+\epsilon/3)\}/\{d(\Llip-\m)\} \eqsp .
\end{equation}
This choice of $\sig2_0$ is justified by the following result,

\begin{lemma}\label{lemma:sigma0}
Under \Cref{assumption:C1GradientLipschitz} and \Cref{assumption:stronglyConvex}($m$) for $m\geq 0$, we have
\begin{equation}\label{eq:Z0bounds}
\Z_0 \leq (2\pi\sigma_0^2)^{d/2}/(1+\sigma_0^2 \m)^{d/2} \leq \Z_0 \left(1+\epsilon/3\right) \eqsp.
\end{equation}
\end{lemma}

\begin{proof}
The proof is postponed to \Cref{sec:proof-lemma-sigma0}.
\end{proof}

Given a choice of $\simu$, define the event
\begin{equation}\label{eq:defEventA}
\boreleanA_{\simu,\epsilon} = \left\{ \left| \prod_{i=0}^{M-1}\pihati(g_i) - \prod_{i=0}^{M-1}\pi_i(g_i) \right|\leq \prod_{i=0}^{M-1}\pi_i(g_i)\frac{\epsilon}{2}  \right\} \eqsp.
\end{equation}
On $\boreleanA_{\simu,\epsilon}$, using \Cref{lemma:sigma0}, \eqref{eq:bridgeSampling} and \eqref{eq:def-Zhat}, we have:
\begin{equation*}
\Z \left( 1-\epsilon/2\right) \leq \Zhat \leq \Z \left( 1+\epsilon\right) \eqsp .
\end{equation*}
It remains to choose $\simu$ to minimize approximately the $\cost$ defined in \eqref{eq:def-cost} under the constraint $\PP(\boreleanA_{\simu,\epsilon}) \geq 1-\mu$.
We define the positive increasing sequence $\suite{\sig2}{0}{M-1}$ recursively, starting from $i=0$. For $i\in\nset$, set
\begin{equation}
  \label{eq:def_sigma_i}
\sig2_{i+1} = \varsigma_s(\sigDi) \eqsp,
\end{equation}
where $\varsigma_s:\rset_{+}^\star \to \rset$ is defined for $t\in\ooint{0,(2d+7)/m}$ by
\begin{equation}
\label{eq:def-recurrence-sigDi}
\varsigma_s(t) =
\parenthese{\frac{1}{t}-\frac{m+(2^{k(t)+1}\sig2_0)^{-1}}{2(d+4)}}^{-1} \eqsp, \quad
k(t) = \left\lfloor \frac{\log(t/\sig2_0)}{\log(2)} \right\rfloor
\end{equation}
and $\varsigma_s(t)=\plusinfty$ otherwise.
The subscript $s$ in $\varsigma_s$ stresses that this choice is valid for the strongly convex case and will be different for the convex case.
With this choice of $(\sigDi)_{i\geq 0}$, the number of phases $M$ is defined by
\begin{equation}\label{eq:def-M}
M = \inf \defEns{ i\geq 1 : \sig2_{i-1} \geq (2d+7)/m } \eqsp .
\end{equation}
By \eqref{eq:def-recurrence-sigDi}, for $t\in\coint{\sig2_0, (2d+7)/m}$, $\varsigma_s(t)\geq t (4d+16)/(4d+15)$, which implies $\Ms <\plusinfty$.
With this definition of $\varsigma_s$, for $i\in\defEns{0,\ldots,\Ms-2}$, we have
\begin{equation}\label{eq:heuristic-set-ai}
\ai = \frac{1}{2}\parenthese{\frac{1}{\sigDi}-\frac{1}{\sig2_{i+1}}}=\frac{\m + (2^{k+1}\sig2_0)^{-1}}{4(d+4)} \eqsp , \quad \mbox {if} \quad
2^k \sig2_0 \leq \sigDi < 2^{k+1} \sig2_0 \eqsp,
\end{equation}
and $a_{\Ms-1}=(2\sig2_{\Ms-1})^{-1}$.
Define $\morc[k] \subset \nset$ for $k\in\nset$ and $\nbDeux \in \nset$ by,
\begin{align}
\label{eq:def-chunk}
\morc[k] &= \defEns{i\in\defEns{0,\ldots,\Ms-2} : 2^k \sig2_0 \leq \sigDi <  2^{k+1} \sig2_0} \eqsp, \\
\label{eq:def-nbDeux}
\nbDeux &= \inf \defEns{ k \geq 0 : \morc[k] = \emptyset} < \plusinfty \eqsp .
\end{align}

The number of phases $\Ms$ and variances $\suite{\sig2}{0}{\Ms-1}$ being defined, we now proceed with the choice of the stepsize $\gami$, the number of samples $\ni$ and the burn-in period $\Ni$ for $i\in\defEns{0,\ldots,\Ms-1}$.

\begin{lemma}\label{lemma:productOfErrors}
Set $\eta = (\epsilon \sqrt{\mu})/8$.
Assume that there exists a choice of the simulation parameters $\suite{N}{0}{M-1}$, $\suite{n}{0}{M-1}$ and $\suite{\gamma}{0}{M-1}$ satisfying,
\begin{enumerate}[label=\roman*)]
\item
\label{eq:conditionsConvexbiasVar}
For all $k\in\defEns{0,\ldots, \nbDeux-1}$, $i\in\morc[k]$,
\begin{equation*}
\absolute{\expe{\pihati(\gi)}-\pii(\gi)} \leq \frac{\eta}{\nbDeux \cmorc[k]}, \quad
\Var{\pihati(\gi)} \leq \frac{\eta^2}{\nbDeux \cmorc[k]} \eqsp ,
\end{equation*}
\item
\label{eq:conditionsConvexFinalbiasVar}
$
\absolute{\expe{\pihatM(\g[M-1])}-\pi_{M-1}(\g[M-1]) } \leq \eta, \quad
\Var{\pihatM(\g[M-1])} \leq \eta^2 \eqsp,
$
\end{enumerate}
where $\pii(\gi)$ is defined in \eqref{eq:oneRatio} and $\pihati(\gi)$ in \eqref{eq:defEstimatorZiPlusUnZi}.
Then $\PP(\boreleanA_{\simu,\epsilon}) \geq 1-\mu$, where $\boreleanA_{\simu,\epsilon}$ is defined in \eqref{eq:defEventA}.
\end{lemma}

\begin{proof}
The proof is postponed to \Cref{sec:proofLemmaProductOfErrors}
\end{proof}

To show the existence of $\gami,\ni,\Ni$ satisfying the conditions of \Cref{lemma:productOfErrors}
, we apply \Cref{prop:bias,prop:bias2,prop:variance} for each $i\in\defEns{0,\ldots,\Ms-1}$. 
We then have the following lemmas,

\begin{lemma}\label{lemma:parametersULA}
Set $\eta = (\epsilon \sqrt{\mu})/8$.
Assume \Cref{assumption:C1GradientLipschitz}, \Cref{assumption:stronglyConvex}($m$) for $m>0$ and,
\begin{enumerate}[label=\roman*)]
\item for all $k\in\defEns{0,\ldots, \nbDeux-1}$, $i\in\morc[k]$,
\begin{align}
\label{eq:parameter-gami}
\gami &\leq \frac{1}{2285}\frac{\eta^2 \kappai^2 \sigma_i^4 \mi}{\nbDeux^2 d^2 \Li^2}  \leq \frac{1}{\mi+\Li} \eqsp, \\
\label{eq:parameter-ni}
\ni &\geq \frac{196 \nbDeux}{\eta^2} \frac{\sqrt{\mi}}{\kappai\sigma_i} \frac{1}{\kappai\gami} \eqsp, \\
\label{eq:parameter-Ni}
\Ni &\geq 2(\kappai\gami)^{-1}\log\parenthese{5\nbDeux d^2} \eqsp,
\end{align}
\item
\begin{align}
\label{eq:gam-M-1}
\gamma_{\Ms-1} &\leq 40^{-1} \eta^2 \Llip_{\Ms-1}^{-2} \m_{\Ms-1} \leq (\m_{\Ms-1}+L_{\Ms-1})^{-1} \eqsp, \\
\label{eq:n-M-1}
n_{\Ms-1} &\geq 19 (\kappa_{\Ms-1} \gamma_{\Ms-1})^{-1} \eta^{-2} \eqsp, \\
\label{eq:N-M-1}
N_{\Ms-1} &\geq (\kappa_{\Ms-1}\gamma_{\Ms-1})^{-1} \eqsp.
\end{align}
\end{enumerate}
Then, the conditions \ref{eq:conditionsConvexbiasVar}-\ref{eq:conditionsConvexFinalbiasVar} of \Cref{lemma:productOfErrors} are satisfied.
\end{lemma}


\begin{proof}
The proof is postponed to \Cref{sec:proofLemmaParametersULA}.
\end{proof}

We have a similar result under the additional assumption \Cref{assumption:hC3}.

\begin{lemma}\label{lemma:parametersULA-UC3}
Set $\eta = (\epsilon \sqrt{\mu})/8$.
Assume \Cref{assumption:C1GradientLipschitz}, \Cref{assumption:stronglyConvex}($m$) for $m>0$, \Cref{assumption:hC3} and,
\begin{enumerate}[label=\roman*)]
\item for all $k\in\defEns{0,\ldots, \nbDeux-1}$, $i\in\morc[k]$,
\begin{equation}
\label{eq:parameter-gami-UC3}
\gami \leq \sqrt{\frac{3}{7}}\frac{\eta\kappai\mi^{1/2}\sigma_i^2}{8\nbDeux d} \parenthese{d\tildL^2 + 10\Li^4\mi^{-1}}^{-1/2} \leq \frac{1}{\mi+\Li} \eqsp,
\end{equation}
and $\ni, \Ni$ as in \eqref{eq:parameter-ni}, \eqref{eq:parameter-Ni},
\item
\begin{equation}
\label{eq:gam-M-1-UC3}
\gamma_{\Ms-1} \leq \sqrt{\frac{3}{7}}\frac{\eta \kappa_{\Ms-1} \m_{\Ms-1}^{-1/2}}{4} \parenthese{d\tildL^2 + 10 L_{\Ms-1}^4\m_{\Ms-1}^{-1}}^{-1/2} \leq \frac{1}{\m_{\Ms-1}+L_{\Ms-1}} \eqsp,
\end{equation}
and $n_{\Ms-1}, N_{\Ms-1}$ as in \eqref{eq:n-M-1}, \eqref{eq:N-M-1}.
\end{enumerate}
Then, the conditions \ref{eq:conditionsConvexbiasVar}-\ref{eq:conditionsConvexFinalbiasVar} of \Cref{lemma:productOfErrors} are satisfied.
\end{lemma}

\begin{proof}
The proof is postponed to the supplementary material \cite[\Cref{sec:proofLemmaParametersULA-UC3-suppl}]{Supplement}.
\end{proof}


\subsection{Convex potential $\U$}
\label{sec:NormalizingconstantConvexGradientLipU}

We now consider the convex case. The annealing process on the variances $\suite{\sig2}{0}{M-1}$ is different from the strongly convex case and is defined in \eqref{eq:def-sigma_i-convex}. In particular,
the stopping criteria for the annealing process 
is distinct from the case where $\U$ is strongly convex and relies on a truncation argument. More precisely, a concentration theorem for log-concave functions \cite[Theorem 3.1]{pereyra2016maximum} states that for $\alpha\in\ooint{0,1}$,
\begin{equation*}
\int_{\rset^d} \1_{\{U\geq d(\tau_\alpha+1) \}}(x) \pi(x) \rmd x \leq \alpha \eqsp , \quad \tau_\alpha = \left(\frac{16\log(3/\alpha)}{d}\right)^{1/2} \eqsp.
\end{equation*}
Let $\epsilon\in\ooint{0,1}$, $\tau=\tau_{\epsilon/2}$ and $\rayon = \rhoUn^{-1} \{d (\tau+1)+\rhoDeux\}$. By \eqref{eq:linear-growth-U}, we have
\begin{equation}\label{eq:concentration}
\int_{\rset^d} \1_{\ball{0}{\rayon}}(x) \pi(x) \rmd x \geq 1-\epsilon/2 \eqsp .
\end{equation}
Given a choice of $M$ and $\sig2_{M-1}$, define $\gbarM : \rset^d \to \rset_{+}$ for all $x\in\rset^d$ by
\begin{equation}\label{eq:truncatedgM}
\gbarM(x) = \exp\left\{ \frac{1}{2\sigma^2_{M-1}}(\norm{x}^2 \wedge \rayon^2) \right\} \eqsp ,
\end{equation}
and $\J$ by,
\begin{equation*}
\J =  \left. \int_{\rset^d} \e^{-\U(x)} \rmd x  \middle /  \int_{\rset^d} \e^{-\U(x)-(\norm{x}^2-\rayon^2)_{+}/(2\sigma^2_{M-1})} \rmd x \right. \eqsp .
\end{equation*}
Note that $\Z/\Z_{M-1} = \J \times \piM (\gbarM)$
and by \eqref{eq:concentration},
\begin{equation}\label{eq:boundsJ}
\J (1-\epsilon/2) \leq 1 \leq \J \eqsp.
\end{equation}
On the event $\Asimu$ defined in \eqref{eq:defEventA} with $g_{M-1}$ replaced by $\gbarM$ and by \eqref{eq:Z0bounds} (with $m=0$), \eqref{eq:boundsJ}, we get
\begin{equation*}
\Z \left( 1-\epsilon/2\right)^2 \leq \Zhat \leq \Z \parenthese{1+\epsilon} \eqsp ,
\end{equation*}
where $\Zhat$ is defined in \eqref{eq:def-Zhat} with $\gM$ replaced by $\gbarM$.
We now state our results in the convex case.

\begin{theorem}\label{thm:costAlgorithmConvex}
Assume \Cref{assumption:C1GradientLipschitz}, \Cref{assumption:stronglyConvex}($m$) for $m\geq 0$. Let $\epsilon,\mu \in\ooint{0,1}$.
There exists an explicit choice of the simulation parameters $\simu$ \eqref{eq:sim-parameters} such that the estimator $\Zhat$ defined in \eqref{eq:def-Zhat} (with $\gM$ replaced by $\gbarM$ defined in \eqref{eq:truncatedgM}) satisfies with probability at least $1-\mu$
\[ (1-\epsilon) \Z \leq \Zhat \leq (1+\epsilon)\Z \eqsp,
\]
and the $\cost$ \eqref{eq:def-cost} of the algorithm is upper-bounded by
\begin{multline}\label{eq:thm-Convex}
\cost \leq \parenthese{\frac{17728 C}{\epsilon^2\mu} + \log(C d^2)} \frac{(487 C)^2 d^2(d+4)}{\epsilon^2\mu} \\
\times \parenthese{C + \frac{6L \{d(\tau +1)+\rhoDeux\}^2}{\rhoUn^{2}} + \frac{8L^2 \{d(\tau +1)+\rhoDeux\}^4}{3\rhoUn^{4}} } \eqsp,
\end{multline}
where $\rhoUn, \rhoDeux$ are defined in \eqref{eq:linear-growth-U}, $\tau = 4d^{-1/2}\{\log(6/\epsilon)\}^{1/2}$ and
\begin{equation}\label{eq:thm-nbDeux-upper-bound-convex}
C = \left\lceil \frac{1}{\log(2)}\log\parenthese{ \frac{d L \{d(\tau +1)+\rhoDeux\}^2}{2\rhoUn^2 \log(1+\epsilon/3)}} \right\rceil \eqsp.
\end{equation}
\end{theorem}

\begin{theorem}\label{thm:costAlgorithmConvex-UC3}
Assume \Cref{assumption:C1GradientLipschitz}, \Cref{assumption:stronglyConvex}($m$) for $m\geq 0$ and \Cref{assumption:hC3}.
Let $\epsilon,\mu \in\ooint{0,1}$.
There exists an explicit choice of the simulation parameters $\simu$ \eqref{eq:sim-parameters} such that the estimator $\Zhat$ defined in \eqref{eq:def-Zhat} (with $\gM$ replaced by $\gbarM$ defined in \eqref{eq:truncatedgM}) satisfies with probability at least $1-\mu$
\[ (1-\epsilon) \Z \leq \Zhat \leq (1+\epsilon)\Z \eqsp,
\]
and the $\cost$ \eqref{eq:def-cost} of the algorithm is upper-bounded by
\begin{multline}\label{eq:thm-Convex-UC3}
\cost \leq 2474 \parenthese{\frac{17728 C}{\epsilon^2\mu} + \log(C d^2)}  \frac{(C+1)d(d+4)}{\epsilon\sqrt{\mu}} \Bigg\{ \frac{8L^2 \{d(\tau +1)+\rhoDeux\}^4}{3\rhoUn^4} \\
+ \frac{d^{1/2}\tildL \{d(\tau +1)+\rhoDeux\}^3}{\sqrt{10}\rhoUn^3}\max\parenthese{\frac{5\rhoUn}{d(\tau +1)+\rhoDeux}, \parenthese{\frac{5}{9}+\frac{\rhoUn^2}{\{d(\tau +1)+\rhoDeux\}^2 L}}^2} \\
+ \frac{6L \{d(\tau +1)+\rhoDeux\}^2}{\rhoUn^2} + C  \Bigg\} \eqsp ,
\end{multline}
where $\rhoUn, \rhoDeux,C$ are defined in \eqref{eq:linear-growth-U}, \eqref{eq:thm-nbDeux-upper-bound-convex} respectively and $\tau = 4d^{-1/2}\{\log(6/\epsilon)\}^{1/2}$.
\end{theorem}

\begin{corollary}
\label{corollary:thms-median-trick-convex}
Let $\epsilon, \tilde{\mu}\in\ooint{0,1}$. Repeat $2\ceil{4\log(\tilde{\mu}^{-1})}+1$ times the algorithm of \Cref{thm:costAlgorithmConvex,thm:costAlgorithmConvex-UC3} with $\mu=1/4$ and denote by $\Zhat$ the median of the output values. We have with probability at least $1-\tilde{\mu}$,
\[ (1-\epsilon) \Z \leq \Zhat \leq (1+\epsilon)\Z \eqsp.
\]
\end{corollary}

The proofs follow the same arguments as
\Cref{thm:costAlgorithmStronglyConvex,thm:costAlgorithmStronglyConvex-UC3,corollary:thms-median-trick-strongly} and are detailed in the supplementary material \cite[\Cref{sec:proofCostAlgorithmConvex}]{Supplement}.

Note that $\gbarM$ \eqref{eq:truncatedgM} is a $\Lip{\gbarM}$-Lipschitz function where,
\begin{equation*}
\Lip{\gbarM} = \frac{\rayon}{\sigma^2_{M-1}} \exp\left( \frac{\rayon^2}{2\sigma^2_{M-1}} \right) \eqsp .
\end{equation*}
The results of \Cref{sec:MSE-loc-lipschitz} give an upper bound on $\MSE_{M-1}$ which is polynomial in the parameters if $\sig2_{M-1}$ is approximately equal to $\rayon^2$.
For $i\in\nset^\star$, we define $(\sigDi)_{i\geq 0}$ recursively. Set $\sig2_0$ as in \eqref{eq:def-sig20} and
\begin{equation}
\label{eq:def-sigma_i-convex}
\sig2_{i+1} = \varsigma_c(\sigDi) \eqsp,
\end{equation}
where $\varsigma_c : \rset_{+}^\star \to \rset$ is defined for $t\in\ooint{0,\rayon^2}$ by,
\begin{equation}
\label{eq:def-recurrence-sigDi-convex}
\varsigma_c(t) =
\parenthese{\frac{1}{t}-\frac{1}{2(d+4)(2^{k(t)+1}\sig2_0)}}^{-1} \eqsp, \quad k(t) = \left\lfloor \frac{\log(t/\sig2_0)}{\log(2)} \right\rfloor \eqsp,
\end{equation}
and $\varsigma_c(t) = \plusinfty$ otherwise.
Define $\Mc$ in this Section by,
\begin{equation}\label{eq:def-M-convex}
\Mc = \inf \defEns{ i\geq 1 : \sig2_{i-1} \geq \rayon^2 } \eqsp .
\end{equation}
By \eqref{eq:def-recurrence-sigDi-convex}, for $t\in\coint{\sig2_0, \rayon^2}$, $\varsigma_c(t)\geq \defEns{(4d+16)/(4d+15)}t$, which implies $\Mc <\plusinfty$.
The following lemmas are the counterparts of \Cref{lemma:parametersULA,lemma:parametersULA-UC3}. They specify the choice of $\suite{\gamma}{0}{\Mc-1}$, $\suite{n}{0}{\Mc-1}$, $\suite{N}{0}{\Mc-1}$ to satisfy the conditions of \Cref{lemma:productOfErrors}.

\begin{lemma}\label{lemma:parametersULA-convex}
Set $\eta = (\epsilon \sqrt{\mu})/8$.
Assume \Cref{assumption:C1GradientLipschitz}, \Cref{assumption:stronglyConvex}($m$) for $m\geq 0$ and,
\begin{enumerate}[label=\roman*)]
\item
for all $k\in\defEns{0,\ldots, \nbDeux-1}$, $i\in\morc[k]$,
\begin{align}
\label{eq:parameter-gami-convex}
\gami &\leq \frac{1}{462}\frac{\eta^2 \Li^{-2} \sigma_i^{-2}}{\nbDeux^2 d^2}  \leq \frac{1}{\mi+\Li} \eqsp, \\
\label{eq:parameter-ni-convex}
\ni &\geq \frac{453 \nbDeux}{\eta^2} \frac{1}{\kappai\gami} \eqsp, \\
\label{eq:parameter-Ni-convex}
\Ni &\geq 2(\kappai\gami)^{-1}\log\parenthese{\nbDeux d^2} \eqsp,
\end{align}
\item
\begin{align}
\label{eq:gam-M-1-convex}
\gamma_{\Mc-1} &\leq (1/26) \eta^2 d^{-1} \LMc^{-2} \kappaMc \leq (\mMc + \LMc)^{-1} \eqsp, \\
\label{eq:n-M-1-convex}
n_{\Mc-1} &\geq 29 \eta^{-2} (\kappa_{\Mc-1} \gamma_{\Mc-1})^{-1} \eqsp, \\
\label{eq:N-M-1-convex}
N_{\Mc-1} &\geq 2(\kappa_{\Mc-1}\gamma_{\Mc-1})^{-1} \log(d) \eqsp.
\end{align}
\end{enumerate}
Then, the conditions \ref{eq:conditionsConvexbiasVar}-\ref{eq:conditionsConvexFinalbiasVar} of \Cref{lemma:productOfErrors} are satisfied, with $\gM$ replaced by $\gbarM$.
\end{lemma}

\begin{proof}
The proof is postponed to the supplementary material \cite[\Cref{sec:proofLemmaParametersULA-convex}]{Supplement}.
\end{proof}

We have a similar result under the additional assumption \Cref{assumption:hC3}.

\begin{lemma}\label{lemma:parametersULA-UC3-convex}
Set $\eta = (\epsilon \sqrt{\mu})/8$.
Assume \Cref{assumption:C1GradientLipschitz}, \Cref{assumption:stronglyConvex}($m$) for $m\geq 0$, \Cref{assumption:hC3} and,
\begin{enumerate}[label=\roman*)]
\item
for all $k\in\defEns{0,\ldots, \nbDeux-1}$, $i\in\morc[k]$,
\begin{equation}
\label{eq:parameter-gami-UC3-convex}
\gami \leq \sqrt{\frac{3}{7}}\frac{\eta\sigma_i^{-1}}{8\nbDeux d} \parenthese{d\tildL^2 + 10\Li^4\sigDi}^{-1/2} \leq \frac{1}{\mi+\Li} \eqsp,
\end{equation}
$\ni, \Ni$ as in \eqref{eq:parameter-ni-convex}, \eqref{eq:parameter-Ni-convex} and,
\item
\begin{equation}
\label{eq:gam-M-1-UC3-convex}
\gamma_{\Mc-1} \leq \sqrt{\frac{3}{8\rme}}\frac{\eta \kappa_{\Mc-1} \sigma_{\Mc-1}}{\sqrt{d}} \parenthese{d\tildL^2 + 10 \LMc^4\sigma_{\Mc-1}^2}^{-1/2} \leq \frac{1}{\m_{\Mc-1}+L_{\Mc-1}} \eqsp,
\end{equation}
$n_{\Mc-1}, N_{\Mc-1}$ as in \eqref{eq:n-M-1-convex}, \eqref{eq:N-M-1-convex}.
\end{enumerate}
Then, the conditions \ref{eq:conditionsConvexbiasVar}-\ref{eq:conditionsConvexFinalbiasVar} of \Cref{lemma:productOfErrors} are satisfied, with $\gM$ replaced by $\gbarM$.
\end{lemma}

\begin{proof}
The proof is postponed to the supplementary material \cite[\Cref{sec:proofLemmaParametersULA-UC3-convex}]{Supplement}.
\end{proof}


\section{Numerical experiments}
\label{sec:numerical-experiments}

\begin{figure}
\begin{center}
\includegraphics[scale=0.7]{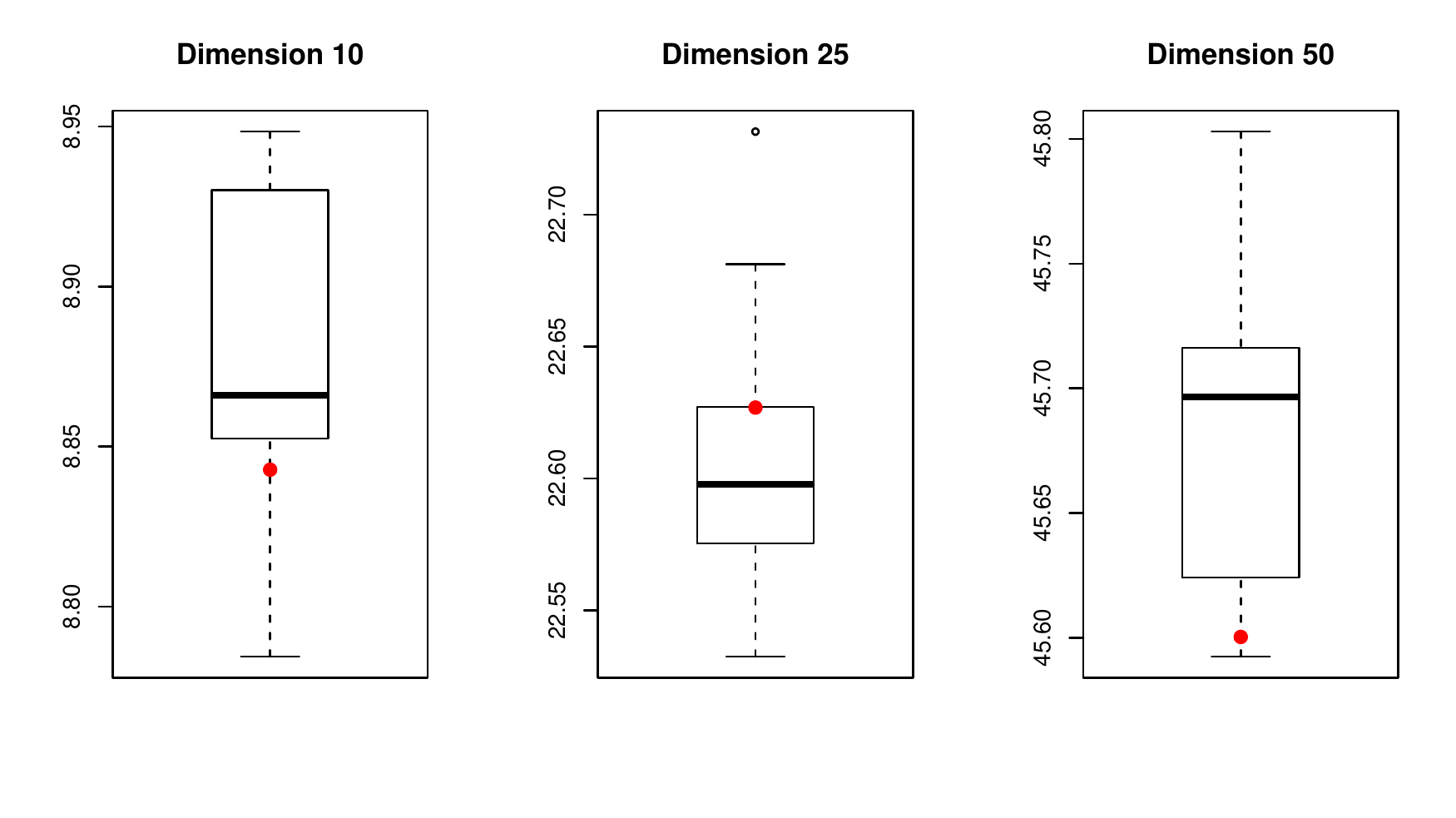}
\end{center}
\caption{\label{figure:boxplot-gaussian-dim} Boxplots of the logarithm of the normalizing constants of a multivariate Gaussian distribution in dimension $d\in\{10,25,50\}$.}
\end{figure}

For the following numerical experiments, the code and data are available at \url{https://github.com/nbrosse/normalizingconstant}.
We first experiment our algorithm to compute the logarithm of the normalizing constant of a multivariate Gaussian distribution in dimension $d\in\{10,25,50\}$, of mean $0$ and inverse covariance matrix $\diag(2,1^{\otimes (d-1)})$. We set $\epsilon = \mu = 0.1$. The number of phases $M$ of the algorithm and the variances $\defEns{\sigma^2_i}_{i=0}^{M-1}$ are chosen accordingly to the formulas \eqref{eq:def_sigma_i} and \eqref{eq:def-M}. For each phase of the algorithm, the step size $\gami$ is set equal to $10^{-2} (\mi + \Li)^{-1}$, the burn-in period $\Ni$ to $10^4$ and the number of samples $\ni$ to $10^5$ where $\mi,\Li$ are defined in \eqref{eq:def-Li-mi}. We carry out $10$ independent runs of the algorithm and compute the boxplots in \Cref{figure:boxplot-gaussian-dim}. The true values of the logarithm of the normalizing constants are known and displayed by the red points in \Cref{figure:boxplot-gaussian-dim}.

We illustrate then our methodology to compute Bayesian model evidence; see  \cite{friel2012estimating} and the references therein.
Let $y\in\rset^p$ be a vector of observations and $\mo_1, \ldots, \mo_l$  be a collection of competing models.
Let $\{p(\mo_i)\}_{i=1}^l$ be a prior distribution on the collection of models.
For $i\in\defEns{0,\ldots,l}$, denote by
$p(y | \theta^{(\mo_i)}, \mo_i)$
the likelihood of the model $\mo_i$. The dominating measure is implicitly considered to be the Lebesgue measure on $\rset^p$. Similarly, for $i\in\defEns{0,\ldots,l}$, denote by $p(\theta^{(\mo_i)} | \mo_i)$ the prior density on the parameters $\theta^{(\mo_i)}$ under the model $\mo_i$ where the dominating measure is implicitly considered to be the Lebesgue measure on $\rset^{d^{(\mo_i)}}$.
The posterior distribution of interest is then for $i\in\defEns{0,\ldots,l}$,
\begin{equation*}
p(\theta^{(\mo_i)}, \mo_i | y) \propto p(y | \theta^{(\mo_i)}, \mo_i) p(\theta^{(\mo_i)} | \mo_i) p(\mo_i)
\end{equation*}
The posterior distribution conditional on model $\mo_i$ can also be considered
\begin{equation}\label{eq:simu-1}
p(\theta^{(\mo_i)} | \mo_i, y) \propto p(y | \theta^{(\mo_i)}, \mo_i) p(\theta^{(\mo_i)} | \mo_i)
\end{equation}
For $i\in\defEns{0,\ldots,l}$, the evidence $p(y | \mo_i)$ of the model $\mo_i$ is defined by the normalizing constant for the posterior distribution \eqref{eq:simu-1}
\begin{equation*}
p(y|\mo_i) = \int_{\rset^{d^{(\mo_i)}}} p(y|\theta^{(\mo_i)}, \mo_i) p(\theta^{(\mo_i)}|\mo_i) \rmd \theta^{(\mo_i)} \eqsp .
\end{equation*}
The Bayes factor $\BF_{12}$ between two models $\mo_i$ and $\mo_j$ is then defined by the ratio of evidences \cite[Section 7.2.2]{robert2007bayesian}, $\BF_{ij} = p(y|\mo_i) / p(y| \mo_j)$.
In the following experiments, we estimate the log evidence $\log(p(y|\mo_i))$.
For ease of notation, the dependence on the model $\mo$ of the parameters $\theta$ and the dimension $d$ of the state space is implicit in the sequel.

Define $\ell^{(\mo)}:\rset^d \to \rset$ by $\ell^{(\mo)}(\theta) = -\log(p(y|\theta, \mo) p(\theta|\mo))$ for $\theta\in\rset^d$.
In the examples we consider, $\ell^{(\mo)}$ satisfies \Cref{assumption:C1GradientLipschitz}, \Cref{assumption:stronglyConvex}, \Cref{assumption:hC3} and has a unique minimum $\theta_\star^{(\mo)}$. Define then $U^{(\mo)}:\rset^d \to \rset_+$ by $U^{(\mo)}(\theta) = \ell^{(\mo)}(\theta + \theta_\star^{(\mo)}) - \ell^{(\mo)}(\theta_\star^{(\mo)})$ for $\theta\in\rset^d$. The algorithm described in \Cref{sec:analysis-algorithm} can be applied to $U^{(\mo)}$. For each example, two different models will be considered and $U^{(\mo)}$ will be written as $U^{(k)}$ for $k=1,2$.

\begin{figure}
\begin{center}
\includegraphics[scale=0.6]{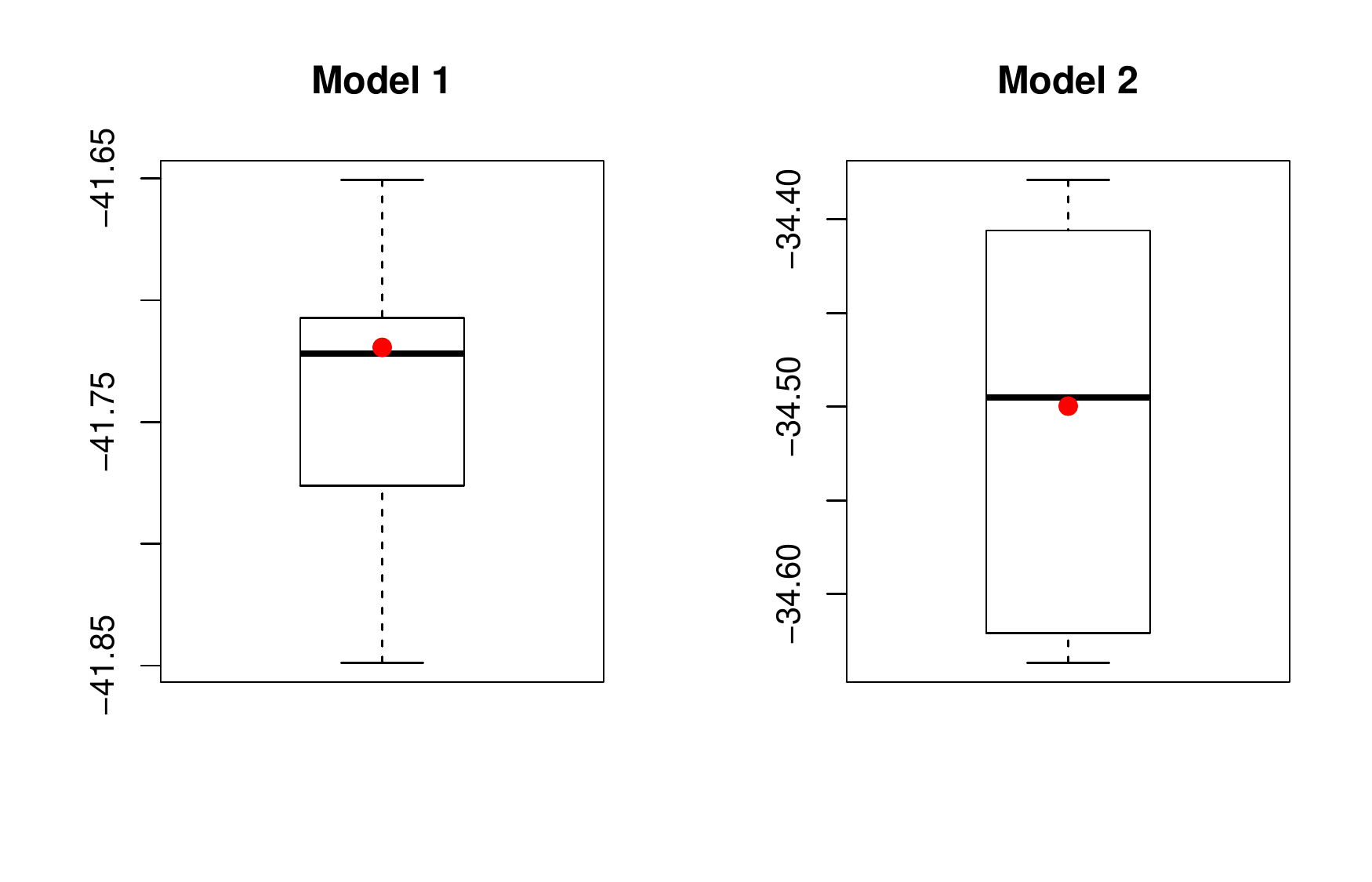}
\end{center}
\caption{\label{figure:boxplots-gaussian} Boxplots of the log evidence for the two models on the Gaussian regression.}
\end{figure}

The numerical experiments are carried out on a Gaussian linear and logistic regression following the experimental setup of \cite[Section 4]{friel2012estimating}, which is now considered as a classical benchmark.
The linear regression is conducted on $p=42$ specimens of radiata pine \cite{williams1959regression}. The response variable $y\in\rset^p$ is the maximum compression strength parallel to the grain. The explanatory variables are $x\in\rset^p$ the density and $z\in\rset^p$ the density adjusted for resin content. $x$ and $z$ are centered.
The covariates of the first model $\mo_1$, $\XUn \in \rset^{p \times 2}$, are composed of an intercept and $x$, while the covariates of the second model $\mo_2$, $\XDeux \in \rset^{p \times 2}$, are composed of an intercept and $z$.
For $k=1,2$, the likelihood is defined by,
\begin{equation*}
p(y|\theta, \mo_k) = \parenthese{\frac{\lambda}{2\pi}}^{d/2} \exp \parenthese{-(\lambda/2)\norm{y - \Xk \theta}^2} \eqsp ,
\end{equation*}
where $\lambda=10^{-5}$.
For the two models, the parameter $\theta$ follows the same Gaussian prior of mean $(3000, 185)$ and inverse covariance matrix $\lambda Q_0 = \lambda \diag(0.06, 6)$ where $\diag$ denotes a diagonal matrix. These values are taken from \cite[section 4.1]{friel2012estimating}.
For $k=1,2$, $U^{(k)}$ is $m^{(k)}$-strictly convex and $L^{(k)}$-gradient Lipschitz, where $m^{(k)}$ (resp. $L^{(k)}$) is the minimal (resp. maximal) eigenvalue of $\lambda([\Xk]^{\transpose}\Xk + Q_0)$. We set $\epsilon=\mu=0.1$.
The number of phases $M$ of the algorithm and the variances $\defEns{\sigma^2_i}_{i=0}^{M-1}$ are chosen accordingly to the formulas \eqref{eq:def_sigma_i} and \eqref{eq:def-M}.
For each phase, the step size $\gami$ is set equal to $10^{-2} (\kappa_i \sigma^2_i m_i)/(d L_i^2)$, the burn-in period $\Ni$ to $10^3 (\kappa_i \gami)^{-1}$ and the number of samples $\ni$ to $10^4 m_i^{1/2} / (\kappa_i^2 \sigma_i \gami)$ where $\mi,\Li,\kappai$ are defined in \eqref{eq:def-Li-mi} and \eqref{eq:defkappa}. The experiments are repeated $10$ times and the boxplots for each model $\mo$ are plotted in \Cref{figure:boxplots-gaussian}. Note that for this Gaussian model, the log evidence is known and displayed by the red points in \Cref{figure:boxplots-gaussian}.

\begin{figure}
\begin{center}
\includegraphics[scale=0.6]{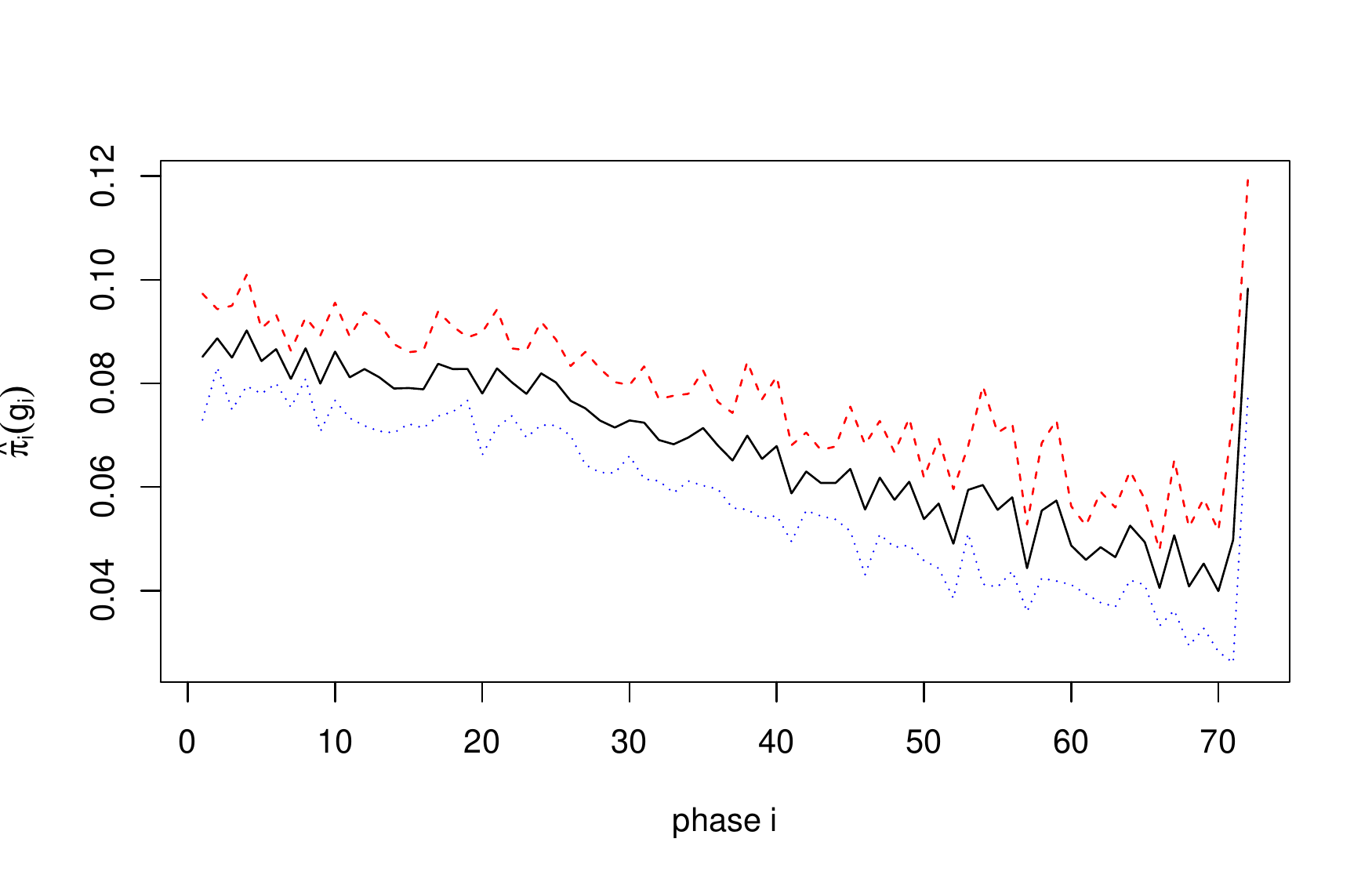}
\end{center}
\caption{\label{figure:errorplot-gaussian} Error plot of $\pihati(\gi)$ for $i\in\defEns{0,\ldots,M-1}$ in the example of the Gaussian regression (model $\mo_1$). The mean of $\pihati(\gi)$ is displayed in black and is spaced apart from the other two curves by the standard deviation of $\pihati(\gi)$.}
\end{figure}

With the same parameters for the algorithm, we run $10$ independent runs at each phase to measure the variability of each estimator $\pihati(\gi)$ defined in \eqref{eq:defEstimatorZiPlusUnZi}. The result is plotted in \Cref{figure:errorplot-gaussian} for the model $\mo_1$. The last estimator $\pihatM(\gM)$ is much higher, which underlines the specificity of the last phase in the algorithm.

\begin{figure}
\begin{center}
\includegraphics[scale=0.5]{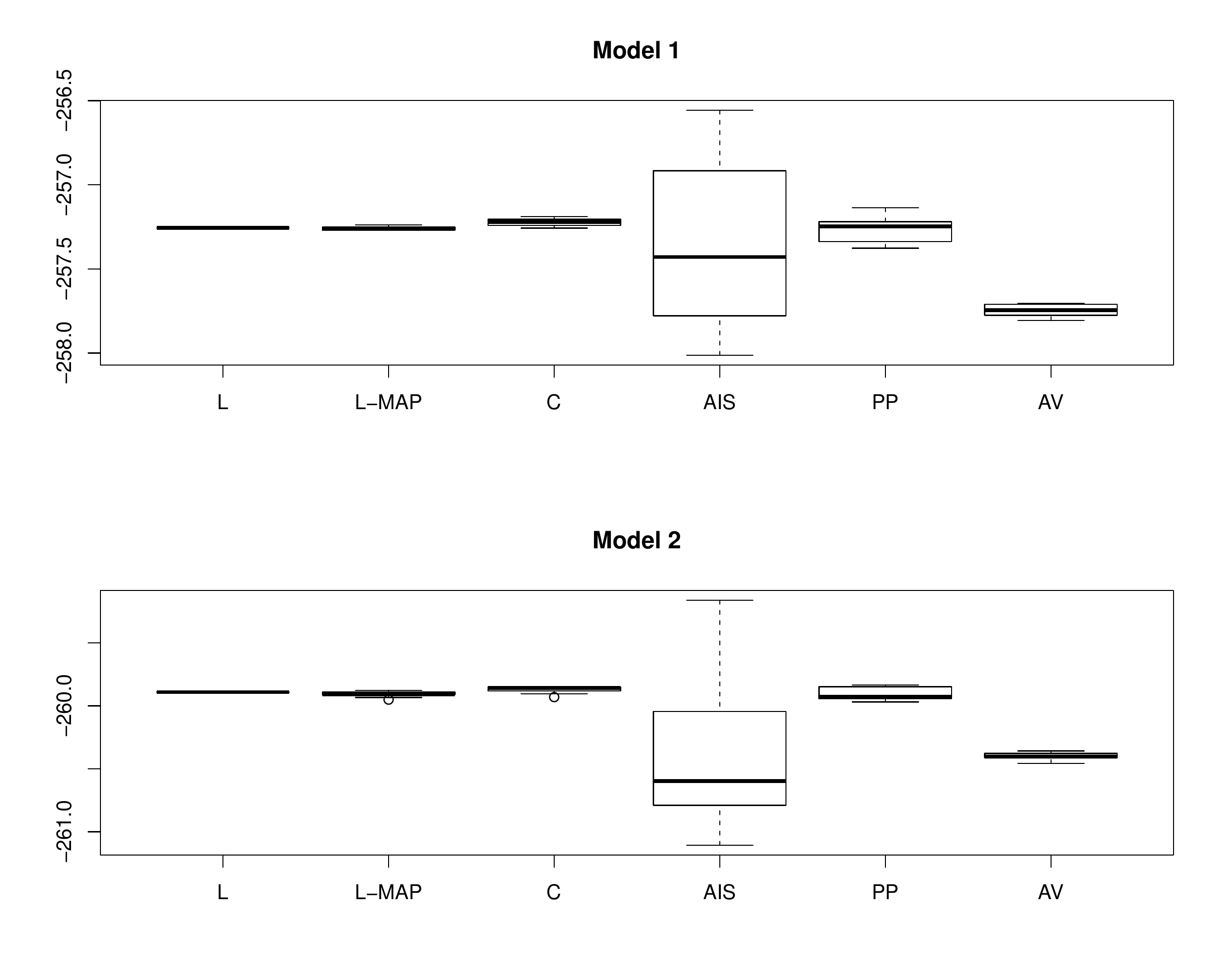}
\end{center}
\caption{\label{figure:boxplots-logistic-1} Boxplots of the log evidence for the two models on the logistic regression. The methods are the Laplace method (L), Laplace at the Maximum a Posteriori (L-MAP), Chib's method (C), Annealed Importance Sampling (AIS), Power Posterior (PP) and our method (AV). }
\end{figure}

The logistic regression is performed on the Pima Indians dataset\footnote{http://archive.ics.uci.edu/ml/datasets/Pima+Indians+Diabetes}. In this case, $y\in\{0,1\}^p$ is a vector of diabetes indicators for $p=532$ Pima Indian women and the potential predictors for diabetes are: number of pregnancies $\NP\in\rset^p$, plasma glucose concentration $\PGC\in\rset^p$, diastolic blood pressure $\BP\in\rset^p$, triceps skin fold thickness $\TST\in\rset^p$, body mass index $\BMI\in\rset^p$, diabetes pedigree function $\DP\in\rset^p$ and age $\AGE\in\rset^p$. These variates are centered and standardized.
The covariates of the first model $\mo_1$ are $\XUn = (\intercept, \NP, \PGC, \BMI, \DP) \in \rset^{p \times 5}$  and the covariates of the second model $\mo_2$ are $\XDeux = (\intercept, \NP, \PGC, \allowbreak \BMI, \DP, \AGE) \in \rset^{p \times 6}$, where $\intercept$ is the intercept of the regressions.
The likelihood is defined for $k=1,2$ by,
\begin{equation*}
p(y|\theta, \mo_k) = \exp \parenthese{ \sum_{i=1}^p \defEns{y_i \theta^{\Tr} \Xk_i - \log\parenthese{1+\rme^{\theta^{\Tr} \Xk_i}}} } \eqsp ,
\end{equation*}
where $\Xk_i$ denotes the $i^{\mathrm{th}}$ row of $\Xk$. For the two models, the prior on $\theta$ is Gaussian, of mean $0$ and inverse covariance matrix $\tau \Id$ where $\tau=0.01$.
For $k=1,2$, $U^{(k)}$ is $\tau$-strongly convex and $L^{(k)}$-gradient Lipschitz, where $L^{(k)}=\lambda_{\max} ([\Xk]^{\transpose} \Xk) /4 + \tau$ and $\lambda_{\max} ([\Xk]^{\transpose} \Xk)$ is the maximal eigenvalue of $[\Xk]^{\transpose}\Xk$.
We set $\epsilon=\mu=0.1$.
The algorithm to estimate $\log(p(y|\mo))$ described in \Cref{sec:normalizingConstantStronglyConvex} is applied with the following modifications. The number of phases is decreased and the recurrence for the variances $\suite{\sig2}{0}{M-1}$ is thus redefined by $\sig2_{i+1} = \varsigma_s^{5} (\sigDi)$ as long as the stopping condition \eqref{eq:def-M} is not fulfilled.
For $i\in\defEns{1,\ldots,30}$, the burn-in period $\Ni$ is set equal to $10^4$, the number of samples $\ni$ to $10^6$ and the step size $\gami$ to $10^{-2} (\mi+\Li)^{-1}$ where $\mi,\Li$ are defined in \eqref{eq:def-Li-mi}; for $i>30$, the number of samples $\ni$ is set equal to $10^5$ and the step size $\gami$ to $10^{-1} (\mi+\Li)^{-1}$.
We compare our results with different methods reviewed in \cite{friel2012estimating} and implemented in \cite{jason-wyse-code}. These are the Laplace method (L), Laplace at the Maximum a Posteriori (L-MAP), Chib's method (C) Annealed Importance Sampling (AIS) and Power Posterior (PP).
The experiments are repeated $10$ times and the boxplots for each model $\mo$ and each method are plotted in \Cref{figure:boxplots-logistic-1}.

\begin{figure}
\begin{center}
\includegraphics[scale=0.6]{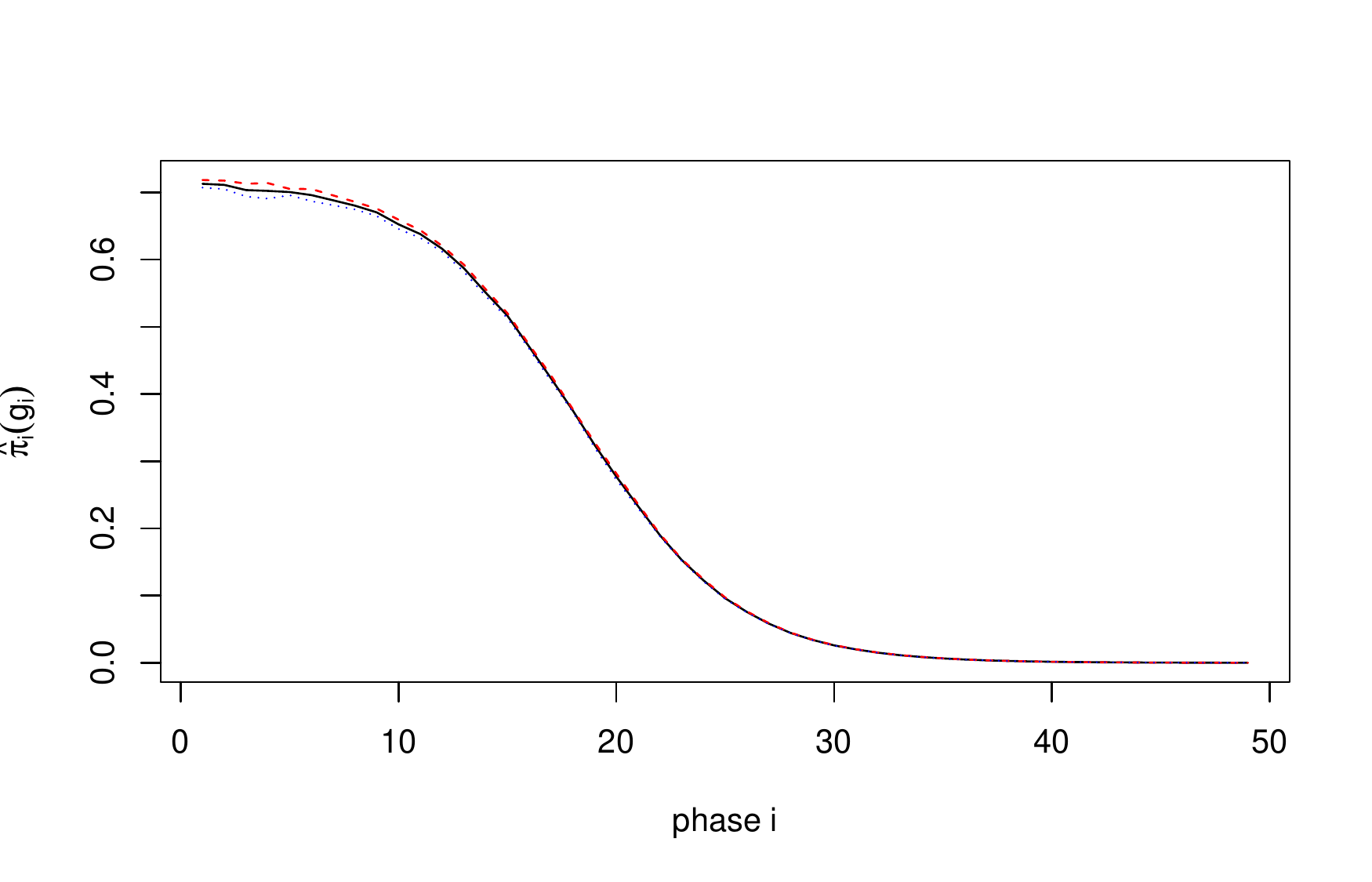}
\end{center}
\caption{\label{figure:errorplot-logistic} Error plot of $\pihati(\gi)$ for $i\in\defEns{0,\ldots,M-1}$ in the example of the logistic regression (model $\mo_1$). The mean of $\pihati(\gi)$ is displayed in black and is spaced apart from the other two curves by the standard deviation of $\pihati(\gi)$.}
\end{figure}

With the same parameters for the algorithm, we run $10$ independent runs at each phase to measure the variability of each estimator $\pihati(\gi)$ defined in \eqref{eq:defEstimatorZiPlusUnZi} and display the result in \Cref{figure:errorplot-logistic} for the model $\mo_1$.

\begin{figure}
\begin{center}
\includegraphics[scale=0.6]{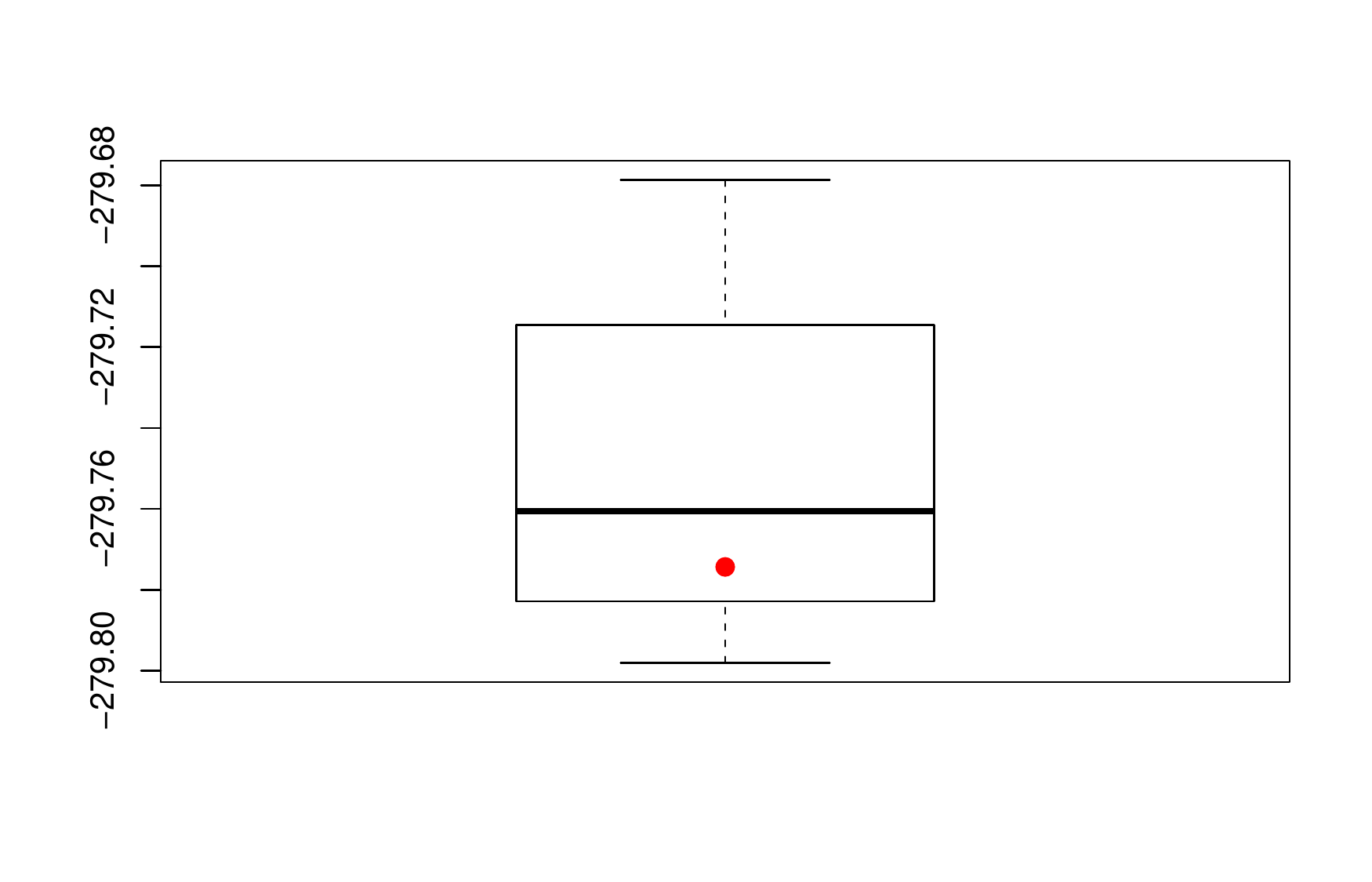}
\end{center}
\caption{\label{figure:boxplots-mixture} Boxplot of the log evidence for the mixture of Gaussian distributions.}
\end{figure}

The final example we address is a Bayesian analysis of a finite mixture of Gaussian distributions, see \cite[Section 4.2]{moral:2006:SMC} and we aim at estimating the log evidence of the posterior distribution. Note that this model does not fit into our assumptions because the potential $U$ is not continuously differentiable on $\rset^d$ and neither convex. Nevertheless, we experiment heuristically our algorithm on a close model given by its likelihood
\begin{equation*}
  p(y | \{\theta_j\}_{j=1}^{4}) = \prod_{i=1}^{p} \left[ \frac{1}{4} \parenthese{\frac{\lambda}{2\uppi}}^{2} \defEns{\sum_{j=1}^{4} \exp\parenthese{-(\lambda/2)(y_i - \theta_j)^2}} \right]
\end{equation*}
for $y=(y_1,\ldots,y_p)\in\rset^p$ a vector of observations. The prior distributions are set following the recommendations of \cite[Section 4.2.1]{moral:2006:SMC} and \cite{richardson:green:1997}. For $j\in\{1,\ldots,4\}$, $\theta_j$ is drawn from a Gaussian distribution of mean $\xi=1.35$ and inverse variance $\varsigma=7.6 \times 10^{-3}$.
$\lambda$ is set equal to $0.03$.
The observations $y\in\rset^{100}$ are $100$ simulated data points from an equally weighted mixture of four Gaussian densities with means $(-3, 0, 3, 6)$ and standard deviations $0.55$, taken from \cite{jasra:holmes:stephens:2005}.
Define for $\theta=(\theta_1,\ldots,\theta_4)\in\rset^4$, $\ell:\rset^4\to\rset$ by $\ell(\theta) = -\log(p(y | \theta)p(\theta))$. The \texttt{optim} function of R \cite{R:2018} gives a local minimum at $\theta^* \approx (1.76562^{\otimes 4})$. Define then the potential $U:\rset^4\to\rset$ for $\theta\in\rset^4$ by $U(\theta) = \ell(\theta + \theta^*) - \ell(\theta^*)$. Set $\epsilon=\mu=0.1$, $m=\varsigma$ and $L=1$. Similarly to the logistic regression, to decrease the running time of the algorithm, the recurrence for the variances $\suite{\sig2}{0}{M-1}$ is defined by $\sig2_{i+1} = \varsigma_s^{5} (\sigDi)$ as long as the stopping condition \eqref{eq:def-M} is not fulfilled.
For each phase, the step size $\gami$ is set equal to $10^{-1} (\kappa_i \sigma^2_i m_i)/(d L_i^2)$, the burn-in period $\Ni$ to $10^4$ and the number of samples $\ni$ to $10^5$ where $\mi,\Li,\kappai$ are defined in \eqref{eq:def-Li-mi} and \eqref{eq:defkappa}.
For comparison purposes, we run the same algorithm using the Metropolis Adjusted Langevin Algorithm (MALA) instead of ULA to estimate $\pihati(\gi)$ at each phase. The step size $\gami$ is set equal to $(\kappa_i \sigma^2_i m_i)/(d L_i^2)$ and the number of samples $\ni$ to $10^6$.
The experiments are repeated $10$ times. The boxplot is plotted in \Cref{figure:boxplots-mixture} and the red point indicates the mean of our algorithm using MALA.

\section{Mean squared error for locally Lipschitz functions}
\label{sec:MSE-loc-lipschitz}

In this Section, we extend the results of \cite[Section
3]{durmusSampling} to locally Lipschitz functions.
This Section is of independent interest and only \Cref{theo:bound_bias,theorem:var-loc-lip} are used in \Cref{sec:proofs}.
Let $\U:\rset^d \to
\rset$ be a continuously differentiable function.
Consider the target
distribution $\pi$ with density $x \mapsto \rme^{-U(x)}/\int_{\rset^d}
\rme^{-U(y)} \rmd y$ \wrt~the Lebesgue measure. We deal with the problem of estimating $\int_{\rset^d} f(x) \rmd \pi(x)$
for locally Lipschitz $f :\rset^d \to \rset$  by the ULA algorithm defined for $k\in\nset$ by,
\begin{equation}
\label{MSE:eq:euler-proposal}
X_{k+1}= X_k - \gaStep_{k+1} \nabla U(X_k) + \sqrt{2 \gaStep_{k+1}} Z_{k+1} \eqsp,
\end{equation}
where $(Z_k)_{k \geq 1}$ is an \iid\ sequence of $d$-dimensional Gaussian vectors with zero mean, identity covariance and $(\gaStep_k)_{k \geq 1}$ is a sequence of positive step sizes, which can either be held constant or be chosen to decrease to $0$. For $n,p \in \nset$, denote by
\begin{equation}
\label{MSE:eq:def_GaStep}
\GaStep_{n,p} \eqdef \sum_{k=n}^p \gaStep_k \eqsp, \qquad \GaStep_n = \GaStep_{1,n} \eqsp,
\end{equation}
and consider the Markov kernel $\RKer_\gaStep$ given for all $\boreleanA \in \mathcal{B}(\rset^d)$
and $x \in \rset^d$ by
\begin{equation}
\label{MSE:eq:definition-RgaStep}
\RKer_\gaStep(x,\boreleanA) =
\int_\boreleanA (4\uppi \gaStep)^{-d/2} \exp \parenthese{-(4 \gaStep)^{-1}\norm[2]{y-x+ \gaStep \nabla U(x)}} \rmd y
\eqsp.
\end{equation}
Define
\begin{equation}
\label{MSE:eq:iterate_kernel}
\QKer^{n,p}_\gaStep = \RKer_{\gaStep_n} \cdots \RKer_{\gaStep_p}  \eqsp, \qquad \QKer^{n}_\gaStep = \QKer^{1,n}_\gaStep \eqsp,
\end{equation}
with the convention that for $n,p \geq 0$, $n < p$, $\QKer^{p,n}_\gaStep$ and $\QKer^{0,0}_\gaStep$ are the identity operator.

For all initial distribution $\mu_0$ on $(\rset^d, \borelSet(\rset^d))$, $\PP_{\mu_0}$ and $\PE_{\mu_0}$ denote the probability and the expectation respectively associated with the sequence of Markov kernels \eqref{MSE:eq:definition-RgaStep} and the initial distribution $\mu_0$ on the canonical space $((\rset^d)^{\nset}, \borelSet(\rset^d)^{\otimes \nset})$ and $(X_k)_{k\in\nset}$ denotes the canonical process.
Let $f : \rset^d \to \rset$ and consider the following assumption,


\begin{assumptionL}\label{assumption:f-loc-lipschitz-max}
\begin{enumerate}
\item
\label{item:loc-lip-1-max}
There exists $\LLip{f} : \rset^d \to \coint{0,\plusinfty}$ a continuous function such that for all $x,y \in\rset^d$, $\absolute{f(y)-f(x)} \leq \norm{y-x} \max \defEns{\LLip{f}(x), \LLip{f}(y)}$.
\item
\label{item:loc-lip-2-max}
There exist $\varepsilon>0$, $C_\pi >0$ and continuous functions $C_Q, C_{Q,\epsilon} : \rset^d \to \coint{0,\plusinfty}$ such that for all $x\in\rset^d$,
\begin{align}
\label{eq:CQ-Cpi-max}
&\pi(\LLip{f}^{2}) \leq C_\pi \eqsp , \quad \sup_{p \geq n \geq 0} \delta_x Q^{n,p}_\gamma \parenthese{\LLip{f}^{2}} \leq C_Q (x) \eqsp , \\
\nonumber
&\sup_{p \geq n \geq 0} \delta_x Q^{n,p}_\gamma \parenthese{\LLip{f}^{2(1+\epsilon)}} \leq C_{Q,\epsilon} (x)
\end{align}
\end{enumerate}
\end{assumptionL}

Under \Cref{assumption:f-loc-lipschitz-max}, we study  the approximation of
 $\int_{\rset^d} f(y) \pi(\rmd y)$ by the weighted average estimator
\begin{equation}
 \label{eq:def_GammaN_plusn}
\hat{\pi}^N_n(f)= \sum_{k=N+1} ^{N+n} \weight{k} f(X_k) \eqsp, \quad  \weight{k}= \gamma_{k+1} \Gamma_{N+2 , N+n+1}^{-1}\eqsp,
 \end{equation}
where  $N \geq 0$ is the length of the burn-in period and  $n \geq 1$ is the number of samples.
The Mean Squared Error ($\MSE$) of $\hat{\pi}_n^N(f)$ is defined by:
\begin{equation}
\label{MSE:eq:decomp_MSE}
\MSE_f(x, N,n) = \expeMarkov{x}{\defEns{\hat{\pi}_n^N(f)- \pi(f) }^2} \eqsp,
\end{equation}
and can be decomposed as,
\begin{equation}\label{MSE:eq:decomp_MSE}
\MSE_f(x, N,n) = \left\{ \PE_x[ \estimateur{f} ] - \pi(f) \right\}^2 + \VarDeux{x}{ \estimateur{f} }\eqsp.
\end{equation}
The analysis of $\MSE_f(x, N,n)$ is similar to \cite[Section 3]{durmusSampling}.
First, the squared bias in \eqref{MSE:eq:decomp_MSE} is bounded. Denote by,
\begin{align}
\label{eq:defAc0}
\Ac{0} &=  2L^2 \kappa^{-1} d \eqsp , \\
\label{eq:defAc1}
\Ac{1} &=  2d L^2 + dL^4(\kappa^{-1} + (\m+L)^{-1})(\m^{-1}+6^{-1}(\m+L)^{-1})  \eqsp, \\
\label{eq:defBc0}
\Bc{0} &= d \left( 2L^2 + \kappa^{-1}\{d\tildL^2/3 + 4L^4/(3\m)\}\right)  \eqsp , \\
\label{eq:defBc1}
\Bc{1} &= d L^4 \parenthese{ \kappa^{-1} + \{6(\m+L)\}^{-1} + \m^{-1}} \eqsp,
\end{align}
where $\kappa$ is given by \eqref{eq:defkappa}. Define then for $n\in\nset^\star$,
\begin{align}
\label{eq:def-u1}
u_n^{(1)}(\gamma) &=  \prod_{k=1}^n (1 - \kappa \gamma_k/2) \eqsp, \\
\label{eq:def-u2}
u_n^{(2)}(\gamma) &=  \sum_{i=1}^n \parenthese{\Ac{0} \gamma_i^2 + \Ac{1}\gamma_i^3} \prod_{k=i+1}^n (1- \kappa \gamma_k/2)
\eqsp, \\
\label{eq:def-u2D}
u_n^{(3)}(\gamma) &= \sum_{i=1}^n \parenthese{\Bc{0}\gamma_i^3 + \Bc{1}\gamma_i^4} \prod_{k=i+1}^n (1- \kappa \gamma_k/2)
\eqsp.
\end{align}

\begin{proposition}
\label{theo:bound_bias}
Assume \Cref{assumption:C1GradientLipschitz} and \Cref{assumption:stronglyConvex}($m$) for $m>0$. Let $f : \rset^d \to \rset$ satisfying \Cref{assumption:f-loc-lipschitz-max}. Let $(\gamma_k)_{k \geq 1}$ be a nonincreasing sequence with $\gamma_1 \leq 1/(m+L)$. Let $x^\star$ be the unique minimizer of $U$. Let $(X_n)_{n \geq 0}$ be given by \eqref{MSE:eq:euler-proposal} and started at $x \in \rset^d$. Then for all $N \geq 0$, $n\geq 1$:
\begin{multline}\label{eq:theo-bound-bias}
\left\{ \PE_x[ \pihat(f) ] - \pi(f) \right\}^2
\leq \defEns{C_\pi + C_Q(x)} \\
\times \sum_{k=N+1}^{N+n} \weight{k} \defEns{2 ( \norm[2]{x-x^\star}+d/m)u_k^{(1)}(\gamma)+ w_k(\gamma)} \eqsp,
\end{multline}
where $u_n^{(1)}(\gamma)$ is given in \eqref{eq:def-u1} and $w_n(\gamma)$ is equal to $u_n^{(2)}(\gamma)$ defined by \eqref{eq:def-u2} and to $u_n^{(3)}(\gamma)$, defined by \eqref{eq:def-u2D}, if \Cref{assumption:hC3} holds.
\end{proposition}

\begin{proof}
For all $k \in \{ N+1, \dots, N+n \}$, let $\xi_k$ be the optimal transference plan between $\delta_x Q_\gamma^k$ and $\pi$ for $W_2$. By the Jensen and the Cauchy Schwarz inequalities, and  \Cref{assumption:f-loc-lipschitz-max}, we have:
\begin{align*}
&\parenthese{\PE_x[\pihat(f)]-\pi(f) }^2
= \parenthese{\sum_{k=N+1} ^{N+n} \weight{k} \int_{\rset^d \times \rset^d} \{f(z) - f(y)\} \xi_k (\rmd z , \rmd y) }^2 \\
&\phantom{-----} \leq \sum_{k=N+1} ^{N+n} \weight{k} \parenthese{\int_{\rset^d \times \rset^d} \norm{z-y} \max\defEns{\LLip{f}(z), \LLip{f}(y)}  \xi_k (\rmd z , \rmd y) }^2 \\
&\phantom{-----}\leq \defEns{C_\pi + C_Q(x)}  \sum_{k=N+1}^{N+n} \weight{k} \int_{\rset^d \times \rset^d}  \norm{z-y}^2 \xi_k (\rmd z , \rmd y) \eqsp.
\end{align*}
The proof follows from \cite[Theorems 5 and 8]{durmusSampling}.
\end{proof}

To deal with the variance term in \eqref{MSE:eq:decomp_MSE},
we adapt the proof of \cite[Theorem 2]{joulin:ollivier:2010} to our setting,
where $f$ is only locally Lipschitz and the Markov chain \eqref{MSE:eq:euler-proposal} is inhomogeneous.
It is based on the Gaussian Poincar{\'e} inequality \cite[Theorem~3.20]{boucheron2013concentration}.
Let $Z=(Z_1,\dots,Z_d)$ be a Gaussian vector with identity covariance
matrix and $f:\rset^d \to \rset$ be a locally Lipschitz function.
Recall that by Rademacher's Theorem \cite[Theorem 3.2]{evans2015measure}, a locally Lipschitz function is almost everywhere differentiable \wrt~Lebesgue measure on $\rset^d$.
The Gaussian Poincar{\'e} inequality states that $\Var{f(Z)} \leq \PE[\norm{\nabla f (Z)}^2]$.
Noticing that for all $x\in \rset^d$,
$\rkerv(x,\cdot)$ defined in \eqref{MSE:eq:definition-RgaStep} is a Gaussian distribution with mean $x - \gamma
\nabla \U(x)$ and covariance matrix $2 \gamma \operatorname{I}_d$, the Gaussian Poincar{\'e} inequality implies:
\begin{equation}\label{eq:Poincare}
0 \leq \int \rkerv (x, \rmd y) \left\{ f(y) - \rkerv f(x) \right\}^2
\leq 2\gamma \int \rkerv(x,\rmd y) \norm{\nabla f(y)}^2 \: .
\end{equation}

First consider the following decomposition of  $\pihat(f)-\PE_x[\pihat(f)]$ as the sum of martingale increments,
\begin{multline*}
\hat{\pi}_n^N(f)- \PE_x[\pihat(f)]= \sum_{k=N}^{N+n-1} \left\{ \CPE[x]{\pihat(f)}{\mcg_{k+1}} - \CPE[x]{\pihat(f)}{\mcg_k} \right\} \\
+ \CPE[x]{\pihat(f)}{\mcg_N} - \PE_x[\pihat(f)] \eqsp,
\end{multline*}
where $(\mcg_n)_{n \geq 0}$ is the natural filtration associated with the Markov chain $(X_n)_{n \geq 0}$.
This implies that the variance may be decomposed as the following sum
\begin{multline}
\label{eq:decomposition-variance}
\VarDeux{x}{\pihat(f)}= \sum_{k=N}^{N+n-1} \PE_x\left[ \parenthese{ \CPE[x]{\pihat(f)}{\mcg_{k+1}} - \CPE[x]{\pihat(f)}{\mcg_k}}^2\right] \\
+
\expeMarkov{x}{\parenthese{\CPE[x]{\pihat(f)}{\mcg_N} - \PE_x[\pihat(f)]}^2}\eqsp.
\end{multline}
Because $\pihat(f)$ is an additive functional, the martingale increment
$\CPE[x]{\pihat(f)}{\mcg_{k+1}} - \CPE[x]{\pihat(f)}{\mcg_k}$ has a simple expression.
For $k = N+n,\dots,N+1$, define backward in time the function
\begin{equation}
\label{eq:def_mart_f}
\martInc^N_{n,k}: x_k \mapsto \omega_{k,n}^N f(x_{k}) +R_{\gamma_{k+1}} \martInc^N_{n,k+1}( x_k ) \eqsp,
\end{equation}
with the convention $\martInc_{n,N+n+1}^N = 0$. Denote finally
\begin{equation}
\label{eq:def_F_f}
\martIncF^{N}_{n}: x_N \mapsto R_{\gamma_{N+1}} \martInc^N_{n,N+1}(x_N)\eqsp.
\end{equation}
Note that for $k \in \{N,\dots,N+n-1\}$, by the Markov property,
\begin{equation}
\label{eq:relation_mart_var_cond}
\martInc^N_{n,k+1}(X_{k+1}) - R_{\gamma_{k+1}}\martInc^N_{n,k+1}(X_{k})= \CPE[x]{\pihat(f)}{\mcg_{k+1}} - \CPE[x]{\pihat(f)}{\mcg_k} \eqsp,
\end{equation}
and $\martIncF^N_{n}(X_N)= \CPE[x]{\pihat(f)}{\mcg_N}$.
With these notations, \eqref{eq:decomposition-variance} may be equivalently expressed as
\begin{multline}
\label{eq:decomposition-variance-alter}
\VarDeux{x}{\pihat(f)}
 = \sum_{k=N}^{N+n-1} \PE_x\left[ R_{\gamma_{k+1}}\defEns{\martInc^N_{n,k+1}(\cdot)
 - R_{\gamma_{k+1}}\martInc^N_{n,k+1}(X_{k}) }^2 (X_k) \right]\\
  + \VarDeux{x}{\martIncF_n^N(X_{N})}  \eqsp.
\end{multline}
Now for $k=N+n,\dots,N+1$, we will use the Gaussian Poincar{\'e} inequality \eqref{eq:Poincare}
to the sequence of function $\martInc_{n,k}^N$. It is required to prove that $\martInc_{n,k}^N$ is locally Lipschitz (see \Cref{lemma:LipQ}).
For the variance of $\martIncF_{n}^N(X_N)$, similar arguments apply using \Cref{lem:var_2}.

\begin{lemma}\label{lemma:LipQ}
Assume \Cref{assumption:C1GradientLipschitz}, \Cref{assumption:stronglyConvex}($m$) for $m>0$ and let $f:\rset^d \to \rset$ satisfying \Cref{assumption:f-loc-lipschitz-max}.
Let  $(\gamma_k)_{k \geq 1}$ be a nonincreasing sequence with $\gamma_1 \leq 2/(m+L)$. Then for all $ \ell \geq n \geq 0$, $Q^{n,\ell}_\gamma f$ is locally Lipschitz and differentiable for almost all $x\in\rset^d$. Its gradient is bounded by,
\begin{equation}\label{eq:nablaQ}
\norm{\nabla Q^{n,\ell}_\gamma f (x)} \leq \prod_{k=n}^\ell (1- \kappa \gamma_k)^{1/2} (\delta_x Q^{n,\ell}_\gamma \LLip{f}^2)^{1/2} \eqsp.
\end{equation}
\end{lemma}

\begin{proof}
Let $\xi_{x,y}$ be the optimal transference plan between $\delta_x Q^{n,\ell}_\gamma$ and $\delta_y Q^{n,\ell}_\gamma$ for $W_2$. By Rademacher's Theorem \cite[Theorem 3.2]{evans2015measure}, $\nabla Q^{n,\ell}_\gamma f (x)$ exists for almost all $x\in\rset^d$. For such $x$,
using Cauchy-Schwarz's inequality and \cite[Theorem 4]{durmusSampling}, we have
\begin{align*}
&\norm{\nabla Q^{n,\ell}_\gamma f (x)}
= \sup_{\norm{u} \leq 1} \lim_{t\to 0} \absolute{\parenthese{Q^{n,\ell}_\gamma f (x+tu)-Q^{n,\ell}_\gamma f (x)}/t} \\
&= \sup_{\norm{u} \leq 1} \lim_{t\to 0} \absolute{ t^{-1}\int_{\rset^d \times \rset^d} \defEns{f(z_2) - f (z_1)} \xi_{x,x+tu}(\rmd z_1, \rmd z_2)} \\
&\leq \sup_{\norm{u} \leq 1} \liminf_{t\to 0} t^{-1}
W_2 (\delta_x Q^{n,\ell}_\gamma, \delta_{x+tu} Q^{n,\ell}_\gamma) \\
&\phantom{----------}\times
\defEns{\int_{\rset^d \times \rset^d} (\LLip{f}^2(z_1) \vee \LLip{f}^2(z_2)) \xi_{x,x+tu}(\rmd z_1, \rmd z_2) }^{1/2} \\
&\leq \sup_{\norm{u} \leq 1} \liminf_{t\to 0} \prod_{k=n}^\ell (1- \kappa \gamma_k)^{1/2}
\defEns{\int_{\rset^d \times \rset^d} (\LLip{f}^2(z_1) \vee \LLip{f}^2(z_2)) \xi_{x, x+tu}(\rmd z_1, \rmd z_2) }^{1/2} \eqsp.
\end{align*}
It is then sufficient to prove that,
\begin{equation*}
\lim_{y\to x} \int_{\rset^d \times \rset^d} \LLip{f}^2(z_1) \vee \LLip{f}^2(z_2) \xi_{x,y}(\rmd z_1, \rmd z_2)  = \int_{\rset^d} \LLip{f}^2(z_1) \delta_x Q^{n,\ell}_\gamma(\rmd z_1) \eqsp.
\end{equation*}
Let $\varepsilon, \eta, \ray>0$ and $y\in\rset^d$. 
Since $a \vee b - a = (b-a)_{+}$, we have
\[ \int_{\rset^d \times \rset^d} (\LLip{f}^2(z_2)-\LLip{f}^2(z_1))_{+} \xi_{x,y}(\rmd z_1, \rmd z_2) = E_1(y) + E_2(y) + E_3(y) \]
where,
\begin{align*}
E_1(y) &= \int_{\rset^d \times \rset^d} \parenthese{\LLip{f}^2(z_2)-\LLip{f}^2(z_1)}_{+} \1_{\defEns{\norm{z_1}+\norm{z_2} \geq 2\ray}} \xi_{x,y}(\rmd z_1, \rmd z_2) \eqsp,\\
E_2(y) &= \int_{\rset^d \times \rset^d} \parenthese{\LLip{f}^2(z_2)-\LLip{f}^2(z_1)}_{+} \1_{\defEns{\norm{z_1}+\norm{z_2} \leq 2\ray}} \1_{\defEns{\norm{z_1-z_2} \leq \eta}} \xi_{x,y}(\rmd z_1, \rmd z_2) \eqsp,\\
E_3(y) &= \int_{\rset^d \times \rset^d} \parenthese{\LLip{f}^2(z_2)-\LLip{f}^2(z_1)}_{+} \1_{\defEns{\norm{z_1}+\norm{z_2} \leq 2\ray}} \1_{\defEns{\norm{z_1-z_2} \geq \eta}} \xi_{x,y}(\rmd z_1, \rmd z_2) \eqsp. \\
\end{align*}
H\"older's inequality gives for $p,q>1$, $1/p+1/q = 1$,
\begin{multline*}
E_1(y) \leq \parenthese{\int_{\rset^d} \LLip{f}^{2q}(z_2)  \delta_y Q^{n,\ell}_\gamma(\rmd z_2)}^{1/q} \\
\times \parenthese{\int_{\rset^d} \1_{\defEns{\norm{z_1} \geq \ray}}  \delta_x Q^{n,\ell}_\gamma(\rmd z_1) + \int_{\rset^d} \1_{\defEns{\norm{z_2} \geq \ray}}  \delta_y Q^{n,\ell}_\gamma(\rmd z_2)}^{1/p} \eqsp.
\end{multline*}
Under \Cref{assumption:f-loc-lipschitz-max}-\ref{item:loc-lip-2-max}, the first term on the right hand side is dominated by a constant for $q$ small enough, and the second term tends to $0$ for $\ray$ large enough, uniformly for $y$ in a compact neighborhood of $x$ by \cite[Theorem 3]{durmusSampling} and
\[ \int_{\rset^d} \1_{\defEns{\norm{z_2} \geq \ray}}  \delta_y Q^{n,\ell}_\gamma(\rmd z_2)
\leq \ray^{-2} \int_{\rset^d} \norm[2]{z_2} \delta_y Q^{n,\ell}_\gamma(\rmd z_2) \eqsp.
\]
We can then choose $\ray$ such that $E_1(y) \leq \varepsilon/3$.
We consider now $E_2(y)$. $\LLip{f}^2$ is a continuous function, uniformly continuous on a compact set and we can then choose $\eta$ such that $E_2(y) \leq \varepsilon/3$. We finally consider $E_3(y)$. By Markov's inequality and $\lim_{y\to x} W_2^2(\delta_x Q^{n,\ell}_\gamma,\delta_y Q^{n,\ell}_\gamma) = 0$, there exists a compact neighborhood $\mathcal{V}(x)$ of $x$ such that $y\in\mathcal{V}(x)$ implies $E_3(y) \leq \varepsilon/3$.

\end{proof}

\begin{lemma}
\label{lem:var_2}
Assume \Cref{assumption:C1GradientLipschitz} and \Cref{assumption:stronglyConvex}($m$) for $m>0$.
Let $(\gamma_k)_{k \geq 1}$ be a nonincreasing sequence with $\gamma_1 \leq  2/(m+L)$ and $N \geq 0$.
Let $f : \rset^d \to \rset$ be such that $Q_\gamma^{k+1,N} f$ is locally Lipschitz for $k\in\defEns{1,\ldots,N}$.
Then for all $x \in \rset^d$,
\[
\int_{\rset^d } Q^{N}_\gamma(x, \rmd y) \defEns{ f(y) - Q^{N}_\gamma f(x)}^2 \leq
2 \sum_{k=1}^N \gamma_k \int_{\rset^d} Q^{k}_\gamma(x, \rmd y) \norm{\nabla Q_\gamma^{k+1,N} f (y)}^2 \eqsp.
\]
\end{lemma}
\begin{proof}
Using $\CPE[x]{f(X_N)}{\mcg_k}=  Q_\gamma^{k+1,N} f(X_k)$, we get
\begin{align*}
\VarDeuxLigne{x}{f(X_N)}
&= \sum_{k=1}^N \expeMarkov{x}{\CPE[x]{\left( \CPE[x]{f(X_N)}{\mcg_k} - \CPE[x]{f(X_N)}{\mcg_{k-1}} \right)^2}{\mcg_{k-1}} } \\
&= \sum_{k=1}^N \expeMarkov{x}{ R_{\gamma_k} \defEns{ Q_\gamma^{k+1,N} f(\cdot) - R_{\gamma_k} Q_\gamma^{k+1,N} f(X_{k-1})}^2(X_{k-1})} \eqsp.
\end{align*}
Eq.~\eqref{eq:Poincare} implies that
\begin{equation*}
\VarDeuxLigne{x}{f(X_N)} \leq 2 \sum_{k=1}^N \gamma_k \int_{\rset^d} Q^{k}_\gamma(x, \rmd y) \norm{\nabla Q_\gamma^{k+1,N} f (y)}^2 \eqsp.
\end{equation*}
\end{proof}

\begin{proposition}\label{theorem:var-loc-lip}
Assume \Cref{assumption:C1GradientLipschitz} and \Cref{assumption:stronglyConvex}($m$) for $m>0$.
Let $f : \rset^d \to \rset$ satisfying \Cref{assumption:f-loc-lipschitz-max} and $(\gamma_k)_{k \geq 1}$ be a nonincreasing sequence with $\gamma_1 \leq 2/(m+L)$.
Then for all $N \geq 0$, $ n \geq 1$, we get
\begin{equation}\label{eq:theorem-var-loc-lip}
\VarDeux{x}{\pihat(f)} \leq  \frac{8 C_Q(x)}{\kappa^{2} \Gamma_{N+2,N+n+1}}\defEns{1+\Gamma_{N+2,N+n+1}^{-1}\parenthese{\kappa^{-1}+\frac{2}{m+L}}} \eqsp .
\end{equation}
\end{proposition}

\begin{proof}
For $k \in \defEns{N,\dots,N+n-1}$ and for all $y,x \in \rset^d$, we have
\begin{multline}
  \label{eq:fixed_x_N_Nplusk_mart}
\abs{\martInc^N_{n,k+1}(y) -\martInc^N_{n,k+1}(x) } =  \bigg| \weight{k+1}\defEns{f(y) - f(x)} \\
+ \sum_{i=k+2}^{N+n} \weight{i} \defEns{Q^{k+2,i}_\gamma f(y) - Q^{k+2,i}_\gamma f(x)} \bigg| \eqsp.
\end{multline}
By \Cref{lemma:LipQ}, $\martInc^N_{n,k+1}$ is locally Lipschitz and for almost all $x\in\rset^d$,
\begin{equation*}
\norm{\nabla \martInc^N_{n,k+1}(x)} \leq \sum_{i=k+1}^{N+n} \weight{i} \defEns{ \prod_{\ell = k+2}^i (1-\kappa\gamma_\ell)^{1/2} } (\delta_x Q^{k+2,i}_\gamma \LLip{f}^2)^{1/2} \eqsp .
\end{equation*}
For $k \in \defEns{N,\dots,N+n-1}$ and $x\in\rset^d$, we have by \eqref{eq:Poincare} and  the Cauchy-Schwarz inequality,
\begin{multline*}
 R_{\gamma_{k+1}}\defEns{\martInc^N_{n,k+1}(\cdot) - R_{\gamma_{k+1}}\martInc^N_{n,k+1}(x) }^2 (x) \\
\leq 2 \gamma_{k+1} \Omega_{k,n}^N
\defEns{\sum_{i=k+1}^{N+n} \weight{i} \prod_{\ell=k+2}^i(1- \kappa \gamma_\ell)^{1/2} (\delta_x Q^{k+1,i}_\gamma \LLip{f}^2)} \eqsp,
 \end{multline*}
where,
\begin{equation}
\Omega_{k,n}^N = \sum_{i=k+1}^{N+n} \weight{i} \prod_{\ell=k+2}^i(1- \kappa \gamma_\ell)^{1/2} \eqsp.
\end{equation}
By \Cref{assumption:f-loc-lipschitz-max}-\ref{item:loc-lip-2-max}, we get for $k \in \defEns{N,\dots,N+n-1}$
\begin{multline*}
 \mathbb{E}_x \left[ R_{\gamma_{k+1}} \left\{ \Phi^N_{n,k+1}(\cdot) - R_{\gamma_{k+1}} \Phi^N_{n,k+1}(X_k) \right\}^2 (X_k) \right] \\
\leq
2 \gamma_{k+1} \Omega_{k,n}^N
\defEns{\sum_{i=k+1}^{N+n} \weight{i} \prod_{\ell=k+2}^i(1- \kappa \gamma_\ell)^{1/2} (\delta_x Q^{i}_\gamma \LLip{f}^2)}
\leq  2 \gamma_{k+1}  C_Q(x) (\Omega_{k,n}^N)^2 \eqsp.
\end{multline*}
Using $(1-t)^{1/2} \leq (1-t/2)$ for $t\in \ccint{0,1}$, we have
\begin{equation}
  \label{eq:bound_Omega}
  \Omega_{k,n}^N \leq (\kappa \Gamma_{N+2,N+n+1}/2)^{-1} \eqsp.
\end{equation}
Using this inequality, we get
\begin{multline}
\label{eq:sumMartingaleVar}
  \sum_{k=N}^{N+n-1} \mathbb{E}_x \left[ R_{\gamma_{k+1}} \left\{ \Phi^N_{n,k+1}(\cdot) - R_{\gamma_{k+1}} \Phi^N_{n,k+1}(X_k) \right\}^2 (X_k) \right]
\\
\leq  8 C_Q(x)  \Gamma_{N+1,N+n} / (\kappa \Gamma_{N+2,N+n+1})^{2} \eqsp .
\end{multline}

We now bound $\VarDeux{x}{\martIncF_n^N(X_{N})}$.  
Since
 for all $x \in \rset^d$, we have
\begin{equation*}
\martIncF_{n}^N(x) = \sum_{i=N+1}^{N+n} \weight{i} Q^{N+1,i}_\gamma f(x)  \eqsp,
\end{equation*}
by \Cref{lemma:LipQ}, $Q^{k+1,N}_\gamma \martIncF_n^N$ is locally Lipschitz for $k\in\defEns{1,\ldots,N}$ with for almost all $x\in\rset^d$,
\begin{equation*}
\norm{ \nabla Q^{k+1,N}_\gamma \martIncF_{n}^N(x)} \leq \sum_{i=N+1}^{N+n} \weight{i} \prod_{\ell=k+1}^i (1-\kappa\gamma_\ell)^{1/2} \parenthese{ \delta_x Q^{k+1,i}_\gamma \LLip{f}^2}^{1/2}  \eqsp.
\end{equation*}
Isolating the term $\prod_{\ell=k+1}^{N} (1-\kappa\gamma_\ell)^{1/2}$ and since $(1-\kappa\gamma_{N+1})^{1/2} \leq 1$, the Cauchy-Schwarz inequality implies
\begin{multline*}
\norm{ \nabla Q^{k+1,N}_\gamma \martIncF_{n}^N(x)}^2 \leq
\defEns{\prod_{\ell=k+1}^{N} (1-\kappa\gamma_\ell)} \\
\times \Omega_{N,n}^N \sum_{i=N+1}^{N+n} \weight{i} \prod_{\ell=N+1}^i (1-\kappa\gamma_\ell)^{1/2} \delta_x Q^{k+1,i}_\gamma \LLip{f}^2  \eqsp.
\end{multline*}
Plugging this inequality in \Cref{lem:var_2}, using \Cref{assumption:f-loc-lipschitz-max}-\ref{item:loc-lip-2-max}, $\sum_{k=1}^N \gamma_k \prod_{i=k+1}^N (1- \kappa \gamma_i) \leq \kappa^{-1}$ and \eqref{eq:bound_Omega}, we get
\begin{equation}
\label{eq:sumMartingaleVarD}
\VarDeux{x}{\martIncF_n^N(X_{N})}  \leq 2 \kappa^{-1} C_Q(x) (\kappa/2)^{-2} \Gamma_{N+2,N+n+1}^{-2} \eqsp.
\end{equation}
Combining \eqref{eq:sumMartingaleVar} and \eqref{eq:sumMartingaleVarD} in \eqref{eq:decomposition-variance} concludes the proof.

\end{proof}


\section{Proofs}\label{sec:proofs}

\subsection{Proofs of propositions \ref{prop:bias}, \ref{prop:bias2}, \ref{prop:variance}}
\label{sec:proofs-sec-OneRatio}

We assume in this Section that \Cref{assumption:C1GradientLipschitz}
and \Cref{assumption:stronglyConvex}($\m$) for some $m \geq 0$ hold.
The proofs rely on the results given in \Cref{sec:MSE-loc-lipschitz}, \Cref{theo:bound_bias,theorem:var-loc-lip}
which establish bounds on the mean squared error for locally Lipschitz
functions. For
$i\in\{0, \ldots, M-1\}$, $\sigDi >0$ and
$\gami>0$, consider the Markov chain $(\XEi{n})_{n \geq 0} $ \eqref{eq:euler-proposal} and its associated
Markov kernel $\Ri$ defined for all $\boreleanA \in \mathcal{B}(\rset^d)$
and $x \in \rset^d$ by
\begin{equation}
\label{eq:definition-RgaStep}
\Ri(x,\boreleanA) =
\int_\boreleanA (4\pi \gami)^{-d/2} \exp \parenthese{-(4 \gami)^{-1}\norm[2]{y-x+ \gami \nabla \Ui(x)}} \rmd y
\eqsp.
\end{equation}
Under \Cref{assumption:C1GradientLipschitz} and \Cref{assumption:stronglyConvex}($m$) for $m\geq 0$, \cite[Theorems 2.1.12, 2.1.9]{nesterov2013introductory} show the following useful inequalities for all $x, y \in \rset^d$,
\begin{align}
\label{eq:KappaNesterov}
\left\langle \nabla \Ui(y) - \nabla \Ui(x), y-x \right\rangle &\geq \frac{\kappai}{2}\norm{y-x}^2 + \frac{1}{\mi+\Li}\norm{\nabla \Ui(y) - \nabla \Ui(x)}^2 \eqsp , \\
\label{eq:mNesterov}
\left\langle \nabla \Ui(y) - \nabla \Ui(x), y-x \right\rangle &\geq \mi \norm{y-x}^2 \eqsp,
\end{align}
where $\Li,\mi$ are defined in \eqref{eq:def-Li-mi} and $\kappai$ in \eqref{eq:defkappa}.
We then check \Cref{assumption:f-loc-lipschitz-max} for $\gi$, where $\gi :\rset^d \to \rset$ is defined in \eqref{eq:defgi}. Note that $\gi$ is continuously differentiable and for $x\in\rset^d$, $\nabla \gi (x) = 2\ai x\e^{\ai \norm{x}^2}$.
Define $\LLip{\gi} :\rset^d \to \rset_+$ for $x\in\rset^d$ by,
\begin{equation}\label{eq:defLLipgi}
\LLip{\gi}(x) = 2\ai \norm{x}\e^{\ai \norm{x}^2}
\end{equation}
We have for all  $x,y \in \rset^d$:
\begin{multline}
\label{eq:bound_gradient_ga}
\abs{\gi(y)-\gi(x)} = \abs{\int_{0}^1 \ps{\nabla \gi (ty +(1-t)x) }{y-x} \rmd t } \\
\leq \norm{y-x} \max(\LLip{\gi}(x), \LLip{\gi}(y))  \eqsp,
\end{multline}
which implies that \Cref{assumption:f-loc-lipschitz-max}-\ref{item:loc-lip-1-max} holds for $\gi$.
The following \Cref{prop:QnpgammaSlow,prop:piLambdaSlow} enable to check \Cref{assumption:f-loc-lipschitz-max}-\ref{item:loc-lip-2-max} for $\gi$.

\begin{lemma}\label{prop:QnpgammaSlow}
  Assume \Cref{assumption:C1GradientLipschitz} and
  \Cref{assumption:stronglyConvex}($m$) for $m\geq 0$. For all
  $\sigDi \in\ooint{0,\plusinfty}$, $n\in\nset$, $\gami \in \ocint{0,
    2/(\mi+\Li)}$, $\ai \in \coint{0,\kappai/8 \wedge
    (2\sig2_i)^{-1}}$ and $x\in\rset^d$, we have:
\begin{equation*}
\sup_{n\in\nset} \Ri^{n} \parenthese{\LLip{\gi}^2} (x) \leq 4 \ai^2 \gi^2(x) \Cci{0} \defEns{\norm{x}^2 + \Cci{1}} \eqsp,
\end{equation*}
where $\LLip{\gi}$ is defined in \eqref{eq:defLLipgi} and $\Cci{0},\Cci{1}$ in \eqref{eq:defC}.
\end{lemma}

\begin{proof}
In the proof, the subscript $i$ is not specified for ease of notation.
Let $ \gamma \in \ocint{0,2/(m+L)}$.
Note that for all  $\alpha \in \coint{0,(4\gamma)^{-1}}$, we have
\begin{align*}
\RKer_\gamma (\rme^{\alpha \norm{\cdot}^2})(x)
&=  \frac{\rme^{- (4\gamma)^{-1} \norm{x - \gamma \nabla \U (x)}^2 }}{(4\pi \gamma )^{d/2}} \int_{\mathbb{R}^d} \rme^{ (\alpha-(4\gamma )^{-1}) \norm{y}^2+   \fracUn{2\gamma } \langle y, x-\gamma \nabla \U (x) \rangle } \rmd y \\
& = \phi(x) \eqsp,
\end{align*}
where $\phi(x) = (1-4\gamma \alpha)^{-d/2} \exp \{
  (\alpha/(1-4\alpha\gamma) ) \norm{x - \gamma \nabla \U (x)}^2
\}$.  By the Leibniz integral rule and \eqref{eq:KappaNesterov},
we obtain:
\begin{align*}
&\RKer_\gamma   (\norm{\cdot}^2 \rme^{\alpha \norm{\cdot}^2})(x) =  \partial_\alpha \RKer_\gamma (\rme^{\alpha \norm{\cdot}^2})(x)  \\
&= (1-4\gamma \alpha)^{-d/2-1}\left\{ 2\gamma  d + \frac{\norm{x - \gamma \nabla \U (x)}^2 }{1-4\alpha\gamma } \right\}
 \exp \left( \frac{\alpha}{1-4\alpha\gamma } \norm{x - \gamma \nabla \U (x)}^2 \right) \\
& \leq (1-4\gamma \alpha)^{-d/2-1}\left\{ 2\gamma  d + \frac{1-\kappa \gamma}{1-4\alpha\gamma } \norm{x}^2 \right\} \exp \left( \frac{\alpha(1-\kappa \gamma)}{1-4\alpha\gamma } \norm{x}^2 \right)  \eqsp.
\end{align*}
Let $a \in \coint{0,\kappa/8}$. Since $a < (4\gamma)^{-1}$, by a straightforward induction we have
\begin{align}
\nonumber
&\delta_x R_\gamma^{p} (\norm{\cdot}^2 \rme^{2a \norm{\cdot}^2}) \leq (1-4\gamma  \alpha_0)^{-d/2-1} \exp \left( \alpha_{p} \norm{x}^2 \right) \\
\nonumber
&\phantom{\delta_x R^{p}_{\gamma}  (\norm{\cdot}^2 \g[2a]) }\times  \sum_{\ell=0}^{p-1}  2\gamma  d \alpha_{\ell} \alpha_{0}^{-1}\left\{ \prod_{k=1}^{\ell}  (1-4\gamma  \alpha_{k})^{-d/2-1} \right\}
\left\{ \prod_{k=\ell+1}^{p-1} (1-4\gamma  \alpha_{k})^{-d/2} \right\} \\
\nonumber
&\phantom{\delta_x R^{p}_{\gamma}  (\norm{\cdot}^2 }+ (1-4\gamma  \alpha_0)^{-d/2-1} \left\{ \prod_{k=1}^{p-1} (1-4\gamma  \alpha_{k})^{-d/2-1}  \right\}
\alpha_{p} \alpha_0^{-1} \norm{x}^2 \exp \left( \alpha_{p} \norm{x}^2 \right)\\
\label{eq:Qnpgammadecreasing1}
&
 \leq
\frac{1}{\alpha_0}\exp \left( \alpha_{p} \norm{x}^2 \right)
\left\{ \prod_{k=0}^{p-1} (1-4\gamma  \alpha_{k})^{-d/2-1} \right\}
\left\{ \alpha_{p} \norm{x}^2 + 2d\gamma \sum_{\ell=0}^{p-1} \alpha_{\ell} \right\}
 \:,
\end{align}
where $(\alpha_{\ell})_{\ell\in\nset}$ is the decreasing sequence defined for $\ell\geq 1$ by:
\begin{equation}\label{eq:defSequenceAlpha}
\alpha_{0} = 2a, \quad \alpha_{\ell}= \alpha_{\ell-1} \frac{(1-\kappa  \gamma)}{1-4\alpha_{\ell-1} \gamma} \:.
\end{equation}
We now bound the right-hand-side of \eqref{eq:Qnpgammadecreasing1}.
First, by using the following inequality,
\begin{align*}
\log(1-4\gamma  \alpha) & = -4\alpha\int_0^{\gamma}  (1-4\alpha t)^{-1} \rmd t \geq -4\alpha\gamma  (1-4\alpha\gamma )^{-1} \:,
\end{align*}
we have:
\begin{align}
\nonumber
\prod_{k=0}^{p-1} (1-4\gamma \alpha_{k})^{-d/2-1} &= \exp \left( -\left(\frac{d}{2}+1\right)\sum_{k=0}^{p-1} \log(1-4\alpha_k \gamma ) \right) \\
\label{eq:boundprodalpha}
&\leq \exp \left( \left(\frac{d}{2}+1\right)\frac{4\gamma }{1-\kappa \gamma }\sum_{k=0}^{p-1} \alpha_k \frac{1-\kappa \gamma }{1-4\alpha_k \gamma } \right) \:.
\end{align}
Second, by a straightforward induction we get for all $ \ell \geq 0$, $\alpha_\ell \leq 2a  \{(1-\kappa \gamma )(1-8 a \gamma)^{-1}\}^\ell$.
Using \eqref{eq:defSequenceAlpha} and this result implies:
\begin{equation*}
\sum_{k=0}^{p-1} \alpha_k \frac{1-\kappa \gamma }{1-4\alpha_k \gamma } = \sum_{k=1}^p \alpha_{k} \leq
2a \frac{1-\kappa\gamma}{\kappa\gamma-8a\gamma} \:, \qquad \sum_{\ell=0}^{p-1} \alpha_\ell \leq 2a \frac{1-8a\gamma}{\kappa\gamma-8a\gamma} \eqsp.
\end{equation*}
Combining these inequalities and \eqref{eq:boundprodalpha}  in \eqref{eq:Qnpgammadecreasing1}   concludes the proof.
\end{proof}

\begin{lemma}\label{prop:piLambdaSlow}
Assume \Cref{assumption:C1GradientLipschitz} and \Cref{assumption:stronglyConvex}($m$) for $m\geq 0$. For all $\sigDi \in\ooint{0,\plusinfty}$ and $\ai \in \ccint{0,\mi/\{4(d+4)\}\wedge (2\sig2_i)^{-1} }$, we have
\[ \piV(\LLip{\gi}^2) \leq 4 \ai^2 \Cci{2} \eqsp, \]
where $\Cci{2}$ is defined in \eqref{eq:defC}.
\end{lemma}

\begin{proof}
In the proof, the subscript $i$ is not specified for ease of notations.
Recall that the generator of the Langevin diffusion \eqref{eq:langevin} associated to $\U$ is defined for any $f$ in $\mathcal{C}^2(\rset^d)$ by
\begin{equation*}
\generator f = - \langle \nabla \U, \nabla f \rangle + \Delta f \eqsp.
\end{equation*}
In particular, for $f(x)=\norm{x}^2 \e^{2a \norm{x}^2}$ and $x\in\rset^d$, we have
\begin{align*}
\nabla f (x) &= 2(1+2a\norm{x}^2)x \e^{2a \norm{x}^2} \:, \\
\Delta f (x) &= \e^{2a \norm{x}^2} \left\{ 16a^2 \norm{x}^4 + 4a(d+4) \norm{x}^2 + 2d \right\} \:.
\end{align*}
Using  \eqref{eq:mNesterov} and $\nabla \U(0) = 0$, we get
\begin{equation*}
\generator (\norm{\cdot}^2 \e^{2a \norm{\cdot}^2})(x) \leq \e^{2a\norm{x}^2} \left\{
2d + 2 \left( 2a(d+4)-\m \right) \norm{x}^2 + 4a\left(4a-\m \right)\norm{x}^4 \right\} \:.
\end{equation*}
Using that $a \in \ccint{0,m/(4(d+4))}$, we have $2a(4a-m) \leq -(8/5) a m$. Then an elementary study of $ t  \mapsto \e^{2at} \left\{ 2d + 4a\left(4a-\m \right)t^2 \right\} $ on $\rset_+$ shows that:
\[ \sup_{x\in\rset^d} \e^{2a\norm{x}^2} \left\{ 2d + 4a\left(4a-\m \right)\norm{x}^4 \right\}
\leq 4d \eqsp .
\]
Therefore we get using $ 2( 2a(d+4)-\m ) \leq -m$,
\begin{equation*}
  \generator (\norm{\cdot}^2 \e^{2a \norm{\cdot}^2})(x) \leq -m  \norm[2]{x}\e^{2a\norm{x}^2} + 4d \eqsp.
\end{equation*}
Finally applying
 \cite[Theorem 4.3-(ii)]{meyn1993stability} shows the result.
\end{proof}

\begin{proof}[Proofs of \Cref{prop:bias,prop:bias2}]
\Cref{prop:QnpgammaSlow,prop:piLambdaSlow,theo:bound_bias} prove the result.
\end{proof}

\begin{proof}[Proof of \Cref{prop:variance}]
The proof follows from \Cref{prop:QnpgammaSlow,theorem:var-loc-lip}.
\end{proof}

\subsection{Proof of \Cref{lemma:productOfErrors}}
\label{sec:proofLemmaProductOfErrors}

The case $\nbDeux=0$ being straightforward, assume $\nbDeux \in \nset^\star$.
%
Using Markov's inequality, we have
\begin{equation}\label{eq:chebyshevPAc}
\PP(\Asimu^{\complementaire}) \leq  \frac{4}{\epsilon^2}\frac{\expe{\left( \prod_{i=0}^{M-1} \pihati(g_i) - \prod_{i=0}^{M-1}\pi_i(g_i)\right)^2}}{\left(\prod_{i=0}^{M-1}\pi_i(g_i)\right)^2}
\eqsp .
\end{equation}
Since $\pihati(g_i)$ for $i\in\{0,\ldots,M-1\}$ are independent, we get
\begin{equation}\label{eq:chebyshevProduct}
\frac{\expe{\left( \prod_{i=0}^{M-1} \pihati(g_i) - \prod_{i=0}^{M-1}\pi_i(g_i)\right)^2}}{\left(\prod_{i=0}^{M-1}\pi_i(g_i)\right)^2}
= \Fc_1^2 (\Fc_2-1) + (\Fc_1-1)^2 \eqsp,
\end{equation}
where
\begin{equation*}
\Fc_1  = \prod_{i=0}^{M-1}\PE\left[\pihati(\gi)\right]/ \pii(\gi)  \eqsp, \quad
\Fc_2 = \prod_{i=0}^{M-1} \PE\left[\{\pihati(\gi)\}^2 \right] / \PE^2\left[\pihati(\gi)\right] \eqsp .
\end{equation*}
In addition, since $\defEns{0,\ldots,M-2} = \cup_{k=0}^{\nbDeux-1} \morc[k]$, we can consider the following decomposition
\begin{align*}
\Fc_1 &= \prod_{k=0}^{\nbDeux-1} \prod_{i\in\morc[k]} \left( 1+\frac{\expe{\pihati(\gi)}-\pii(\gi)}{\pii(\gi)}\right) \\
& \phantom{--------} \times \left( 1+\frac{\expe{\pihatM(\g[M-1])}-\pi_{M-1}(\g[M-1])}{\pi_{M-1}(\g[M-1])}\right) \eqsp ,\\
\Fc_2 &= \prod_{k=0}^{\nbDeux-1} \prod_{i\in\morc[k]} \left( 1+\frac{\Var{\pihati(\gi)}}{\expe{\pihati(\gi)}^2}\right)
\left( 1+\frac{\Var{\pihatM(\g[M-1])}}{\expe{\pihatM(\g[M-1])}^2}\right) \eqsp .
\end{align*}
We now bound $\Fc_1, \Fc_2$ separately.
Using $1+t \leq \exp(t)$ for $t\in\rset$ with $t = \eta/(\nbDeux\cmorc[k])$ and leaving the term $i=M-1$ out, we get by conditions \ref{eq:conditionsConvexbiasVar}-\ref{eq:conditionsConvexFinalbiasVar}
\begin{equation}\label{eq:bound-F1}
\Fc_1 \leq (1+\eta) \exp \left( \eta \right) \eqsp .
\end{equation}
Since $\pihati(\gi) \geq 1$, we have
$\Var{\pihati(\gi)}/\expe{\pihati(\gi)}^2 \leq \eta^2/\nbDeux\cmorc[k]$.
Therefore using $1+t \leq \exp(t)$ for $t\in\rset$ with $t = \eta^2/(\nbDeux\cmorc[k])$  leaving the term $i=M-1$ out, we obtain by conditions \ref{eq:conditionsConvexbiasVar}-\ref{eq:conditionsConvexFinalbiasVar}
\begin{equation}\label{eq:bound-F2}
\Fc_2 \leq \left(1+\eta^2\right) \exp(\eta^2) \eqsp .
\end{equation}
By combining \eqref{eq:chebyshevPAc}, \eqref{eq:chebyshevProduct}, \eqref{eq:bound-F1} and \eqref{eq:bound-F2}, we get:
\begin{equation*}
(\epsilon^2/4)\PP(\Asimu^{\complementaire})\leq
(1+\eta)^2 \rme^{2\eta} \parenthese{(1+\eta^2)\rme^{\eta^2}-1} +
\parenthese{(1+\eta)\rme^{\eta} - 1}^2 \eqsp.
\end{equation*}
With $\eta \leq 1/8$ and $\rme^t -1 \leq t\rme^t$ for $t\geq0$, we have $(\epsilon^2/4)\PP(\Asimu^{\complementaire}) \leq 9 \eta^2$.

\subsection{Proofs of \Cref{sec:normalizingConstantStronglyConvex}}
\label{sec:proofsStronglyConvex}

We preface the proofs by a technical lemma which gathers useful bounds and inequalities.
We recall that in this Section the number of phases $M$ is defined by \eqref{eq:def-M}
\begin{equation*}
M = \inf \defEns{ i\geq 1 : \sig2_{i-1} \geq (2d+7)/m } \eqsp .
\end{equation*}

\begin{lemma}\label{lemma:tech-s}
Assume \Cref{assumption:C1GradientLipschitz} and \Cref{assumption:stronglyConvex}($m$) for
$m > 0$. Let $\suite{\sig2}{0}{M-1}$ defined by \eqref{eq:def_sigma_i} for $\sig2_0$ given in \eqref{eq:def-sig20} and $M$ in \eqref{eq:def-M}.
\begin{enumerate}
\item \label{item:tech-s-1}
$\nbDeux \leq \left\lceil (1/\log(2))\log \{ (2d+7)/(m\sig2_0)\} \right\rceil$ where $\nbDeux$ is defined in \eqref{eq:def-nbDeux}.
\item \label{item:tech-s-2}
For all $k\in\defEns{0,\ldots,\nbDeux-1}$ and $i\in\morc[k]$,
$2^{k+1} \sig2_0 \ai \cmorc[k] \leq 1$, where $\ai$ is defined in \eqref{eq:heuristic-set-ai} and $\morc[k]$ in \eqref{eq:def-chunk}.
\item \label{item:tech-s-3}
For all $i\in\{0,\ldots,\Ms-1\}$ and $\gami\leq 1/(\mi+\Li)$,
there exist $\alphai\in\ccint{4,14}$ and $\betai\in\ccint{1,10}$ such that $\Cci{2}+\Cci{0} \Cci{1} = \alphai d \mi^{-1}$ and $\Cci{0}\Cci{1} = \betai d \kappai^{-1}$ where $\Cci{0},\Cci{1},\Cci{2}$ and $\kappai$ are given in \eqref{eq:defC} and \eqref{eq:defkappa} respectively.
\item \label{item:tech-s-4}
For all $i\in\{0,\ldots,\Ms-1\}$,
$0< \Aci{1} \leq 4d\Li^4 \kappai^{-1} \mi^{-1}$, where $\Li,\mi$ and $\kappai$ are given in \eqref{eq:def-Li-mi} and \eqref{eq:defkappa} respectively.
\item \label{item:tech-s-5}
For all $i\in\{0,\ldots,\Ms-2\}$,
$ \kappai\sigDi \leq 4d+16$.
\item \label{item:tech-s-6}
For all $i\in\{0,\ldots,\Ms-1\}$,
$\sqrt{\mi}/(\kappai\sigma_i) \leq 1$
\item \label{item:tech-s-7}
For all $i\in\{0, \ldots, \Ms-1\}$,
\begin{equation*}
\frac{\mi+\Li}{2\mi} \leq \frac{m+\Llip}{2m} \eqsp , \quad
\frac{\Li^2}{\kappai^3 \sigma_i^4 \mi} \leq \parenthese{\frac{\m+\Llip}{2\m}}^3 \eqsp.
\end{equation*}
\item \label{item:tech-s-8}
For all $k\in\defEns{0,\ldots,\nbDeux-1}$ and $i\in\morc[k]$,
\begin{equation*}
\kappai^{-2} \mi^{-1/2} \sigma_i^{-2} \leq \frac{(2^{k+1}\sig2_0)^{3/2}}{(1+\m 2^k \sig2_0)^{5/2}} \eqsp, \quad
\frac{\Li^2 \mi^{-1/2}}{\kappai^2 \sigDi \mi^{1/2}} \leq \parenthese{\frac{m+L}{2m}}^2 \frac{1}{1+m 2^k \sig2_0} \eqsp.
\end{equation*}
\end{enumerate}
\end{lemma}

\begin{proof}
\begin{enumerate}
\item By \eqref{eq:def-M} and \eqref{eq:def-chunk},
\begin{equation*}
\nbDeux \leq \inf \defEns{k\geq 0 : 2^k \sig2_0 \geq \frac{2d+7}{m}} = \left\lceil \frac{1}{\log(2)}\log \left( \frac{2d+7}{m\sig2_0}\right) \right\rceil \eqsp.
\end{equation*}
\item
Let $k\in\defEns{0,\ldots,\nbDeux-1}$.
Denote by $i_0 = \inf   \morc[k]$.
By \eqref{eq:heuristic-set-ai} and \eqref{eq:def-chunk},
\begin{equation*}
  \cmorc[k]\ai   = \sum_{i\in\morc[k]} \ai \leq \frac{1}{2\sig2_{i_0}} \leq \frac{1}{2^{k+1} \sig2_0} \eqsp,
\end{equation*}
and the proof follows.
\item
Let $k\in\defEns{0,\ldots,\nbDeux-1}$ and $i\in\morc[k]$,
Since $\mi \geq \m+ (2^{k+1}\sig2_0)^{-1}$, $\ai \leq \mi/\{4(d+4)\}$. Therefore using in addition that  $\gami\leq 1/(\mi+\Li)$, we have $\kappai - 8\ai \geq \kappai (d+2)/(d+4)$ and $1-8\ai\gami \geq (d+3)/(d+4)$.
The definition of $\Cci{0}, \Cci{1}, \Cci{2}$ \eqref{eq:defC} completes the proof.
\item The upper bound is a straightforward consequence of $(1+\kappai(\mi+\Li)^{-1})(1+6^{-1}\mi(\mi+\Li)^{-1}) \leq 2$.
\item The bound follows using that $\sig2_{\Ms-2} \leq (2d+7)/m$ by \eqref{eq:def-M} and the sequence $ \suiteD{\kappai\sigDi}{i=0}{M-2}$ is non-decreasing since
\begin{equation*}
\kappai\sigDi = 2 \defEns{1+\frac{mL\sigDi - 1/\sigDi}{m + L + 2/\sigDi}} \eqsp,
\end{equation*}
and $\suiteD{\sigDi}{i=0}{M-2}$ is non-decreasing.
\item The proof is a direct consequence of the fact that the sequence $i
  \mapsto \sqrt{\mi}/(\kappai\sigma_i)$ is non-increasing since $m <
  L$, $\suiteD{\sigDi}{i=0}{M-2}$ is non-decreasing and
\begin{equation*}
\frac{\sqrt{\mi}}{\kappai\sigma_i} = \frac{1}{2}\frac{1}{\sqrt{1+m\sigDi}}\defEns{1+\frac{1+m\sigDi}{1+L\sigDi}} \eqsp .
\end{equation*}
\item Using that $\suiteD{\sigDi}{i=0}{M-2}$ is non-decreasing and
\begin{equation*}
\frac{\mi+\Li}{2\mi} = \frac{2+(m+L)\sigDi}{2+2m\sigDi} \eqsp , \quad
\frac{\Li^2}{\kappai^3 \sigma_i^4 \mi} \leq \parenthese{\frac{(\m+\Llip)\sigDi + 2}{(2\m)\sigDi + 2}}^3 \eqsp,
\end{equation*}
concludes the proof.
\item
Let $k\in\defEns{0,\ldots,\nbDeux-1}$ and $i\in\morc[k]$.
 Since $\kappai \geq \mi$ and $2^{k} \sig2_0 \leq \sigDi \leq 2^{k+1} \sig2_0$, we have
\begin{equation*}
\kappai^{-2} \mi^{-1/2} \sigma_i^{-2} \leq \frac{\sigma_i^{3}}{(\mi\sigDi)^{5/2}} \leq \frac{(2^{k+1}\sig2_0)^{3/2}}{(1+\m 2^k \sig2_0)^{5/2}} \eqsp,
\end{equation*}
and
\begin{equation*}
\frac{\Li^2}{\kappai^2} \leq \parenthese{\frac{m+L}{2m}}^2 \eqsp, \quad
\frac{1}{\mi\sigDi} \leq \frac{1}{1+m 2^k \sig2_0} \eqsp.
\end{equation*}
\end{enumerate}
\end{proof}

\subsubsection{Proof of \Cref{lemma:sigma0}}
\label{sec:proof-lemma-sigma0}

Because $\U$ satisfies \Cref{assumption:C1GradientLipschitz}, \Cref{assumption:stronglyConvex}($m$) for $m\geq 0$ and $\U(0)=0$, $\nabla \U(0)=0$, we have:
\[ \exp(-(\Llip/2)\norm{x}^2) \leq \exp(-\U(x)) \leq \exp(-(\m/2)\norm{x}^2) \eqsp , \]
which implies by integration that,
\begin{equation*}
(2\pi\sigma_0^2)^{d/2} / (1+\sigma_0^2 \Llip)^{d/2} \leq \Z_0 \leq (2\pi\sigma_0^2)^{d/2} / (1+\sigma_0^2 \m)^{d/2} \eqsp,
\end{equation*}
where $\Z_0 = \int_{\rset^d} \e^{-\U[0]}$ and $\U[0]$ is defined in \eqref{eq:defUi}.
The proof follows from the expression of $\sig2_0$ and the bound,
\begin{equation*}
\parenthese{\frac{1+\Llip\sig2_0}{1+\m\sig2_0}}^{d/2} \leq \exp\left( \frac{d}{2}\sig2_0(\Llip-\m)\right) \eqsp .
\end{equation*}

\subsubsection{Proof of \Cref{lemma:parametersULA}}
\label{sec:proofLemmaParametersULA}

Let $k\in\defEns{0,\ldots,\nbDeux-1}$ and $i\in\morc[k]$. Assume that $\gami \leq (\mi+\Li)^{-1}$.
By \Cref{prop:bias}, \Cref{prop:variance}, \Cref{lemma:tech-s}-\ref{item:tech-s-2} and $\sigDi \leq 2^{k+1} \sig2_0$, to check condition-\ref{eq:conditionsConvexbiasVar} of \Cref{lemma:productOfErrors}, it is then sufficient for $\gami,\ni,\Ni$ to satisfy,
\begin{align}
\label{eq:conditionsBias}
\frac{4d}{\ni\mi\kappai\gami}\exp\left(-\Ni\frac{\kappai\gami}{2}\right)+2\kappai^{-1}\left( \Aci{0}\gami+\Aci{1}\gami^2 \right) &\leq
\frac{\eta^2}{4\nbDeux^2}\frac{\sigma_i^4}{\Cci{2}+\Cci{0} \Cci{1}} \eqsp, \\
\label{eq:conditionsVar}
\frac{32\ai\Cci{0}\Cci{1}}{\kappai^2 \ni\gami}\left(1+\frac{2}{\kappai \ni \gami}\right) &\leq \frac{\sigDi\eta^2}{\nbDeux} \eqsp .
\end{align}
By \eqref{eq:defA0}, \Cref{lemma:tech-s}-\ref{item:tech-s-3} and \Cref{lemma:tech-s}-\ref{item:tech-s-4}, there exist $\alphai\in\ccint{4,14}$ and $\betai\in\ccint{1,10}$ such that these two inequalities hold if $\gami,\ni,\Ni$ satisfy
\begin{align}
\label{eq:conditions-gami}
2\Li^2 \kappai^{-1} d \gami+ 4d\Li^4 \kappai^{-1} \mi^{-1}\gami^2 &\leq
\frac{\eta^2}{16\nbDeux^2}\frac{\kappai \mi \sigma_i^4}{\alphai d} \eqsp, \\
\label{eq:conditions-ni}
\frac{1}{\ni}\left(1+\frac{2}{\kappai \ni \gami}\right) &\leq
\frac{\eta^2 \sigDi \kappai^3 \gami}{32\nbDeux\ai \betai d} \eqsp, \\
\label{eq:conditions-Ni}
\Ni \geq \Nibar &= -2(\kappai\gami)^{-1}\log\parenthese{\frac{\eta^2\sigma_i^4 \mi^2 \ni\kappai\gami}{32\nbDeux^2 \alphai d^2}} \eqsp.
\end{align}
These inequalities are shown to be true successively for $\gami,\ni$
and $\Ni$ chosen as in the statement of the Lemma.
Denote by
$\gamibar$ and $\nibar^{-1}$ the positive roots associated to
\eqref{eq:conditions-gami} and \eqref{eq:conditions-ni} seen as
equalities and given by
\begin{align}
\label{eq:exact-gamibar}
\gamibar &= 4^{-1} \Li^{-2} \mi \parenthese{-1 + \sqrt{1+\frac{\eta^2 \kappai^2 \sigma_i^4}{4\alphai \nbDeux^2 d^2}}} \eqsp, \\
\label{eq:exact-nibar}
\nibar^{-1} &= 4^{-1} \kappai\gami \parenthese{-1 + \sqrt{1+\frac{\eta^2 \kappai^2 \sigDi}{4\nbDeux\ai\betai d}}} \eqsp.
\end{align}
Note that for \eqref{eq:conditions-gami} and \eqref{eq:conditions-ni}
to hold, it suffices that $\gami \leq \gamibar$ and $\ni \geq
\nibar$.
We now lower bound $\gamibar$ and upper bound $\nibar$.

 Using that
$\sqrt{1+t}\geq 1 + 2^{-1} t (1+t)^{-1/2}$ for $t=\eta^2\kappai^2
\sigma_i^4/(4\alphai \nbDeux^2 d^2)$ and $(\eta^2\kappai^2
\sigma_i^4)/(4\alphai \nbDeux^2 d^2) \leq 25$ by $\alphai\geq 4$ and
\Cref{lemma:tech-s}-\ref{item:tech-s-5}, concludes that
if \eqref{eq:parameter-gami} holds then $\gami \leq \gamibar$.
The fact that $\gami \leq (\mi+\Li)^{-1}$ can be checked by simple algebra.

First, by \eqref{eq:heuristic-set-ai} and the definition of $\morc[k]$, $\ai \leq \mi/\{4(d+4)\}$,
\begin{equation*}
  \nibar^{-1} \geq  4^{-1} \kappai\gami \parenthese{-1 + \sqrt{1+\frac{\eta^2 \kappai^2 \sigDi(d+4)}{\nbDeux\mi\betai d}}} \eqsp.
\end{equation*}
Then using that $\sqrt{1+t}\geq 1 + 2^{-1} t (1+t)^{-1/2}$ for $t=\eta^2 \kappai^2 \sigDi(d+4)/(\nbDeux\mi\betai d)$ and $\betai \geq 1$ concludes that
if \eqref{eq:parameter-ni} holds then $\ni \geq \nibar$.
Finally, we have by \eqref{eq:parameter-ni}, $(\ni\kappai\gami)^{-1} \leq \eta^2\kappai\sigma_i/(196\sqrt{\mi}\nbDeux)$, which gives with $\kappai \geq \mi$,
\begin{align*}
\Nibar &\leq
2(\kappai\gami)^{-1}\log\defEns{ \frac{64\alphai}{196}\nbDeux d^2(1+\m \sigDi)^{-3/2} } \\
&\leq 2(\kappai\gami)^{-1}\log\parenthese{5\nbDeux d^2} \eqsp,
\end{align*}
which concludes that \eqref{eq:parameter-Ni} implies \eqref{eq:conditions-Ni}.

The same reasoning applies to check condition-\ref{eq:conditionsConvexFinalbiasVar} of \Cref{lemma:productOfErrors}. The details are gathered in the supplementary material \cite[\Cref{sec:proofLemmaParametersULA-suppl}]{Supplement}.

\subsubsection{Proof of \Cref{thm:costAlgorithmStronglyConvex,thm:costAlgorithmStronglyConvex-UC3,corollary:thms-median-trick-strongly}}
\label{sec:proofCostAlgorithmStronglyConvex}

For $i\in\defEns{0,\ldots,\Ms-1}$, set $\gami, \ni, \Ni$ such that \eqref{eq:parameter-gami}, \eqref{eq:parameter-ni}, \eqref{eq:parameter-Ni}, \eqref{eq:gam-M-1}, \eqref{eq:n-M-1} and \eqref{eq:N-M-1} are equalities.
By \eqref{eq:def-cost}, we consider the following decomposition for the $\cost=A+B$ where $A =\sum_{i=0}^{\Ms-2} \{ \Ni+\ni \}$ and $B =n_{\Ms-1} + N_{\Ms-1}$.
We bound $A$ and $B$ separately.

First \Cref{lemma:tech-s}-\ref{item:tech-s-6} implies that for all $i\in\defEns{0,\ldots,\Ms-2}$, $\ni \leq (196 \nbDeux)/(\eta^2\kappai\gami)$ and  therefore using \Cref{lemma:tech-s}-\ref{item:tech-s-7}
\begin{equation}\label{eq:thm5-1}
A  \leq \parenthese{\frac{196\nbDeux}{\eta^2} + 2\log(5\nbDeux d^2)} \frac{2285\nbDeux^2 d^2}{\eta^2}\parenthese{\frac{m+L}{2m}}^3 (\Ms-1) \eqsp.
\end{equation}
We now give a bound on $\Ms-1$.
Define
\begin{equation}\label{eq:def-nbDeux-inte}
\nbDeux[\inte] = \sup \defEns{ k\geq 1 : \m 2^k \sig2_0 \leq 1} \wedge \nbDeux \leq \left\lfloor -\frac{\log(\m\sig2_0)}{\log(2)} \right\rfloor \eqsp.
\end{equation}
By \Cref{lemma:tech-s}-\ref{item:tech-s-2} and \eqref{eq:heuristic-set-ai}, we have
\begin{equation}\label{eq:thm5-3}
 \frac{\Ms-1}{4(d+4)} =\sum_{k=0}^{\nbDeux-1} \frac{ \cmorc[k]}{4(d+4)}
\leq \nbDeux[\inte] + 2 \eqsp.
\end{equation}
Note that $\nbDeux[\inte] \leq \nbDeux \leq C$ by \eqref{eq:def-nbDeux-inte} and \Cref{lemma:tech-s}-\ref{item:tech-s-1}.
Combining \eqref{eq:thm5-1}, \Cref{lemma:tech-s}-\ref{item:tech-s-7} and \eqref{eq:thm5-3}, we get
\begin{equation}\label{eq:thm5-4}
\sum_{i=0}^{\Ms-2} \Ni+\ni \leq
\parenthese{\frac{98\nbDeux}{\eta^2}+\log(5\nbDeux d^2)} \frac{4570\nbDeux^2 d^2}{\eta^2}\parenthese{\frac{m+L}{2m}}^3 4(d+4) (C+2) \eqsp.
\end{equation}
Regarding the term $i=\Ms-1$, we have
\begin{equation}\label{eq:cost-M-1-strictly-convex}
n_{\Ms-1} + N_{\Ms-1} \leq \parenthese{\frac{19}{\eta^2}+1}\frac{40}{\eta^2}\frac{m+L}{2m}\frac{L}{m} \eqsp.
\end{equation}
Replacing $\eta$ by $(\epsilon\sqrt{\mu})/8$ and combining \eqref{eq:thm5-4} and \eqref{eq:cost-M-1-strictly-convex} gives \eqref{eq:thm-stronglyConvex}.


Assume \Cref{assumption:hC3}. We now prove \Cref{thm:costAlgorithmStronglyConvex-UC3} and use \Cref{lemma:parametersULA-UC3} instead of \Cref{lemma:parametersULA}.
For $i\in\defEns{0,\ldots,\Ms-1}$, set $\gami, \ni, \Ni$ such that \eqref{eq:parameter-gami-UC3}, \eqref{eq:parameter-ni}, \eqref{eq:parameter-Ni}, \eqref{eq:gam-M-1-UC3}, \eqref{eq:n-M-1} and \eqref{eq:N-M-1} are equalities.
By \eqref{eq:def-cost}, we have the decomposition $\cost=A+B$ where $A =\sum_{i=0}^{\Ms-2} \{ \Ni+\ni \}$ and $B =n_{\Ms-1} + N_{\Ms-1}$.
\Cref{lemma:tech-s}-\ref{item:tech-s-6} implies that for all $i\in\defEns{0,\ldots,\Ms-2}$, $\ni \leq (196 \nbDeux)/(\eta^2\kappai\gami)$,
and using that for $a,b\geq 0$, $\sqrt{a+b}\leq \sqrt{a}+\sqrt{b}$, we have
\begin{equation*}
A \leq \parenthese{\frac{196\nbDeux}{\eta^2} + 2\log(5\nbDeux d^2)} \sqrt{\frac{7}{3}} \frac{8\nbDeux d}{\eta} \sum_{i=0}^{\Ms-2} \frac{d^{1/2}\tildL + \sqrt{10} \Li^2 \mi^{-1/2}}{\kappai^2 \sigDi \mi^{1/2}} \eqsp.
\end{equation*}
Then, by \Cref{lemma:tech-s}-\ref{item:tech-s-2} and \Cref{lemma:tech-s}-\ref{item:tech-s-8}, and splitting the sum in two parts $k\leq \nbDeux[\inte]$ and $k>\nbDeux[\inte]$,
\begin{align*}
\sum_{i=0}^{\Ms-2} \frac{1}{\kappai^2 \sigDi \mi^{1/2}}
&\leq \sum_{k=0}^{\nbDeux-1} \sum_{i\in\morc[k]} \frac{(2^{k+1}\sig2_0)^{3/2}}{(1+\m 2^k \sig2_0)^{5/2}} \\
&\leq 4(d+4)\frac{2^{3/2}}{m^{3/2}} \sum_{k=0}^{\nbDeux-1} \frac{(m 2^k \sig2_0)^{3/2}}{(1+ m 2^k \sig2_0)^{7/2}} \\
&\leq 4(d+4)\frac{2^{3/2}}{m^{3/2}} \parenthese{\nbDeux[\inte]+\sum_{k=\nbDeux[\inte]+1}^{\nbDeux-1} (m 2^k \sig2_0)^{-2}} \\
&\leq 4(d+4) \frac{2^{3/2}}{m^{3/2}} \parenthese{\nbDeux[\inte]+\frac{4}{3}} \eqsp.
\end{align*}
We have similarly by \Cref{lemma:tech-s}-\ref{item:tech-s-2} and \Cref{lemma:tech-s}-\ref{item:tech-s-8},
\begin{equation*}
\sum_{i=0}^{\Ms-2} \frac{\Li^2 \mi^{-1/2}}{\kappai^2 \sigDi \mi^{1/2}} \leq 4(d+4) \parenthese{\frac{m+L}{2m}}^2 \parenthese{\nbDeux[\inte]+2} \eqsp .
\end{equation*}
Combining these inequalities with
\begin{equation}\label{eq:cost-M-1-strictly-convex-UC3}
n_{\Ms-1} + N_{\Ms-1} \leq \parenthese{\frac{19}{\eta^2}+1} \sqrt{\frac{7}{3}}\frac{4}{\eta} \defEns{\frac{d^{1/2}\tildL}{m^{3/2}} + \sqrt{10}\parenthese{\frac{m+L}{2m}}^2 } \eqsp,
\end{equation}
and replacing $\eta$ by $(\epsilon\sqrt{\mu})/8$ establish \eqref{eq:thm-stronglyConvex-UC3}.

\begin{proof}[Proof of \Cref{corollary:thms-median-trick-strongly}]
 Let $N=\ceil{4\log(\tilde{\mu}^{-1})}$ and $(\Zhat_i)_{i  \in \{1,\ldots,2N+1\}}$ be $2N+1$ independent outputs of the algorithms of \Cref{thm:costAlgorithmStronglyConvex,thm:costAlgorithmStronglyConvex-UC3} with $\mu=1/4$, sorted by increasing order. Denote by $\Zhat = \Zhat_{N+1}$ the median of $(\Zhat_i)_{i  \in \{1,\ldots,2N+1\}}$.  In addition, define the independent Bernoulli random variables $(W_i)_{i  \in \{1,\ldots,2N+1\}}$ by
 \begin{equation*}
 W_i= \1_{\mathsf{A}_i} \eqsp, \text{ where } \mathsf{A}_i = \defEns{\abs{\Zhat_i/\Z -1} \geq \epsilon} \eqsp.
 \end{equation*}
  Since $\Zhat$ is the median of $(\Zhat_i)_{i  \in \{1,\ldots,2N+1\}}$, we have
 \begin{equation*}
\PP\parenthese{\absolute{\Zhat/\Z - 1} > \epsilon}\leq \proba{\sum_{i=1}^{2N+1} W_i \geq N+1} \eqsp.
\end{equation*}
In addition since $\PP(W_i=1) \leq 1/4$, we have by \cite[Corollary 5.2]{proschan:sethuraman:1976}
\begin{equation*}
\proba{\sum_{i=1}^{2N+1} W_i \geq N+1} \leq \proba{\sum_{i=1}^{2N+1} \tilde{W}_i \geq N+1} \eqsp,
\end{equation*}
where $(\tilde{W}_i)_{i  \in \{1,\ldots,2N+1\}}$ are \iid~Bernoulli random variables with parameter $1/4$.
Then by Hoeffding's inequality \cite[Theorem 2.8]{boucheron2013concentration} and using for all $t \geq 1$, $8(t/2+3/4)^2/\{t(2t+1)\} \geq 1$, we get
\begin{align*}
\proba{\sum_{i=1}^{2N+1} \tilde{W}_i \geq N+1} &\leq \proba{\sum_{i=1}^{2N+1} \tilde{W}_i - (1/4)(2N+1) \geq N/2+3/4} \\
&\leq \exp\parenthese{\frac{-2(N/2+3/4)^2}{2N+1}}  \\
&\leq \exp(-N/4)\eqsp,
\end{align*}
which concludes the proof.
\end{proof}

\section*{Acknowledgements}

This work was supported by the Ecole Polytechnique Data Science Initiative. The authors thank the anonymous referees for useful suggestions which improved the manuscript.

\appendix
\section{Additional proofs of \Cref{sec:normalizingConstantStronglyConvex}}
\label{sec:proofsStronglyConvex-suppl}

\subsection{Proof of \Cref{lemma:parametersULA}}
\label{sec:proofLemmaParametersULA-suppl}

In this Section, the proof for the case $i=\Ms-1$ of \Cref{lemma:parametersULA} is dealt with.
Note that $a_{\Ms-1} = (2\sigma_{\Ms-1}^2)^{-1}$. By \Cref{prop:bias,prop:variance}, to check condition-\ref{eq:conditionsConvexFinalbiasVar} of \Cref{lemma:productOfErrors}, it is then sufficient for $\gamMs,\nMs,\NMs$ to satisfy,
\begin{align}
\nonumber
& \frac{4d}{\nMs\mMs\kappaMs\gamMs}\exp\left(-\NMs\frac{\kappaMs\gamMs}{2}\right) \\
\label{eq:conditionsBiasM-1}
& +2\kappaMs^{-1}\left( \AcMs{0}\gamMs+\AcMs{1}\gamMs^2 \right) \leq \frac{\eta^2 \sigma_{\Ms-1}^4}{\CcMs{2}+\CcMs{0} \CcMs{1}} \eqsp, \\
\label{eq:conditionsVarM-1}
& \frac{8\CcMs{0}\CcMs{1}}{\kappaMs^2 \nMs\gamMs}\left(1+\frac{2}{\kappaMs \nMs \gamMs}\right) \leq \sigma^4_{\Ms-1}\eta^2 \eqsp .
\end{align}
Then \eqref{eq:conditionsBiasM-1} and \eqref{eq:conditionsVarM-1} are satisfied if,
\begin{align}
\label{eq:conditions-gamM-1}
& 2\LMs^2 \kappaMs^{-1} d \gamMs+ 4d\LMs^4 \kappaMs^{-1} \mMs^{-1}\gamMs^2 \leq
\frac{\eta^2}{4}\frac{\kappaMs \mMs \sigma_{\Ms-1}^4}{\alphaMs d} \eqsp, \\
\label{eq:conditions-nM-1}
& \frac{1}{\nMs}\left(1+\frac{2}{\kappaMs \nMs \gamMs}\right) \leq
\frac{\eta^2 \sigma^4_{\Ms-1} \kappaMs^3 \gamMs}{8\betaMs d} \eqsp, \\
\label{eq:conditions-NM-1}
& -2(\kappaMs\gamMs)^{-1}\log\parenthese{\frac{\eta^2\sigma_{\Ms-1}^4 \mMs^2 \nMs\kappaMs\gamMs}{8 \alphaMs d^2}} = \NMsbar \leq \NMs \eqsp.
\end{align}
Denote by $\gamMsbar$ and $\nMsbar^{-1}$ the positive roots associated to \eqref{eq:conditions-gamM-1} and \eqref{eq:conditions-nM-1} seen as equalities. We have:
\begin{align}
\label{eq:exact-gamMbar}
\gamMsbar &= 4^{-1} \LMs^{-2} \mMs \parenthese{-1 + \sqrt{1+\frac{\eta^2 \kappaMs^2 \sigma_{\Ms-1}^4}{\alphaMs d^2}}} \eqsp, \\
\label{eq:exact-nMbar}
\nMsbar^{-1} &= 4^{-1} \kappaMs\gamMs \parenthese{-1 + \sqrt{1+\frac{\eta^2 \kappaMs^2 \sigma^4_{\Ms-1}}{\betaMs d}}} \eqsp.
\end{align}
Note that for \eqref{eq:conditions-gamM-1} and \eqref{eq:conditions-nM-1} to hold, it suffices that $\gamMs \leq \gamMsbar$ and $\nMs \geq \nMsbar$. We now lower bound $\gamMsbar$ and upper bound $\nMsbar$.

Using that $t\geq 0$, $\sqrt{1+t}\geq 1 + 2^{-1} t (1+t)^{-1/2}$ for $t=(\eta^2\kappa^2_{\Ms-1}\sigma^4_{\Ms-1})/(\alpha_{\Ms-1} d^2)$ and $\kappa_{\Ms-1}\sigma_{\Ms-1}^2d^{-1} \geq 2$, $\alphaMs \geq 4$ concludes that if \eqref{eq:gam-M-1} holds then $\gamMs \leq \gamMsbar$. The fact that $\gamMs \leq (\m_{\Ms-1}+L_{\Ms-1})^{-1}$ can be checked by simple algebra.

Then using that $\sqrt{1+t}\geq 1 + 2^{-1} t (1+t)^{-1/2}$ for $t = (\eta^2 \kappa_{\Ms-1}^2\sigma_{\Ms-1}^4)/(\beta_{\Ms-1} d)$ and $\kappa_{\Ms-1}\sigma_{\Ms-1}^2 \geq 10$, $\betaMs \geq 1$ concludes that if \eqref{eq:n-M-1} holds then $\nMs \geq \nMsbar$.
Finally, by \eqref{eq:n-M-1}, we get
\begin{equation*}
\NMsbar \leq (\kappa_{\Ms-1}\gamma_{\Ms-1})^{-1} \log(7/3) \eqsp,
\end{equation*}
which concludes that \eqref{eq:N-M-1} implies \eqref{eq:conditions-NM-1}.


\subsection{Proof of \Cref{lemma:parametersULA-UC3}}
\label{sec:proofLemmaParametersULA-UC3-suppl}

Let $k\in\defEns{0,\ldots,\nbDeux-1}$ and $i\in\morc[k]$. Assume that $\gami \leq (\mi+\Li)^{-1}$. The proof of \Cref{lemma:parametersULA} only needs to be slightly adapted. More precisely, \Cref{prop:bias2} is applied instead of \Cref{prop:bias}. By \eqref{eq:defB0} and \eqref{eq:defB1}, we have
\begin{equation}\label{eq:bounds-Bci}
\Bci{0} \leq 3^{-1} d \kappai^{-1}(d\tildL^2 + 10\Li^4\mi^{-1}) \eqsp , \quad
\Bci{1} \leq (25/12) d\Li^{4}\mi^{-1} \eqsp.
\end{equation}
It is sufficient for $\gami, \ni, \Ni$ to satisfy \eqref{eq:conditionsBias} and \eqref{eq:conditionsVar} with $\Aci{0}\gami+\Aci{1}\gami^2$ replaced by $\Bci{0}\gami^2+\Bci{1}\gami^3$. The counterpart of \eqref{eq:conditions-gami} is then
\begin{equation}
\label{eq:conditionsB0B1}
\frac{1}{3\kappai} \parenthese{d\tildL^2 + 10\Li^4\mi^{-1}} \gami^2 + \frac{25}{12} \Li^{4}\mi^{-1} \gami^3 \leq \frac{\eta^2 \kappai \mi \sigma_i^4}{16 \nbDeux^2 \alphai d^2} \eqsp.
\end{equation}
Since $\gami \leq 1/(\mi+\Li)$ and $\kappai \leq \Li$, we have
\[ (3\kappai)^{-1} \parenthese{d\tildL^2 + 10\Li^4\mi^{-1}} \geq (25/12) \Li^{4}\mi^{-1} \gami \eqsp,
\]
which establishes that if \eqref{eq:parameter-gami-UC3} holds, then \eqref{eq:conditionsB0B1} is satisfied. $\gami \leq (\mi+\Li)^{-1}$ can be checked by simple algebra. For $i=M-1$, the conclusion follows from $\m_{M-1}\sigma_{M-1}^2 d^{-1} \geq 2$ because $\sig2_{M-1} \geq (2d+7)/m$.

\section{Additional proofs of \Cref{sec:NormalizingconstantConvexGradientLipU}}
\label{sec:proofsConvex}

First, we state a technical lemma that gathers useful bounds.
We recall that $\Mc$ is defined in this Section by \eqref{eq:def-M-convex},
\begin{equation*}
\Mc = \inf \defEns{ i\geq 1 : \sig2_{i-1} \geq \rayon^2 } \eqsp .
\end{equation*}

\begin{lemma}\label{lemma:tech-c}
Assume \Cref{assumption:C1GradientLipschitz} and \Cref{assumption:stronglyConvex}($m$) for $m \geq 0$. Let $\suite{\sig2}{0}{M-1}$ defined by \eqref{eq:def-sigma_i-convex} for $\sig2_0$ given in \eqref{eq:def-sig20} and $M$ in \eqref{eq:def-M-convex}.
\begin{enumerate}
\item \label{item:tech-c-1}
$\nbDeux \leq \left\lceil (1/\log(2))\log \left( \sigma_0^{-2} \rho^{-2}d^2(\tau+1)^2\right) \right\rceil$ where $\nbDeux$ is defined in \eqref{eq:def-nbDeux}.
\item \label{item:tech-c-2}
For $k\in\defEns{0,\ldots,\nbDeux-1}$ and $i\in\morc[k]$,
$2^{k+1}\sig2_0 \ai\cmorc[k] \leq 1$
where $\ai$ is defined in \eqref{eq:heuristic-set-ai} (with $m=0$) and $\morc[k]$ in \eqref{eq:def-chunk}. As a consequence, $\cmorc[k] \leq 4(d+4)$.
\item \label{item:tech-c-3}
For $i\in\{0,\ldots,\Mc-1\}$, $\kappai\sigDi \in \ccint{1,2}$.
\item \label{item:tech-c-4}
$\sig2_{\Mc-1} \in \ccint{\rayon^2, (10/9)\rayon^2}$.
\item \label{item:tech-c-5}
For all $i\in\{0,\ldots,\Mc-1\}$ and $\gami\leq 1/(\mi+\Li)$,
there exist $\alphai\in\ccint{4,14}$ and $\betai\in\ccint{1,10}$ such that $\Cci{2}+\Cci{0} \Cci{1} = \alphai d \mi^{-1}$ and $\Cci{0}\Cci{1} = \betai d \kappai^{-1}$ where $\Cci{0},\Cci{1},\Cci{2}$ and $\kappai$ are given in \eqref{eq:defC} and \eqref{eq:defkappa} respectively.
\item \label{item:tech-c-6}
For all $i\in\{0,\ldots,\Mc-1\}$,
$0< \Aci{1} \leq 4d\Li^4 \kappai^{-1} \mi^{-1}$, where $\Li,\mi$ and $\kappai$ are given in \eqref{eq:def-Li-mi} and \eqref{eq:defkappa} respectively.
\end{enumerate}
\end{lemma}

\begin{proof}
The proofs of 1,2,5,6 are identical to the ones of \Cref{lemma:tech-s}.
\begin{itemize}
\item[3.] $\kappai\sigDi = (2\Li)/(\mi+\Li)$.
\item[4.]
By definition of $\Mc$, $\sigma_{\Mc-2}^2 \leq \rayon^2$ and $a_{\Mc-2} \leq \sigma_{\Mc-2}^{-2} / \{4(d+4)\}$. By \eqref{eq:defgi}, we get:
\begin{equation*}
\sigma_{\Mc-1}^{-2} = \sigma_{\Mc-2}^{-2} - 2a_{\Mc-2} \geq \sigma_{\Mc-2}^{-2} \parenthese{1-\frac{1}{2(d+4)}}
\end{equation*}
that is $\sigma_{\Mc-1}^2 \leq (10/9) \sigma_{\Mc-2}^2 \leq (10/9) \rayon^2$.
\end{itemize}
\end{proof}

\subsection{Proof of \Cref{lemma:parametersULA-convex}}
\label{sec:proofLemmaParametersULA-convex}

Let $k\in\defEns{0,\ldots,\nbDeux-1}$ and $i\in\morc[k]$. Assume that $\gami \leq (\mi+\Li)^{-1}$. The proof follows the same lines as the one in \Cref{sec:proofLemmaParametersULA}. By \Cref{lemma:tech-c}-\ref{item:tech-c-5} and \Cref{lemma:tech-c}-\ref{item:tech-c-6}, to check condition-\ref{eq:conditionsConvexbiasVar} of \Cref{lemma:productOfErrors}, it suffices that $\gami \leq \gamibar$, $\ni \geq \nibar$ and $\Ni$ satisfies \eqref{eq:conditions-Ni}, where $\gamibar$ is defined in \eqref{eq:exact-gamibar} and $\nibar$ in \eqref{eq:exact-nibar}.

Using that $\sqrt{1+t}\geq 1 + 2^{-1} t (1+t)^{-1/2}$ for $t=(\eta^2\kappai^2 \sigma_i^4)/(4\alphai \nbDeux^2 d^2)$ and by \Cref{lemma:tech-c}-\ref{item:tech-c-3}, concludes that if \eqref{eq:parameter-gami-convex} holds then $\gami \leq \gamibar$. $\gami \leq (\mi+\Li)^{-1}$ can be checked by simple algebra.

By \eqref{eq:heuristic-set-ai} (with $m=0$) and the definition of $\morc[k]$, $\ai \leq \sigma_i^{-2} /\{4(d+4)\}$,
\begin{equation*}
  \nibar^{-1} \geq  4^{-1} \kappai\gami \parenthese{-1 + \sqrt{1+\frac{\eta^2 \kappai^2 \sigma_i^4 (d+4)}{\nbDeux\betai d}}} \eqsp.
\end{equation*}
Using that $\sqrt{1+t}\geq 1 + 2^{-1} t (1+t)^{-1/2}$ for $t=(\eta^2\kappai^2 \sigma_i^4)/(4\alphai \nbDeux^2 d^2)$ and by \Cref{lemma:tech-c}-\ref{item:tech-c-3}, concludes that if \eqref{eq:parameter-ni-convex} holds then $\ni \geq \nibar$. Finally, by \eqref{eq:parameter-ni-convex}, if \eqref{eq:parameter-Ni-convex} holds, \eqref{eq:conditions-Ni} is satisfied.


The case $i=\Mc-1$ is different because $\gbarMc$ is Lipschitz. Assume $\gamMc \leq (\mMc+\LMc)^{-1}$.
\cite[section 2.1]{durmusSampling} entails that condition-\ref{eq:conditionsConvexFinalbiasVar} of \Cref{lemma:productOfErrors} is satisfied if
\begin{align}
\nonumber
\Lip{\gbarMc}^2 \bigg \{
\frac{4d}{\nMc\mMc\kappaMc\gamMc}\exp\left(-\NMc\frac{\kappaMc\gamMc}{2}\right) & \\
\label{eq:conditions-gMbar-biais}
+ 2\kappaMc^{-1}\left( \AcMc{0}\gamMc+\AcMc{1}\gamMc^2 \right) \bigg \} &\leq \eta^2  \eqsp, \\
\nonumber
\frac{8 \Lip{\gbarMc}^2}{\kappaMc^2 \nMc\gamMc}
\left\{ 1+\frac{2}{\nMc\kappaMc\gamMc}\right\} &\leq \eta^2 \eqsp.
\end{align}
Using $\Lip{\gbarMc}^2 \leq (\sigma_{\Mc-1}^{-2} \rme)$ and \eqref{eq:defA0}, \Cref{lemma:tech-c}-\ref{item:tech-c-6} for $i=\Mc-1$, it is sufficient for $\gamMc, \nMc,\nMc$ to satisfy
\begin{align}
\label{eq:conditions-gamM}
2\LMc^2\kappaMc^{-1} d \gamMc + 4 d \LMc^4 \kappaMc^{-1} \mMc^{-1} \gamMc^2 &\leq (4\rme)^{-1} \kappaMc \eta^2 \sigma_{\Mc-1}^2  \eqsp, \\
\label{eq:conditions-nM}
\nMc^{-1} \parenthese{1+2(\kappaMc\gamMc\nMc)^{-1}} &\leq \frac{\eta^2\kappaMc^2\sigma_{Mc-1}^2 \gamMc}{8\rme}  \eqsp, \\
\label{eq:conditions-NM}
-2\frac{\log \parenthese{(8\rme d)^{-1}\nMc\kappaMc\gamMc\eta^2}}{\kappaMc\gamMc} &= \NMcbar \leq \NMc \eqsp.
\end{align}
Denote by $\gamMcbar, \nMcbar^{-1}$ the roots of \eqref{eq:conditions-gamM}, \eqref{eq:conditions-nM} seen as equalities. We have
\begin{align}
\label{eq:exact-gamMbar}
\gamMcbar &= 4^{-1} \LMc^{-2} \sigma_{Mc-1}^{-2} \defEns{-1+\sqrt{1+\frac{\eta^2 \kappaMc^2 \sigma_{\Mc-1}^4}{\rme d}}}  \eqsp, \\
\label{eq:exact-nMbar}
\nMcbar^{-1} &= 4^{-1} \kappaMc \gamMc \defEns{-1+\sqrt{1+ \rme^{-1} \eta^2 \kappaMc \sigma_{\Mc-1}^2}} \eqsp.
\end{align}

Using that $\sqrt{1+t}\geq 1 + 2^{-1} t (1+t)^{-1/2}$ for $t = (\eta^2 \kappaMc^2 \sigma_{\Mc-1}^4)/(\rme d)$ and by \Cref{lemma:tech-c}-\ref{item:tech-c-3}, concludes that if \eqref{eq:gam-M-1-convex} holds then $\gamMc \leq \gamMcbar$.
$\gamMc \leq (\mMc+\LMc)^{-1}$ can be checked by simple algebra.

Using that $\sqrt{1+t}\geq 1 + 2^{-1} t (1+t)^{-1/2}$ for $t = \rme^{-1} \eta^2 \kappaMc \sigma_{\Mc-1}^2$ and by \Cref{lemma:tech-c}-\ref{item:tech-c-3}, concludes that if \eqref{eq:n-M-1-convex} holds then $\nMc \geq \nMcbar$.

Finally by \eqref{eq:n-M-1-convex}, if \eqref{eq:N-M-1-convex} holds, \eqref{eq:conditions-NM} is satisfied.


\subsection{Proof of \Cref{lemma:parametersULA-UC3-convex}}
\label{sec:proofLemmaParametersULA-UC3-convex}

The proof is identical to the one of \Cref{lemma:parametersULA-UC3}. For $k\in\defEns{0,\ldots,\nbDeux-1}$ and $i\in\morc[k]$, it is sufficient for $\gami$ to satisfy \eqref{eq:parameter-gami-UC3-convex} by \eqref{eq:parameter-gami-UC3} and \Cref{lemma:tech-c}-\ref{item:tech-c-3}.

Regarding the case $i=\Mc-1$, assuming that $\gamMc \leq (\mMc+\LMc)^{-1}$, it is sufficient for $\gamMc,\nMc,\NMc$ to satisfy \eqref{eq:conditions-gMbar-biais} with $\AcMc{0}\gamMc+\AcMc{1}\gamMc^2$ replaced by $\BcMc{0}\gamMc^2+\BcMc{1}\gamMc^3$. The counterpart of \eqref{eq:conditions-gamM} is then,
\begin{equation*}
\frac{1}{3\kappaMc} \parenthese{d\tildL^2 + 10\LMc^4\mMc^{-1}} \gamMc^2 + \frac{25}{12} \frac{\LMc^{4}}{\mMc} \gamMc^3 \leq \frac{\eta^2 \kappaMc \sigma_{\Mc-1}^2}{4 \rme d} \eqsp.
\end{equation*}
This concludes the proof with the same argument as in \Cref{sec:proofLemmaParametersULA-UC3-suppl}.

\subsection{Proof of \Cref{thm:costAlgorithmConvex,thm:costAlgorithmConvex-UC3,corollary:thms-median-trick-convex}}
\label{sec:proofCostAlgorithmConvex}

For $i\in\defEns{0,\ldots,\Mc-1}$, set $\gami, \ni, \Ni$ such that \eqref{eq:parameter-gami-convex}, \eqref{eq:parameter-ni-convex}, \eqref{eq:parameter-Ni-convex}, \eqref{eq:gam-M-1-convex}, \eqref{eq:n-M-1-convex} and \eqref{eq:N-M-1-convex} are equalities.
By \eqref{eq:def-cost}, we have
\begin{equation*}
\cost = \parenthese{\frac{453\nbDeux}{\eta^2} + 2\log(\nbDeux d^2)} \frac{462\nbDeux^2 d^2}{\eta^2} \sum_{i=0}^{\Mc-2} \kappai^{-1}\Li^{2}\sigDi
+ n_{\Mc-1} + N_{\Mc-1}  \eqsp.
\end{equation*}
Note that for $i\in\defEns{0, \ldots, \Mc-2}$,
\begin{equation*}
\kappai^{-1}\Li^{2}\sigDi = 1+(3/2)\Llip \sigDi + (\Llip^2/2)\sigma_i^4 \eqsp.
\end{equation*}
By \Cref{lemma:tech-c}-\ref{item:tech-c-2}, for $k\in\defEns{0,\ldots,\nbDeux-1}$, $\cmorc[k] \leq 4(d+4)$ and for $i\in\morc[k]$, $\sigDi \leq 2^{k+1}\sig2_0$. We then have
\begin{align*}
\sum_{i=0}^{\Mc-2} \frac{\Li^{2}\sigDi}{\kappai} &\leq 4(d+4) \sum_{k=0}^{\nbDeux-1} \defEns{ 1+\frac{3L}{2}2^{k+1}\sig2_0 + \frac{L^2}{2}(2^{k+1}\sig2_0)^2 } \\
&\leq 4(d+4) \defEns{ \nbDeux +3L(2^{\nbDeux} \sig2_0) +\frac{2L^2}{3}(2^{\nbDeux} \sig2_0)^2} \eqsp.
\end{align*}
By \eqref{eq:def-M-convex} and the definition of $\nbDeux$, \eqref{eq:def-nbDeux}, $2^{\nbDeux} \sig2_0 \leq 2 \rayon^2$. The expressions of $\gamMc,\nMc,\NMc$ give
\begin{equation*}
n_{\Mc-1} + N_{\Mc-1} = \parenthese{\frac{29}{\eta^2}+2\log(d)}\frac{26 d}{\eta^2} \kappaMc^{-2} \LMc^2 \eqsp,
\end{equation*}
with $\kappaMc^{-2}\LMc^2 = (1+2^{-1}L\sigma_{\Mc-1}^2)^2$. By \Cref{lemma:tech-c}-\ref{item:tech-c-4}, we then have
\begin{equation}\label{eq:cost-M-1-convex}
n_{\Mc-1} + N_{\Mc-1} \leq \parenthese{\frac{29}{\eta^2}+2\log(d)}\frac{26 d}{\eta^2} \parenthese{1+\frac{5L}{9}\rayon^2}^2 \eqsp,
\end{equation}
and \eqref{eq:thm-Convex} is established.


Assume \Cref{assumption:hC3}. We now prove \Cref{thm:costAlgorithmConvex-UC3} and use \Cref{lemma:parametersULA-UC3} instead of \Cref{lemma:parametersULA}.
For $i\in\defEns{0,\ldots,\Mc-1}$, set $\gami, \ni, \Ni$ such that \eqref{eq:parameter-gami-UC3-convex}, \eqref{eq:parameter-ni-convex}, \eqref{eq:parameter-Ni-convex}, \eqref{eq:gam-M-1-UC3-convex}, \eqref{eq:n-M-1-convex} and \eqref{eq:N-M-1-convex} are equalities.
By \eqref{eq:def-cost} and using that for $a,b\geq 0$, $\sqrt{a+b}\leq \sqrt{a}+\sqrt{b}$, we have
\begin{equation*}
\cost \leq \parenthese{\frac{453\nbDeux}{\eta^2} + 2\log(\nbDeux d^2)} \sqrt{\frac{7}{3}} \frac{8\nbDeux d}{\eta} \sum_{i=0}^{\Mc-2} \frac{\sigma_i}{\kappai} \parenthese{d^{1/2}\tildL + \sqrt{10} \Li^2 \sigma_i}
+ n_{\Mc-1} + N_{\Mc-1}  \eqsp.
\end{equation*}
For $k\in\defEns{0,\ldots,\nbDeux-1}$ and $i\in\morc[k]$, note that
\begin{equation*}
\kappai^{-1} \sigma_i \leq \sigma_i^3 \eqsp , \quad
\kappai^{-1} \sigDi \Li^2 = 1 + \frac{3L}{2}\sigDi + \frac{L^2}{2}\sigma_i^4 \eqsp.
\end{equation*}
Using for $k\in\defEns{0,\ldots,\nbDeux-1}$, $\cmorc[k] \leq 4(d+4)$ by \Cref{lemma:tech-c}-\ref{item:tech-c-2} and for $i\in\morc[k]$, $\sigDi \leq 2^{k+1}\sig2_0$, we get
\begin{align*}
\sum_{i=0}^{\Mc-2} \frac{\sigma_i}{\kappai} \parenthese{d^{1/2}\tildL + \sqrt{10} \Li^2 \sigma_i} &\leq 4(d+4) \sum_{k=0}^{\nbDeux-1} \bigg\{ d^{1/2}\tildL(2^{k+1}\sig2_0)^{3/2} \\
&+\sqrt{10}\parenthese{1+\frac{3L}{2}(2^{k+1}\sig2_0) + \frac{L^2}{2}(2^{k+1}\sig2_0)^2} \bigg\}  \\
&\leq 4(d+4) \defEns{ 5d^{1/2}\tildL\rayon^3 + \sqrt{10}\parenthese{\nbDeux + 6L\rayon^2 + \frac{8L^2}{3}\rayon^4} } \eqsp,
\end{align*}
with $2^{\nbDeux} \sig2_0 \leq 2\rayon^2$. The expressions of $\gamma_{\Mc-1}, n_{\Mc-1}, N_{\Mc-1}$ give
\begin{equation*}
\nMc+\NMc \leq \parenthese{2\log(d) + \frac{29}{\eta^2}}\sqrt{\frac{8\rme}{3}}\frac{\sqrt{d}}{\eta}\frac{d^{1/2}\tildL + \sqrt{10}\LMc^2\sigma_{\Mc-1}}{\kappaMc^2\sigma_{\Mc-1}} \eqsp.
\end{equation*}
By \Cref{lemma:tech-c}-\ref{item:tech-c-4}, $\sigma_{\Mc-1}^2 \in \ccint{\rayon^2, (10/9)\rayon^2}$. We get then
\begin{equation*}
\kappaMc^{-2}\sigma_{\Mc-1}^{-1} = \frac{\parenthese{1+(L/2)\sigma_{\Mc-1}^2}^2}{L^2\sigma_{\Mc-1}} \leq \frac{1}{\rayon L^2}\parenthese{1+\frac{5L}{9}\rayon^2}^2 \eqsp ,
\end{equation*}
and,
\begin{equation*}
\kappaMc^{-2}\LMc^{2} = \parenthese{1+\frac{L}{2}\sigma_{\Mc-1}^2}^2 \leq \parenthese{1+\frac{5L}{9}\rayon^2}^2 \eqsp ,
\end{equation*}
which gives,
\begin{equation}\label{eq:cost-M-1-convex-UC3}
\nMc+\NMc \leq \parenthese{2\log(d) + \frac{29}{\eta^2}}\sqrt{\frac{8\rme}{3}}\frac{\sqrt{d}}{\eta}\parenthese{1+\frac{5L}{9}\rayon^2}^2
\parenthese{\frac{d^{1/2}\tildL}{\rayon L^2} + \sqrt{10}} \eqsp .
\end{equation}
\eqref{eq:thm-Convex-UC3} is established. The proof of \Cref{corollary:thms-median-trick-convex} is the same as the one of \Cref{corollary:thms-median-trick-strongly}.

\def\sectionautorefname{Section}
\def\subsectionautorefname{Section}
\def\subsubsectionautorefname{Section}
\def\corollaryautorefname{Corollary}

\bibliographystyle{abbrvnat}
\bibliography{bibliographie}

\begin{thebibliography}{49}
\providecommand{\natexlab}[1]{#1}
\providecommand{\url}[1]{\texttt{#1}}
\expandafter\ifx\csname urlstyle\endcsname\relax
  \providecommand{\doi}[1]{doi: #1}\else
  \providecommand{\doi}{doi: \begingroup \urlstyle{rm}\Url}\fi

\bibitem[{Andrieu} et~al.(2016){Andrieu}, {Ridgway}, and
  {Whiteley}]{christophe:2016}
C.~{Andrieu}, J.~{Ridgway}, and N.~{Whiteley}.
\newblock {Sampling normalizing constants in high dimensions using
  inhomogeneous diffusions}.
\newblock \emph{ArXiv e-prints}, Dec. 2016.

\bibitem[Ardia et~al.(2012)Ardia, Ba{\c{s}}t{\"u}rk, Hoogerheide, and
  Van~Dijk]{ardia:2012}
D.~Ardia, N.~Ba{\c{s}}t{\"u}rk, L.~Hoogerheide, and H.~K. Van~Dijk.
\newblock A comparative study of {M}onte {C}arlo methods for efficient
  evaluation of marginal likelihood.
\newblock \emph{Computational Statistics \& Data Analysis}, 56\penalty0
  (11):\penalty0 3398--3414, 2012.

\bibitem[Balian(2007)]{balian2007microphysics}
R.~Balian.
\newblock \emph{From microphysics to macrophysics: methods and applications of
  statistical physics}, volume~1.
\newblock Springer Science \& Business Media, 2007.

\bibitem[Behrens et~al.(2012)Behrens, Friel, and Hurn]{Friel:2012}
G.~Behrens, N.~Friel, and M.~Hurn.
\newblock Tuning tempered transitions.
\newblock \emph{Statistics and Computing}, 22\penalty0 (1):\penalty0 65--78,
  2012.
\newblock \doi{10.1007/s11222-010-9206-z}.
\newblock URL \url{http://dx.doi.org/10.1007/s11222-010-9206-z}.

\bibitem[Beskos et~al.(2014)Beskos, Crisan, Jasra, and
  Whiteley]{beskos:cste:2014}
A.~Beskos, D.~O. Crisan, A.~Jasra, and N.~Whiteley.
\newblock Error bounds and normalising constants for sequential {M}onte {C}arlo
  samplers in high dimensions.
\newblock \emph{Adv. in Appl. Probab.}, 46\penalty0 (1):\penalty0 279--306, 03
  2014.
\newblock \doi{10.1239/aap/1396360114}.
\newblock URL \url{http://dx.doi.org/10.1239/aap/1396360114}.

\bibitem[Boucheron et~al.(2013)Boucheron, Lugosi, and
  Massart]{boucheron2013concentration}
S.~Boucheron, G.~Lugosi, and P.~Massart.
\newblock \emph{Concentration inequalities: a nonasymptotic theory of
  independence}.
\newblock Oxford university press, 2013.

\bibitem[Brazitikos et~al.(2014)Brazitikos, Giannopoulos, Valettas, and
  Vritsiou]{brazitikos2014geometry}
S.~Brazitikos, A.~Giannopoulos, P.~Valettas, and B.-H. Vritsiou.
\newblock \emph{Geometry of isotropic convex bodies}, volume 196.
\newblock American Mathematical Society Providence, 2014.

\bibitem[Brosse et~al.(2017)Brosse, Durmus, and Moulines]{Supplement}
N.~Brosse, A.~Durmus, and E.~Moulines.
\newblock {Supplement to ''Normalizing constants of log-concave densities''}.
\newblock working paper or preprint, 2017.

\bibitem[Chen et~al.(2000)Chen, Shao, and Ibrahim]{chenmonte}
M.~Chen, Q.~Shao, and J.~Ibrahim.
\newblock \emph{Monte Carlo methods in Bayesian computation}.
\newblock Springer, New York, 2000.

\bibitem[Cousins and Vempala(2015)]{cousins2015bypassing}
B.~Cousins and S.~Vempala.
\newblock Bypassing {KLS}: Gaussian cooling and an $\text{O}^{*}(n^3)$ volume
  algorithm.
\newblock In \emph{Proceedings of the Forty-Seventh Annual ACM on Symposium on
  Theory of Computing}, pages 539--548. ACM, 2015.

\bibitem[Dalalyan(2016)]{dalalyan2016theoretical}
A.~S. Dalalyan.
\newblock Theoretical guarantees for approximate sampling from smooth and
  log-concave densities.
\newblock \emph{Journal of the Royal Statistical Society: Series B (Statistical
  Methodology)}, 2016.

\bibitem[Del~Moral(2004)]{delmoral:2004:book}
P.~Del~Moral.
\newblock \emph{Feynman-{K}ac formulae}.
\newblock Probability and its Applications (New York). Springer-Verlag, New
  York, 2004.
\newblock ISBN 0-387-20268-4.
\newblock URL \url{https://doi.org/10.1007/978-1-4684-9393-1}.
\newblock Genealogical and interacting particle systems with applications.

\bibitem[{Del Moral} et~al.(2016){Del Moral}, {Jasra}, {Law}, and
  {Zhou}]{Moral:Jasra:2016}
P.~{Del Moral}, A.~{Jasra}, K.~{Law}, and Y.~{Zhou}.
\newblock {Multilevel sequential Monte Carlo samplers for normalizing
  constants}.
\newblock \emph{ArXiv e-prints}, Mar. 2016.

\bibitem[{Durmus} and {Moulines}(2015)]{durmusNonAsymp}
A.~{Durmus} and E.~{Moulines}.
\newblock {Non-asymptotic convergence analysis for the unadjusted Langevin
  algorithm}.
\newblock July 2015.

\bibitem[{Durmus} and {Moulines}(2016)]{durmusSampling}
A.~{Durmus} and E.~{Moulines}.
\newblock {High-dimensional Bayesian inference via the unadjusted Langevin
  algorithm}.
\newblock May 2016.

\bibitem[Dutta et~al.(2013)Dutta, Ghosh, et~al.]{dutta2013bayes}
R.~Dutta, J.~K. Ghosh, et~al.
\newblock Bayes model selection with path sampling: factor models and other
  examples.
\newblock \emph{Statistical Science}, 28\penalty0 (1):\penalty0 95--115, 2013.

\bibitem[Dyer and Frieze(1991)]{dyer1991computing}
M.~Dyer and A.~Frieze.
\newblock Computing the volume of convex bodies: a case where randomness
  provably helps.
\newblock \emph{Probabilistic combinatorics and its applications}, 44:\penalty0
  123--170, 1991.

\bibitem[Ermak(1975)]{ermak:1975}
D.~L. Ermak.
\newblock A computer simulation of charged particles in solution. i. technique
  and equilibrium properties.
\newblock \emph{The Journal of Chemical Physics}, 62\penalty0 (10):\penalty0
  4189--4196, 1975.

\bibitem[Evans and Gariepy(2015)]{evans2015measure}
L.~C. Evans and R.~F. Gariepy.
\newblock \emph{Measure theory and fine properties of functions}.
\newblock CRC press, 2015.

\bibitem[Friel and Wyse(2012)]{friel2012estimating}
N.~Friel and J.~Wyse.
\newblock Estimating the evidence--a review.
\newblock \emph{Statistica Neerlandica}, 66\penalty0 (3):\penalty0 288--308,
  2012.

\bibitem[Friel et~al.(2014)Friel, Hurn, and Wyse]{Friel2014}
N.~Friel, M.~Hurn, and J.~Wyse.
\newblock Improving power posterior estimation of statistical evidence.
\newblock \emph{Statistics and Computing}, 24\penalty0 (5):\penalty0 709--723,
  2014.
\newblock ISSN 1573-1375.
\newblock \doi{10.1007/s11222-013-9397-1}.
\newblock URL \url{http://dx.doi.org/10.1007/s11222-013-9397-1}.

\bibitem[Gelman and Meng(1998)]{gelman1998simulating}
A.~Gelman and X.-L. Meng.
\newblock Simulating normalizing constants: From importance sampling to bridge
  sampling to path sampling.
\newblock \emph{Statistical science}, pages 163--185, 1998.

\bibitem[Huber(2015)]{huber2015}
M.~Huber.
\newblock Approximation algorithms for the normalizing constant of {G}ibbs
  distributions.
\newblock \emph{Ann. Appl. Probab.}, 25\penalty0 (2):\penalty0 974--985, 04
  2015.
\newblock \doi{10.1214/14-AAP1015}.
\newblock URL \url{http://dx.doi.org/10.1214/14-AAP1015}.

\bibitem[Jarzynski(1997)]{jarzynski1997equilibrium}
C.~Jarzynski.
\newblock Equilibrium free-energy differences from nonequilibrium measurements:
  A master-equation approach.
\newblock \emph{Physical Review E}, 56\penalty0 (5):\penalty0 5018, 1997.

\bibitem[Jasra et~al.(2005)Jasra, Holmes, and
  Stephens]{jasra:holmes:stephens:2005}
A.~Jasra, C.~C. Holmes, and D.~A. Stephens.
\newblock Markov chain {M}onte {C}arlo methods and the label switching problem
  in {B}ayesian mixture modeling.
\newblock \emph{Statist. Sci.}, 20\penalty0 (1):\penalty0 50--67, 02 2005.
\newblock \doi{10.1214/088342305000000016}.
\newblock URL \url{https://doi.org/10.1214/088342305000000016}.

\bibitem[Jasra et~al.(2016)Jasra, Kamatani, Osei, and Zhou]{Jasra2016}
A.~Jasra, K.~Kamatani, P.~P. Osei, and Y.~Zhou.
\newblock Multilevel particle filters: normalizing constant estimation.
\newblock \emph{Statistics and Computing}, pages 1--14, 2016.
\newblock ISSN 1573-1375.
\newblock \doi{10.1007/s11222-016-9715-5}.
\newblock URL \url{http://dx.doi.org/10.1007/s11222-016-9715-5}.

\bibitem[Jerrum et~al.(1986)Jerrum, Valiant, and Vazirani]{JERRUM1986169}
M.~R. Jerrum, L.~G. Valiant, and V.~V. Vazirani.
\newblock Random generation of combinatorial structures from a uniform
  distribution.
\newblock \emph{Theoretical Computer Science}, 43:\penalty0 169 -- 188, 1986.
\newblock ISSN 0304-3975.
\newblock \doi{http://dx.doi.org/10.1016/0304-3975(86)90174-X}.
\newblock URL
  \url{http://www.sciencedirect.com/science/article/pii/030439758690174X}.

\bibitem[Joulin and Ollivier(2010)]{joulin:ollivier:2010}
A.~Joulin and Y.~Ollivier.
\newblock Curvature, concentration and error estimates for {M}arkov chain
  {M}onte {C}arlo.
\newblock \emph{Ann. Probab.}, 38\penalty0 (6):\penalty0 2418--2442, 11 2010.
\newblock \doi{10.1214/10-AOP541}.
\newblock URL \url{http://dx.doi.org/10.1214/10-AOP541}.

\bibitem[Knuth et~al.(2015)Knuth, Habeck, Malakar, Mubeen, and
  Placek]{Knuth2015}
K.~H. Knuth, M.~Habeck, N.~K. Malakar, A.~M. Mubeen, and B.~Placek.
\newblock Bayesian evidence and model selection.
\newblock \emph{Digital Signal Processing}, 47:\penalty0 50 -- 67, 2015.
\newblock ISSN 1051-2004.
\newblock \doi{http://dx.doi.org/10.1016/j.dsp.2015.06.012}.
\newblock URL
  \url{http://www.sciencedirect.com/science/article/pii/S1051200415001980}.
\newblock Special Issue in Honour of William J. (Bill) Fitzgerald.

\bibitem[Leli{\`e}vre et~al.(2010)Leli{\`e}vre, Stoltz, and
  Rousset]{lelievre2010free}
T.~Leli{\`e}vre, G.~Stoltz, and M.~Rousset.
\newblock \emph{Free energy computations: A mathematical perspective}.
\newblock World Scientific, 2010.

\bibitem[Marin and Robert(2009)]{marin2009importance}
J.-M. Marin and C.~P. Robert.
\newblock Importance sampling methods for {B}ayesian discrimination between
  embedded models.
\newblock \emph{arXiv preprint arXiv:0910.2325}, 2009.

\bibitem[Meyn and Tweedie(1993)]{meyn1993stability}
S.~P. Meyn and R.~L. Tweedie.
\newblock Stability of {M}arkovian processes iii: Foster-{L}yapunov criteria
  for continuous-time processes.
\newblock \emph{Advances in Applied Probability}, pages 518--548, 1993.

\bibitem[Moral et~al.(2006)Moral, Doucet, and Jasra]{moral:2006:SMC}
P.~D. Moral, A.~Doucet, and A.~Jasra.
\newblock Sequential {M}onte {C}arlo samplers.
\newblock \emph{Journal of the Royal Statistical Society. Series B (Statistical
  Methodology)}, 68\penalty0 (3):\penalty0 411--436, 2006.
\newblock ISSN 13697412, 14679868.
\newblock URL \url{http://www.jstor.org/stable/3879283}.

\bibitem[Neal(2001)]{Neal2001}
R.~M. Neal.
\newblock Annealed importance sampling.
\newblock \emph{Statistics and Computing}, 11\penalty0 (2):\penalty0 125--139,
  2001.
\newblock ISSN 1573-1375.
\newblock \doi{10.1023/A:1008923215028}.
\newblock URL \url{http://dx.doi.org/10.1023/A:1008923215028}.

\bibitem[Nesterov(2013)]{nesterov2013introductory}
Y.~Nesterov.
\newblock \emph{Introductory lectures on convex optimization: A basic course},
  volume~87.
\newblock Springer Science \& Business Media, 2013.

\bibitem[Niemiro and Pokarowski(2009)]{niemiro2009}
W.~Niemiro and P.~Pokarowski.
\newblock Fixed precision {MCMC} estimation by median of products of averages.
\newblock \emph{J. Appl. Probab.}, 46\penalty0 (2):\penalty0 309--329, 06 2009.
\newblock \doi{10.1239/jap/1245676089}.
\newblock URL \url{http://dx.doi.org/10.1239/jap/1245676089}.

\bibitem[Oates et~al.(2016)Oates, Papamarkou, and Girolami]{girolami:2016}
C.~J. Oates, T.~Papamarkou, and M.~Girolami.
\newblock The controlled thermodynamic integral for {B}ayesian model evidence
  evaluation.
\newblock \emph{Journal of the American Statistical Association}, 111\penalty0
  (514):\penalty0 634--645, 2016.
\newblock \doi{10.1080/01621459.2015.1021006}.
\newblock URL \url{http://dx.doi.org/10.1080/01621459.2015.1021006}.

\bibitem[Parisi(1981)]{parisi:1981}
G.~Parisi.
\newblock Correlation functions and computer simulations.
\newblock \emph{Nuclear Physics B}, 180:\penalty0 378--384, 1981.

\bibitem[Pereyra(2016)]{pereyra2016maximum}
M.~Pereyra.
\newblock Maximum-a-posteriori estimation with {B}ayesian confidence regions.
\newblock \emph{arXiv preprint arXiv:1602.08590}, 2016.

\bibitem[Proschan and Sethuraman(1976)]{proschan:sethuraman:1976}
F.~Proschan and J.~Sethuraman.
\newblock Stochastic comparisons of order statistics from heterogeneous
  populations, with applications in reliability.
\newblock \emph{Journal of Multivariate Analysis}, 6\penalty0 (4):\penalty0 608
  -- 616, 1976.
\newblock ISSN 0047-259X.
\newblock \doi{https://doi.org/10.1016/0047-259X(76)90008-7}.
\newblock URL
  \url{http://www.sciencedirect.com/science/article/pii/0047259X76900087}.

\bibitem[{R Core Team}(2018)]{R:2018}
{R Core Team}.
\newblock \emph{R: a language and environment for statistical computing}.
\newblock R Foundation for Statistical Computing, Vienna, Austria, 2018.
\newblock URL \url{https://www.R-project.org/}.

\bibitem[Richardson and Green(1997)]{richardson:green:1997}
S.~Richardson and P.~J. Green.
\newblock On {B}ayesian analysis of mixtures with an unknown number of
  components (with discussion).
\newblock \emph{Journal of the Royal Statistical Society: Series B (Statistical
  Methodology)}, 59\penalty0 (4):\penalty0 731--792, 1997.
\newblock ISSN 1467-9868.
\newblock \doi{10.1111/1467-9868.00095}.
\newblock URL \url{http://dx.doi.org/10.1111/1467-9868.00095}.

\bibitem[Robert(2007)]{robert2007bayesian}
C.~Robert.
\newblock \emph{The Bayesian choice: from decision-theoretic foundations to
  computational implementation}.
\newblock Springer Science \& Business Media, 2007.

\bibitem[Roberts and Tweedie(1996)]{roberts:tweedie:1996}
G.~O. Roberts and R.~L. Tweedie.
\newblock Exponential convergence of {L}angevin distributions and their
  discrete approximations.
\newblock \emph{Bernoulli}, 2\penalty0 (4):\penalty0 341--363, 1996.
\newblock ISSN 1350-7265.
\newblock \doi{10.2307/3318418}.
\newblock URL \url{http://dx.doi.org/10.2307/3318418}.

\bibitem[Valleau and Card(1972)]{Valleau:1972}
J.~P. Valleau and D.~N. Card.
\newblock Monte {C}arlo estimation of the free energy by multistage sampling.
\newblock \emph{The Journal of Chemical Physics}, 57\penalty0 (12):\penalty0
  5457--5462, 1972.
\newblock \doi{10.1063/1.1678245}.
\newblock URL \url{http://dx.doi.org/10.1063/1.1678245}.

\bibitem[Villani(2008)]{villani2008optimal}
C.~Villani.
\newblock \emph{Optimal transport: old and new}, volume 338.
\newblock Springer Science \& Business Media, 2008.

\bibitem[Williams and Williams(1959)]{williams1959regression}
E.~J. Williams and E.~Williams.
\newblock \emph{Regression analysis}, volume~14.
\newblock wiley New York, 1959.

\bibitem[Wyse(2011)]{jason-wyse-code}
J.~Wyse.
\newblock Estimating the statistical evidence - a review, 2011.
\newblock URL \url{https://sites.google.com/site/jsnwyse/code}.

\bibitem[Zhou et~al.(2015)Zhou, Johansen, and Aston]{zhou2015towards}
Y.~Zhou, A.~M. Johansen, and J.~A. Aston.
\newblock Towards automatic model comparison: an adaptive sequential {M}onte
  {C}arlo approach.
\newblock \emph{Journal of Computational and Graphical Statistics}, \penalty0
  (just-accepted), 2015.

\end{thebibliography}

\end{document}